\documentclass[a4paper,authoryear,1p,round]{elsarticle}
\newcommand{\appendixref}[1]{\ref{#1}}

\usepackage[english]{babel}
\usepackage{amssymb}
\usepackage[dvipsnames]{xcolor}
\usepackage{graphicx}
\usepackage{graphbox} 

\usepackage{mathtools}
\usepackage[font=normalsize]{caption}
\usepackage[font=normalsize,margin=2mm]{subfig}
\usepackage{float}
\usepackage{hyperref}
\hypersetup{colorlinks=true,
	linkcolor=blue,
	filecolor=blue,
	urlcolor=blue,
	citecolor=blue
}
\usepackage{xfrac}
\usepackage{enumitem}
\usepackage{gensymb}
\usepackage{tikz}
\usetikzlibrary{patterns}		
\usepackage{listings} 
\input{listingsEnvirorment.sty}		

\usepackage{calc}

\usepackage[authoryear,round]{natbib}

\newcommand{\mr}{\mathrm}
\newcommand{\mc}{\mathcal}

\renewcommand{\S}{^\mr{s}}
\newcommand{\ii}{\mr{i}\,}
\newcommand{\ee}{\mr{e}}
\let\underscore\_
\renewcommand{\_}[1]{_\mr{#1}}
\let\slashHat\^
\renewcommand{\^}[1]{^\mr{#1}}
\let\Re\relax
\let\Im\relax
\DeclareMathOperator\Re{Re}
\DeclareMathOperator\Im{Im}

\newcommand{\h}{\hat}
\newcommand{\br}[3]{\left#1#2\right#3}
\newcommand{\rbr}[1]{\left(#1\right)}
\newcommand{\sbr}[1]{\left[#1\right]}
\newcommand{\cbr}[1]{\left\{#1\right\}}
\newcommand{\dd}{\mr d}
\newcommand{\ddfrac}[2]{\frac{\dd #1}{\dd #2}}

\newcommand{\mean}[1]{\br\langle{#1}\rangle}

\newcommand{\zzvar}{\bar}
\newcommand{\zzzvar}[1]{\bar{\bar{#1}}}

\newcommand{\zzmap}{\zzzvar f}
\newcommand{\zmap}{\zzvar f}

\newcommand{\zDomain}{\mc Z}
\newcommand{\zBoundary}{\partial\zDomain}

\newcommand{\zSurface}{\zBoundary\^s}

\newcommand{\zzDomain}{\zzvar{\zDomain}}
\newcommand{\zzBoundary}{\partial\zzDomain}

\newcommand{\zzSurface}{\zzBoundary\^s}

\newcommand{\zzzDomain}{\zzzvar{\zDomain}}

\newcommand{\x}{x}
\newcommand{\y}{y}
\newcommand{\z}{z}
\newcommand{\xx}{{\zzvar x}}
\newcommand{\yy}{{\zzvar y}}
\newcommand{\zz}{{\zzvar z}}
\newcommand{\xxx}{{\zzzvar x}}
\newcommand{\yyy}{{\zzzvar y}}
\newcommand{\zzz}{{\zzzvar z}}

\renewcommand{\h}{h}
\renewcommand{\L}{L}
\newcommand{\w}{w}
\newcommand{\W}{W}

\newcommand{ \hh}{{\zzvar \h}}
\newcommand{\LL}{{\zzvar \L}}
\newcommand{\ww}{\zzvar \w}
\newcommand{\WW}{\zzvar \W}

\newcommand{\UU}{\zzvar{\mc U}}%

\newcommand{\hhh}{{\zzzvar \h}}

\newcommand{\www}{\zzzvar \w}
\newcommand{\ppphi}{\zzzvar \phi}
\newcommand{\UUU}{\zzzvar{\mc{U}}}
\newcommand{\etazzOfxxx}{\zzzvar\eta}
\newcommand{\fun}{\mu}

\newcommand{\mus}{\zzzvar \mu\^s}
\newcommand{\muw}{\zzvar \mu\^w}

\newcommand{\FF}{\mc F}
\newcommand{\FFInv}{\FF^{-1}}
\newcommand{\kk}{k} 
\newcommand{\MM}{M} 
\newcommand{\mapC}{\mc C}
\newcommand{\mapS}{\mc S}
\newcommand{\CC}[2]{\sbr{{\mapC_{#1}\!*#2}}}
\renewcommand{\SS}[2]{\sbr{\mapS_{#1}\!*#2}}
\newcommand{\kd}{k\_d}
\newcommand{\kMax}{k\_{max}}
\newcommand{\rDamping}{r}
\newcommand{\N}{n}

\newcommand{\wbl}{d}
\newcommand{\Wbl}{D}
\newcommand{\ffNull}{\zmap_0}
\newcommand{\HH}{\zzvar H}
\newcommand{\DD}{\zzvar \Wbl}

\newcommand{\X}{ X}
\newcommand{\XX}{\zzvar X}

\begin{document}

	\begin{frontmatter}	
		\title{A conformally mapped numerical wave tank supporting piston and flap wavemakers}
		\author{Andreas H. Akselsen}
		\affiliation{organization={SINTEF Ocean, Department of Ship and Ocean Structures},
				addressline={Paul Fjermstads vei 59}, 
				city={Trondheim},
				postcode={7052}, 
				state={Tr{\o}nderlag},
				country={Norway}}
		
	\begin{abstract}
This paper advances the development of the conformally mapped model for accurate simulation of two-dimensional water waves, here with emphasis on mapping boundaries that represent piston- and flap-type wavemakers.
With this, a complete numerical representation of wave flumes is provided---the first of its kind based on conformal mapping. 
The model is validated both theoretically and experimentally, with special attention devoted to wavemaker characteristics and the generation of spurious waves.
It is further demonstrated that the method accurately predicts the spectral evolutions of generated wave fields.
The model is computationally efficient, with beyond real-time computation but for the smallest tested periods, making it ideal for numerical wave calibration and for replicating experiments.
	\end{abstract}
	
	\end{frontmatter}

	\section{Introduction}

Conformal mapping techniques have historically featured in a wide range of hydrodynamic and aerodynamic topics  \citep[e.g,][]{lamb1932hydrodynamics,milneThomson1962theoreticalHydrodynamics}. 
Under the assumption of  two-dimensional potential flows, such approaches offer elegant analytical solutions to yield pressure and velocity fields, along with integral properties such as drags and lifts.
In  recent decades, however, they have largely been substituted by numerically discretised approaches, including CFD. 
Nonetheless, combining conformal mapping with numerical methods presents distinct advantages as it allows all linear aspects of the problem---such as solid boundaries and the fluid interior---can be treated analytically, reserving computational resources solely for the nonlinear free surface.
In contrast, fully discretised approaches, like  finite difference, finite volume or finite element methods, require computation throughout the fluid domain.

Compared to boundary element methods, the conformally mapped approach does not require computation along solid boundaries and that the scheme remains entirely explicit, relying on fast Fourier transformations.
The method bearing closest resemblance is the Higher Order Spectral (HOS) method \citep{west1987_originalHOS,bonnefoy2006A_BM}, but there are key distinctions: the free surface is represented precisely,  without truncating the order of accuracy. 
This ensures precise long-term nonlinear behaviour, regardless of water depths and spectral widths.
Moreover, as will be shown, this framework allows for exact representations of wavemaker motion.
\\

The present paper is a continuation of \citet{AHA2025_CMM}, where a numerical conformal mapping  framework was introduces for modelling two-dimensional surface waves over arbitrarily shaped transient bathymetries in continuous and closed domains.
The model  builds primarily on the work of  \citet{chalikov1996,chalikov2005modeling,chalikov2016Book,chalikov2020}, adopting a different formulation. 
Other works building on similar modelling include \citet{zakharov2002_conformalHOS,ruban2004,ruban2005,viotti2014conformal}.
A key novel feature of the present framework is the intermediate mapping layer introduced to separate all prescribed, pre-defined mapping, such as bathymetries and wall boundaries, from the dynamic mapping of the water surface. 
This separation enables  prescribed motions to be pre-computed and allows the evolution of the free surface to be resolved explicitly during the simulation.
\\

This paper focuses on numerical wave tanks, specifically considering flat-bottomed  flumes equipped with piston- or paddle-type wavemakers.
While often viewed as substitutes for physical experiments, numerical wave tanks offer significant potential as complementary tools in laboratory experiments.
Ideally, such tools will be capable of reproducing, for a given wavemaker signal, the experimental observations down to the detailed level of individual wave crests---so-called \emph{phase-resolved} prediction.
Attempts at achieving this have frequently been made, notably using the Higher-Order Spectral (HOS) method \citep{houtani2018_experimentFromHOS, ducrozet2020_timeReversalExperimentalRouge}. 
Broadly speaking, good agreement has been obtained within certain limits of wave steepness and measurement distance, beyond which the accuracy tends to diminish.
A particularly promising application of phase-resolving numerical wave tanks is extreme event screening, whereby long-duration wave sequences are simulated numerically to identify significant wave events---typically large, steep crests of low probability that may induce extreme structural loads. 
Physical testing can then, as far as feasible, be focused on recreating these pre-screened events.

Even in the absence of precise phase-resolved predictive capability, numerical wave tanks can offer considerable utility. 
One compelling application is \emph{numerical wave calibration}, in which the wave calibration process is simulated rather than executed physically, thereby reducing time spent in the laboratory. 
Wave calibration is usually targeted on power spectral density of surface elevation, termed the  wave spectrum, observed at the location of measurement.

Another application is the estimation of statistical distributions at specific locations in the wave tank. 
This allows for inference of the underlying statistical properties to which any smaller set of experimental seed realisations conform. 
Naturally, both applications require the numerical wave tank to accurately capture the spectral and statistical characteristics of the wave field, but  do not demand exact phase-resolved agreement between simulation and experiment.
\\

The following begins with a summary of the conformally mapped model presented in \citet{AHA2025_CMM}, restricted to flat bathymetries (\autoref{sec:model}). 
\autoref{sec:map:piston} and \ref{sec:map:flap}  detail the conformal mapping of piston- and flap-type wavemakers, respectively.
Numerical validation  is presented in \autoref{sec:validation}, followed by a comparison with laboratory experiments in \autoref{sec:exp}, where the method's potential for event reconstruction and numerical wave calibration is explored.
Concluding remarks and a summary are provided in \autoref{sec:closingRemarks} and \autoref{sec:summary}, respectively.

	\section{The double-layered Conformally Mapped Model}
	\label{sec:model}
	
	\subsection{Mapping description}
	The conformal mapping approach simplifies evaluation of complex  boundaries, such as sketched in \autoref{fig:fSketch:z}.
	Consider  a moving domain boundary, $\zBoundary(t)$, along which we aim to enforce a kinematic condition of impermeability. 
	A general means of doing so is through 
the trajectory of a fluid particle, $z\^p(t)$, that is advected by the flow.
 The condition states that a particle initially on the boundary remains on there:
	\begin{equation}
		z\^p(t) \in  \zBoundary(t) \;\text{ provided }\;z\^p(0) \in  \zBoundary(0).
		\label{eq:kinBC:z}%
	\end{equation}
	While valid, this formulation is impractical to work with directly, motivating the use of conformal mappings
	that can project an arbitrarily complex transient domain into a fixed rectangle $\zzzDomain$ (\autoref{fig:fSketch:zzz}) where \eqref{eq:kinBC:z} can be explicitly evaluated.

Wall and bed boundaries are prescribed by time-dependent coordinates, specified prior to the simulation. 
In contrast, the free surface coordinates are unknown and ever changing.
Mapping both simultaneously results in an implicit mapping problem, which would be computationally expensive to solve at each time step. 
To circumvent this, a \emph{double-layered} mapping is proposed in which prescribed boundaries are mapped to straight lines in an intermediate plane $\zz$, while the free surface is mapped to a straight line in a final rectangular plane $\zzz$; $\z\mapsto\zz\mapsto\zzz$ .
For clarity, we introduce the mapping functions $\zmap$ and $\zzmap$,
	\begin{equation}
		\z = \zmap(\zz,t),\qquad \zz = \zzmap(\zzz,t),
	\end{equation}
	and the corresponding conversion of marking functions that take $\zz$ and $\zzz$-coordinate arguments with the same single and double bar notation, respectively. 
	Simulation is performed entirely from the $\zzz$-plane, not requiring inverse mapping and allowing the procedure to remain explicit.

	\begin{figure}[H]
		\centering
		\subfloat[Physical $\z$-plane.]{
%
%
%
%
%


\begin{tikzpicture}[scale=.75]

	%
	%
	
	\def\L{10}
	\def\hd{1/4*\L} 
	\def\tWidth{.025*\L} 
	\def\d{.5*\hd}
	\def\LDelta{.75*\d} 
	
	\def\LL{1.75*\d}	
	\def\hX{.25*\LDelta+.25*\d}
	
	\def\lam{5} 
	\def\wy{.035*\lam} 
	\def\wx{.25*\lam} 

	\def\xClip{.4*\hd}
	\clip (-\xClip,\LDelta) rectangle (\L+\tWidth,-\hd-\tWidth);


\draw[thick]  (0,-\d)--(0,-\hd)   --(\L,-\hd)--(\L,{\LDelta});
\fill[pattern=north east lines, pattern color=black]  (0,-\hd)--(\L,-\hd)--(\L,\LDelta)--(\L+\tWidth,\LDelta)--(\L+\tWidth,-\hd-\tWidth) --(-\tWidth,-\hd-\tWidth)--
(-\tWidth,-.5*\hd)--	(0,-.5*\hd);

	\draw[dashed] (-\xClip,0)--(\L,0) node[anchor=base west] at (-\xClip,0) {$\z=\x$};
\draw[dotted] (0,\LDelta)--(\L,\LDelta);
\draw[dotted] (0,-\d)--(\L,-\d);

\begin{scope}
	\clip  (.4*\d,-\d) rectangle (\L,3*\wy);
	\draw (0,2.5*\wy) cos +(\wx,-.75*\wy)  sin +(\wx,-.75*\wy) cos +(\wx,.75*\wy) sin +(\wx,.75*\wy) cos +(\wx,-\wy) sin +(\wx,-1.2*\wy) cos +(\wx,1.2*\wy) sin +(\wx,\wy);
\end{scope}
\node[anchor=base west] at (.6*\L,2.*\wy) {$\zSurface(t)$};

\def\sinth{0.3420}
\def\costh{0.9397}
\def\qq{.2}
\draw[thick,fill=gray] (0,-\d)--++(\LL*\sinth,\LL*\costh) --++(-\costh*\qq,\sinth*\qq)--++(-\LL*\sinth,-\LL*\costh) ;
\draw[thick,fill=gray] (-\qq/2,-\d) circle (\qq);

	\draw[<->] (-.55*\d,.4*\d) arc (180-45:45:.75*\d);

	\node [anchor=base west] at (\L/2,-\hd) {$\z=\x-\ii\h$} ;
	
	\draw [<->] (.3*\L,-\d)--(.3*\L,0)node[midway,anchor=west] {$\wbl$};
	\draw[<->]  (.4*\L,0)-- (.4*\L,-\hd) node[midway,anchor=west] {$\h$};
	\node[anchor=center] at (.6*\L,-.4*\hd) {$\zDomain(t)$};
	\draw [<->] (.2*\L,\LDelta)--(.2*\L,0)node[midway,anchor=west] {$\Delta$};
	
	\node[anchor=west] (eq) at (0,-.5*\d-.5*\hd) {$\z=\X(\y)+\ii\y$};
	\path[->] (eq.north) edge[out=90, in=-20] (.4,-.5*\d);

\end{tikzpicture}%


\label{fig:fSketch:z}}\\
		\subfloat[Prescribed intermediate $\zz$-plane.]{
%
%
%
%
%
%
\begin{tikzpicture}[scale=.75]
	
	\def\L{10} 
	\def\lam{5} 
	\def\hd{1/4*\L}
	\def\Lh{1/2*\L} 
	\def\wy{.05*\Lh} 
	\def\wx{.25*\lam} 
	\def\tWidth{.05*\Lh} 
	
		\def\d{.5*\hd}	\def\LDelta{.75*\d} \def\xClip{.4*\hd}
	\clip (-\xClip,\LDelta) rectangle (\L+\tWidth,-\hd-\tWidth);
	
	\draw[thick]  (0,{1.5*\wy})--(0,-\hd)--(2*\Lh,-\hd)--(2*\Lh,{1.5*\wy});
	\fill[pattern=north east lines, pattern color=black]  (0,-\hd)--(2*\Lh,-\hd)--(2*\Lh,1.5*\wy)--(2*\Lh+\tWidth,1.5*\wy)--(2*\Lh+\tWidth,-\hd-\tWidth)--(-\tWidth,-\hd-\tWidth)--(-\tWidth,1.5*\wy)--(0,1.5*\wy);

	\draw[dashed] (0,0)--(2*\Lh,0);

	\def\wxh{\lam/2}
	\draw[<->] (.4*\L,0)--(.4*\L,-\hd) node[midway,anchor=west]  {$\hh$};
		\node[anchor=base west] at (.6*\L,\wy) {$\zzSurface(t)$};
	
	\node[anchor=center] at(.6*\L,-.4*\hd)  {$\zzDomain(t)$};
	
	\draw (0,+\wy) cos +(\wx,-.9*\wy)  sin +(\wx,-\wy) cos +(\wx,\wy) sin +(\wx,+\wy) cos +(\wx,-.75*\wy) sin +(\wx,-.75*\wy) cos +(\wx,.75*\wy) sin +(\wx,.75*\wy);
	\draw[dotted] (0,0) (2*\Lh,0);
	
	\node[anchor=base west] at (\Lh,-\hd) {$\zz=\xx-\ii\hh$}; 
	\node[anchor=base west] (eq) at (0,-.75*\hd) {$\zz=\xx\^w_1+\ii\yy$}; \path[->] (eq.north) edge[out=100, in=-10] (.2,-.3*\hd);

\end{tikzpicture}%
%
%
\label{fig:fSketch:zz}}\hfill%
		\subfloat[Rectangular $\zzz$-plane.]{
%
%
%
%


\begin{tikzpicture}[scale=.75]
	
	\def\L{10} 
	\def\lam{5} 
	\def\hd{1/4*\L}
	\def\Lh{1/2*\L} 
	\def\wy{.07*\Lh} 
	\def\wx{.25*\lam} 
	\def\tWidth{.05*\Lh} 
	
	\def\d{.5*\hd}	\def\LDelta{.75*\d} \def\xClip{.4*\hd}
	\clip (-\xClip,\LDelta) rectangle (\L+\tWidth,-\hd-\tWidth);

	\draw[thick]  (0,{1.5*\wy})--(0,-\hd)--(2*\Lh,-\hd)--(2*\Lh,{1.5*\wy});
	\fill[pattern=north east lines, pattern color=black]  (0,-\hd)--(2*\Lh,-\hd)--(2*\Lh,1.5*\wy)--(2*\Lh+\tWidth,1.5*\wy)--(2*\Lh+\tWidth,-\hd-\tWidth)--(-\tWidth,-\hd-\tWidth)--(-\tWidth,1.5*\wy)--(0,1.5*\wy);

	\draw (0,0)--(2*\Lh,0);

	\def\wxh{\lam/2}
	\draw[<->] (.4*\L,0)--(.4*\L,-\hd) node[midway, anchor=west] {$\hhh(t)$};	
	\node[anchor=center] at(.6*\L,-.4*\hd)  {$\zzzDomain(t)$};

	\node[anchor =base west] at (\Lh,0) {$\zzz = \xxx$};
	\node[anchor =base west] at (\Lh,-\hd) {$\zzz = \xxx-\ii\hhh$};

\end{tikzpicture}%


\label{fig:fSketch:zzz}}%
		\caption{Sketch of the $\zDomain$, $\zzDomain$ and $\zzzDomain$-domains.
		}
		\label{fig:fSketch}
	\end{figure}

	Fluid velocities are represented using the complex potential  $\w = \phi+\ii\psi$, $\phi$ being the fluid velocity potential and $\psi$ its harmonic conjugate, the stream function. 
	By virtue of conformality, these functions remain solutions of the Laplace equation after the mapping 
	\begin{equation}
		\ww(\zz,t)=\w[\zmap(\zz,t),t], \qquad \www(\zzz,t)=\ww[\zzmap(\zzz,t),t].
		\label{eq:w}
	\end{equation}
	In anticipation of wall conditions to come, we also introduce an arbitrary 
	pre-determined background potential  $\W(\z,t)$, the total potential then being  $\w+\W$. 
	The  background potential is imposed in the $\zz$-plane, where wall boundaries are steady and the mapping predetermined; $\WW(\zz,t)=\W[\zmap(\zz,t),t]$.

	\subsection{Model}
	\label{sec:model:core}
	The kinematic boundary condition \eqref{eq:kinBC:z} can now be reformulated in the  rectangular $\zzz$-plane where evaluation is more convenient.
	The fluid particle trajectory  $\z\^p(t)$ in \eqref{eq:kinBC:z}   maps directly between the different coordinate systems:
	\begin{equation}
		\z\^p(t) = \zmap[\zz\^p(t),t], \qquad	\zz\^p(t) = \zzmap[\zzz\^p(t),t].
		\label{eq:zp}
	\end{equation}
	This leads us to the notion of a particle velocity $\mc U$ as viewed in the $\z$-plane, and  \textit{apparent particle velocities} $\UU$ and $\UUU$ as viewed in the intermediate $\zz$  and final $\zzz$-plane, respectively.
	Differentiation yields 
	\begin{subequations}
		\begin{align}	
			\mc U &\equiv	\ddfrac{\z\^p}{t} = \w_\z^* + \W_\z^*,\\
			\UU &\equiv \ddfrac{\zz\^p}{t} = 	\frac{\ww_\zz^*+\WW_\zz^*}{|\zmap_\zz|^{2}}-\frac{\zmap_t}{\zmap_\zz},\label{eq:UU}\\
			\UUU &\equiv \ddfrac{\zzz\^p}{t} =	\frac{\www_\zzz^*}{ |\zzmap_\zzz|^2   |\zmap_\zz|^2}  +  \frac{1}{\zzmap_\zzz}\rbr{\frac{\WW_\zz^*}{|\zmap_\zz|^2}-\frac{\zmap_t}{\zmap_\zz}-\zzmap_t  },	\label{eq:UUU}
		\end{align}%
	\end{subequations}%
	asterisk denoting complex conjugation.
	In the $\zzz$-plane, kinematic boundary condition  \eqref{eq:kinBC:z}  
	simplifies to 
	\begin{subequations}
		\begin{alignat}{2}
			\Im\,\UUU(\xxx,t) &= 0,	
			&\quad
			\Im\,\UUU(\xxx-\ii\hhh,t) &= -\hhh_t,	
			\label{eq:ImUUU}
\\
			\Re\,\UUU(\xxx\_1\^w+\ii \yyy,t) &=0,
			& \quad 
			\Re\,\UUU(\xxx_2\^w+\ii \yyy,t) &= 0.
			\label{eq:ReUUU} 	
		\end{alignat}%
		\label{eq:BCUUU}%
	\end{subequations}%
	We further  have that $\zzmap_t$ is real at the bed and  pure imaginary at walls, and that $\zzmap_\zzz$ is real along both bed and walls.

	Assuming a flat, stationary bathymetry, the kinematic condition \eqref{eq:ImUUU}, evaluated along  the free surface and the bed,
	leads to the following conditions that describe the velocity potential and the evolution of the surface mapping: 
\begin{subequations}
		\begin{alignat}{2}
			\Im\rbr{\frac{\zzmap_t}{\zzmap_\zzz}}&=  -|\zzmap_\zzz\zmap_\zz|^{-2} \Im\sbr{ \www_\zzz +  \zzmap_\zzz\rbr{\WW_\zz- \zmap_\zz\zmap_t^* } }    \ \equiv \mus(\xxx,t) \qquad &\text{at } \yyy&=0, \label{eq:zzzEvolution}   \\
		\qquad\Im\rbr{\frac{\zzmap_t}{\zzmap_\zzz}} &= \hhh_t(t) \qquad &\text{at } \yyy&=-\hhh, \label{eq:zzzEvolution_h}  \\
		\ppphi &= \ppphi\^s(x,t)  &\text{at } \yyy&=0,\label{eq:ppphi0} \\
		\ppphi_\yyy &=0 \qquad &\text{at } \yyy&=-\hhh. \label{eq:ppphiyyy_h} 
	\end{alignat}%
	\label{eq:horizontalConditions}%
\end{subequations}%
	Here, $\ppphi\^s=\ppphi(\xxx,0,t) = \Re\www(\xxx,t)$ is the surface potential obtained form the previous time step.

	To satisfy \eqref{eq:horizontalConditions}, the following projection kernels are introduced: 
	\begin{subequations}
		\begin{align}
			\CC{h}{\fun}(z) &
			= \sum_{j=-\MM}^\MM \FF_{\!j}(\fun) \frac{\ee^{\ii \kk_j(z+\ii h)}}{\cosh(\kk_jh)} 
			=  \FF_{\!i}^{-1} \sbr{ \FF_{\!j}(\fun) \frac{2\, \ee^{- \kk_j y}}{1+\ee^{2\kk_jh}}},
			\label{eq:mapC}\\
			\SS{h}{\fun}(z) &
			= \sum_{j=-\MM}^\MM \FF_{\!j}(\fun) \frac{\ee^{\ii \kk_j(z+\ii h)}}{\delta_j-\sinh(\kk_jh)} 
			=  \FF_{\!i}^{-1} \sbr{\FF_{\!j}(\fun) \frac{2\,\ee^{-\kk_j y}}{1-\ee^{2\kk_jh}+2\delta_j}},
		\end{align}%
		\label{eq:mapSC}%
	\end{subequations}%
		$  \FF_{\!j}\,(\fun) = \frac{1}{\N} \sum_{i=1}^\N \fun(x_i,t) \ee^{-\ii k_j x_i}$  being the discrete Fourier transform operator, 
	while  $\delta_j$  equals one if $j=0$ and zero otherwise (see \autoref{list:CS}). 
	Assuming $\fun(x)$ is real, these kernels have the following properties along an upper and lower line:%
	\begin{subequations}
		\begin{align}
			\Re \CC{h}{\fun}(x) &= \Re  \SS{h}{\fun}(x) =  \fun(x),  \\
			\Im \CC{h}{\fun}(x-\ii h) &= 0,\\
			\Re	\SS{h}{\fun}(x-\ii h) &=\mean{\fun},\label{eq:LProps:Sh}%
		\end{align}%
		\label{eq:LProps}%
	\end{subequations}%
	$\mean{\fun} = \FF_{\!0}(\fun)$ being the spatial mean. In the limit $h\to\infty$, both $\mapC$ and $\mapS$ gives a conjugate pair related through the Hilbert transform (the so-called  analytical signal).
	
		Using  \eqref{eq:mapSC}, the solutions to \eqref{eq:horizontalConditions} is
		\begin{subequations}
			\begin{align}
			\zzmap_t& = 	\ii  \zzmap_\zzz \SS{\hhh}{\mus}(\zzz),	\label{eq:zzztMap} \\ 
			\www &= \CC{\hhh}{\ppphi\^s}(\zzz).	\label{eq:www}%
		\end{align}%
		\label{eq:ffft}%
	\end{subequations}%
		Following \citet{chalikov2005modeling}, we integrate only the vertical components 
			\begin{equation}
			\etazzOfxxx_t(\xxx,t)=\Im\cbr{ \zzmap_t(\xxx,t)}
			\label{eq:etat}
		\end{equation}
		in time and reconstruct $\zzmap(\xxx,t)$ at the next time level with the projection 
					\begin{equation}
			\zzmap(\zzz,t) = \zzz + \ii \SS{\hhh}{\etazzOfxxx}(\zzz).
			\label{eq:zzzMap}
		\end{equation}
				The kernel property \eqref{eq:LProps:Sh} now reveals  the map depth to be
		\begin{equation}
			\hhh =  \hh + \mean{\etazzOfxxx},
			\label{eq:hhh}%
		\end{equation}
		and one finds that  \eqref{eq:zzzEvolution_h} is consistent with  projection \eqref{eq:zzztMap} and that  
		\begin{align}
			 \hhh_t = 	\mean{\mus}  =  \mean{\etazzOfxxx_t}.
		\end{align}

			Having now established surface coordinates and velocities, the final step is to impose the dynamic boundary condition.
			In the  $\z$-plane, this is given the Bernoulli equation 
			evaluated along  $\zSurface$ where $p=0$.
		It is straightforward to rewrite this in terms of $\zzz$-variables:
			\begin{align}
				\ppphi_t = \Re\sbr{\frac{\www_\zzz}{\zzmap_\zzz}\rbr{\zzmap_t+\frac{\zmap_t}{\zmap_\zz}}  -  \WW_t+\WW_\zz\frac{\zmap_t}{\zmap_\zz}} - \frac12\br|{\frac{\www_\zzz}{\zmap_\zz \zzmap_\zzz}+\frac{\WW_\zz}{\zmap_\zz}}|^2 - g\y, 
				\label{eq:BC_dyn_zzz}
			\end{align}
			which is to eb evaluated along $\yyy=0$ with coordinates mapped accordingly. 
			Time integration is here carried out using a dynamic time-stepping ODE solver on $\ppphi_t\S$ and $\etazzOfxxx_t$.
			One can alternatively march the corresponding Fourier components forwards in time, which alleviates some Fourier operations but gives a different time stepping response from the ODE solver.

		More details are presented in \citet{AHA2025_CMM}, including a code example illustrating how the model may be implemented.
		It should be noted that equations \eqref{eq:ffft}, \eqref{eq:zzzMap} and \eqref{eq:BC_dyn_zzz} reduces to the Dyachenko equations in traditional variables \citep{dyachenkoZakharov1996_confMap,Dyachenko_2019_confMap,zakharov2002_conformalHOS,zakharov2006_conformalHOS} when omitting the intermediate mapping and assuming deep water.

			\subsection{Wall boundaries}
		Wall impermeability is enforced by horizontally mirroring the domain. 
		A variable $\fun$,  described along positions $\{\x_i\}$ through a vector $[\fun_1,\fun_2,\ldots,\fun_\N]$, is mirrored by extending it to 
		\begin{equation}
			[\fun_i] \coloneqq  [\fun_1,\ldots,\fun_{ \N},\fun_{ \N-1},\ldots,\fun_2].
			\label{eq:mirrorWall}
		\end{equation}
		This effectively enforces $\fun_x=0$ at $x=x_1$ and $x_n$.
		Similar to a padding routine, one can mirror the inputs to every discrete Fourier transformations and then  remove the mirrored half after performing the subsequent inverse transformation. 

		As mirroring enforces $\ppphi_\xxx=0$ along the lateral boundaries, 
		the background potential $\WW$ can be pre-computed to satisfy condition \eqref{eq:ReUUU}.
	 From condition \eqref{eq:UU}, we find
		\begin{subequations}
			\begin{equation}
			\Re\WW_\zz = \Re(\zmap_\zz\zmap_t^*) \equiv \muw_1 \text{ and } \muw_2
				\label{eq:ReWWzz}
		\end{equation}
		along the two walls $\xx=\xx_1\^w$ and $\xx_2\^w$, respectively. 
		A last condition on $\WW$, implicitly assumed in \eqref{eq:ppphiyyy_h}, is
		\begin{equation}
			\Im\WW_\zz = 0  \text{ along } \yy=-\hh.
				\label{eq:ImWWzz}%
		\end{equation}%
		\label{eq:conditionsWW}%
		\end{subequations}%
		When mapping a basin wavemaker, one must  first find a map $\zmap(\zz,t)$ that matches the wavemaker profile while maintaining the assigned  basin length and depth. 
		Then, a similar  function $\WW(\zz,t)$ must be determined from \eqref{eq:conditionsWW}. 
		Both $\zmap$ and $\WW$ are prescribed only along the wall and bed boundaries, meaning that they are not uniquely defined and may be assigned according to preference.
		In contrast, the total potential $\ww+\WW$ is prescribed along a closed boundary and will therefore be unique.

	\subsection{Stabilisation, anti-aliasing and numerical beaches}
	As with other spectral methods, stabilisation measures are necessary to suppress the buildup of high-frequency noise.
	Following \citet{chalikov2005modeling}, 
	we adopt the modal damping approach
	\begin{equation}
		\etazzOfxxx_t \coloneqq	\FFInv\sbr{ \FF(\etazzOfxxx_t) -\hat \nu \FF(\eta)}, 
		\qquad
		\ppphi\S_t \coloneqq 	\FFInv\sbr{  \FF(\ppphi\S_t) -\hat \nu	\FF(\ppphi\S) }
		\label{eq:damping}%
	\end{equation}%
	with  damping coefficient
	\begin{equation}
		\hat\nu(k) = 	rM\, \sqrt{\frac{2\pi g}{\zzzvar{L}}}  \rbr{\frac{\max(|k|-\kd,0)}{\kMax-\kd}}^2.
		\label{eq:damping_nu}
	\end{equation}
	Here,  $\kd$ is an intermediate wavenumber smaller than the largest wavenumber $\kMax\approx \pi \N/\zzzvar{L}$,  $\zzzvar{L}$ being the horizontal length of the $\zzzDomain$-domain.
	The constant $M$, which was neglected in the expression given in \citet{AHA2025_CMM}, is the number of Fourier modes.
	\\

	Anti-aliasing is needed to avoid energy folding. 
	Because the model nonlinearity is of `infinite order', aliasing cannot be formally eliminated. 
	However,   \citet{chalikov2005modeling} observed that simulations become insensitive to additional zero-padding beyond cubic nonlinearity (i.e., adding zero-modes above wavenumber $3\,\kMax$).
	Simulations should be monitored for aliasing artifacts to assess the need for zero-padding. 
	In many cases, such as those presented here, energy is concentrated in the lower wavenumber range, and folding has little or no impact on the results.
	\\

A \textit{numerical beach}, which emulates a radiation boundary, is also essential in a numerical wave tank. 
	An approximate variant of the absorption layer method of \citet{bonnefoy2010}, 
	\begin{align}
		\ppphi_t &\coloneq  \ppphi_t - \zzzvar\nu(\xxx) \y\^s_t,  &  
		\y\^s_t &\approx \Im \rbr{ \zmap_t + \zmap_\zz\zzmap_t }, & 
		\zzzvar\nu & =  \nu_0 u^2(3-2u);\; u = \max(\x-\x\_b,0)/L\_b,
		\label{eq:beach}
	\end{align}
	is here adopted. Absorption intensity $\nu_0$ and beach length $L\_b$ should be chosen such that reflections are sufficiently suppressed or such that the numerical beach resembles a physical beach.

	\section{Mapping piston wavemakers}
	\label{sec:map:piston}
	The piston wavemaker can be represented explicitly with a mapping that expands and contracts about the far corner  $\z=\LL-\ii \h$:
		\begin{equation}
			\zmap(\zz,t) = \zz + \rbr{1-\frac{\zz+\ii\hh}{\LL}} X(t);
			\qquad
			X(t)\in\mathbb R.
			\label{eq:BM2:zmap}
		\end{equation}
		Here,  $\hh$ and $\LL$ respectively equals the water depth $h$ and the wave tank length $\L$ when the wavemaker is in its neutral position $X=0$. 
		The map expansion factor is
		\begin{equation}
			\alpha(t) = 1-\frac{X(t)}{\LL} = \zmap_\zz,
		\end{equation}
		and the background velocity, satisfying  \eqref{eq:UU},  becomes
		\begin{equation}
			\WW_\zz =\rbr{1-\frac{\zz+\ii\hh}{\LL}}  \alpha X_t.
			\label{eq:BM2:WW}
		\end{equation}
		If inserted back into the boundary condition \eqref{eq:zzzEvolution}, one finds
		\begin{equation}
			\mus= \sbr{\frac{\ppphi_\yyy}{ \alpha ^2 |\zzmap_\zzz|^2 }  + 2\frac{\yy+\hh}{\alpha \LL}   \frac{\Re \zzmap_\zzz}{ |\zzmap_\zzz|^2 } X_t}_{\yyy=0}.
			\label{eq:BM2:mu}
		\end{equation}

\section{Mapping flap wavemakers}
	\label{sec:map:flap}
	\subsection{Flap wavemaker using projection kernels \eqref{eq:mapSC}}
	\label{sec:map:flapProjKernel}
	Two methods have been developed for mapping the paddle wavemakers. One, presented in \appendixref{sec:map:SC},  is  based on the Schwarz-Christoffel transformation theorem; the other, presented below, relies on the projection kernels from the previous section.

	The projection kernels \eqref{eq:mapSC} can be  rotated 90 degrees by using $-\ii \zz$ as coordinate argument, understanding that Fourier transformation is carried out along the real input---now aligned with the $\y$-axis. 
	In the rotated frame, mirroring as described in \eqref{eq:mirrorWall} is again employed, this time to enforce a flat bed analogue to the vertical walls in the $\zzz$-plane.
	For orientation, we begin by considering a domain between $\zz=\xx$ and  $\zz=\xx-\ii\HH$, which the mirroring extends up to $\zz=\xx+\ii\HH$. 
	The mapping function 
		\begin{equation}
		\ffNull(\zz,t) = \alpha \zz + \SS{\LL}{\XX}(-\ii \zz)
		\label{eq:ffPadleNull}%
	\end{equation}
	thereby obeys
		\begin{align*}
		\Re \ffNull(\ii\yy,t) &= \X(\yy,t),&
		\Re \ffNull(\LL+\ii\yy,t) &= L,&
		\Im \ffNull(\xx,t) &= 0, &
		\Im \ffNull(\xx-\ii \HH,t) &= -H.
	\end{align*}
Iteration is required on $\XX(\yy,t)$ since it is a function of $\yy$ rather than $\y$. 
	As with the undulating bathymetry mapping described in \citet{AHA2025_CMM}, a straightforward fixed-point iteration scheme is here adopted. 

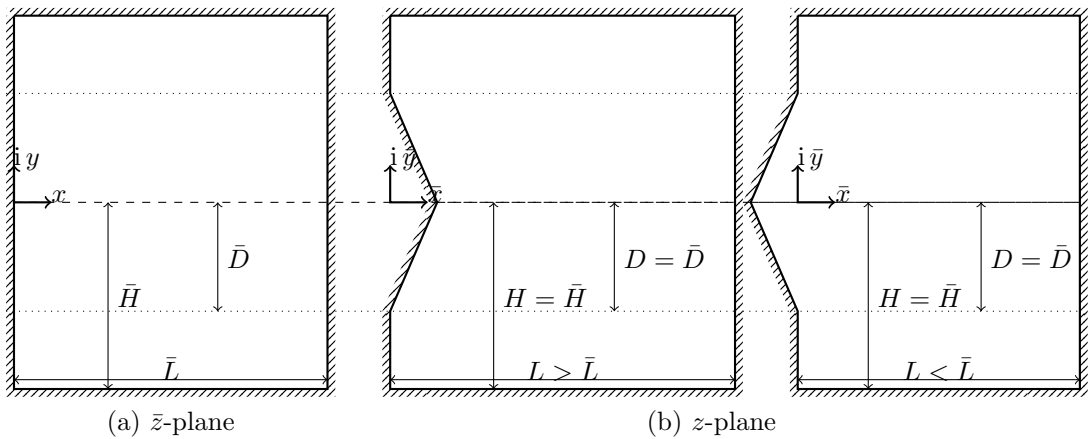
\begin{figure}[H]
		\centering
		\subfloat[$\zz$-plane]{
%
%
%
%
\begin{tikzpicture}[scale=.75]
	
	\def\LL{5.5}
	\def\HH{.6*\LL}
	\def\LLDelta{.25*\HH} 
	\def\tWidth{.025*\LL} 

	\def\HminusD{.25*\LL}
	\def\D{\HH-\HminusD}
	\def\mD{-\HH+\HminusD}
	
		\def\cropHeight{\HH+\tWidth}
	\clip (-\tWidth,\cropHeight) rectangle (\LL+\tWidth,-\HH-\tWidth);

	\draw[thick]   (0,\HH) -- (0,-\HH)   --(\LL,-\HH)--(\LL,\HH)--(0,\HH);
	\fill[pattern=north east lines, pattern color=black]   (0,\HH) -- (0,-\HH)   --(\LL,-\HH)-- (\LL,\HH) -- (0,\HH) -- (-\tWidth,\HH+\tWidth) -- (\LL+\tWidth,\HH+\tWidth) --(\LL+\tWidth,-\HH-\tWidth) --(0-\tWidth,-\HH-\tWidth) --  (0-\tWidth,\HH);
	
	\draw[dashed] (0,0)--(\LL,0) ;
	\draw[dotted] (-\tWidth,\mD)--(\LL,\mD);
	\draw[dotted] (-\tWidth,\D)--(\LL,+\D);
	
	\draw [<->] (.3*\LL,0)--(.3*\LL,-\HH)node[midway,anchor=west] {$\zzvar H$};
\draw [<->] (.65*\LL,0)--(.65*\LL,\mD)node[midway,anchor=west] {$\zzvar D$};
	\draw [<->] (0,.05*\HH-\HH)--(\LL,.05*\HH-\HH)node[midway,anchor=base] {$\zzvar L$};

\def\Lax{.2*\HH}
\draw [thick,<->] (\Lax,0) node[anchor=text] {$\x$} --(0,0)-- (0,\Lax) node[anchor=text] {$\ii\y$} ;

\end{tikzpicture}%
%
%
%
}%
		\subfloat[$\z$-plane]{
%
%
%
%
\begin{tikzpicture}[scale=.75]
	
	
	\def\LL{5.5}
	\def\HH{.6*\LL}
	
	\def\H{\HH}
	\def\L{1.1*\LL}

	\def\tWidth{.025*\LL} 
	\def\xClip{.4*\H}

	\def\HminusD{.25*\LL}
	\def\D{\H-\HminusD}
	\def\mD{-\H+\HminusD}
	\def\hX{.25*\H}

		\def\cropHeight{\HH+\tWidth}
	\clip (-\tWidth-\hX,\cropHeight) rectangle (\L+\tWidth,-\H-\tWidth);

	\draw[thick]   (0,\H) -- (0,\D) --  (\hX,0)-- (0,\mD)--(0,-\H)   --(\L,-\H)--(\L,\H)--(0,\H);
	\fill[pattern=north east lines, pattern color=black]   (0,\H) -- (0,\D) --  (\hX,0)-- (0,\mD)--(0,-\H)   --(\L,-\H)-- (\L,\H) -- (0,\H)-- (-\tWidth,\H+\tWidth) -- (\L+\tWidth,\H+\tWidth) --(\L+\tWidth,-\H-\tWidth) --(0-\tWidth,-\H-\tWidth) -- (0-\tWidth,\mD)  -- (\hX-\tWidth,0) -- (0-\tWidth,\D)-- (0-\tWidth,\H);

	\draw[dashed] (\hX,0)--(\L,0) ;
	\draw[dotted] (-\tWidth-\hX,\mD)--(\L,\mD);
	\draw[dotted] (-\tWidth-\hX,+\D)--(\L,+\D);
	\draw[dashed] (-\tWidth-\hX,\HH-\H)--(\L,\HH-\H) ;

	\draw [<->] (.3*\L,0)--(.3*\L,-\H)node[midway,anchor=west] {$H=\zzvar H$};
	\draw [<->] (.65*\L,0)--(.65*\L,\mD)node[midway,anchor=west] {$D=\zzvar D$};
	\draw [<->] (0,.05*\HH-\H)--(\L,.05*\HH-\H)node[midway,anchor=base] {$L>\zzvar L$};

\def\Lax{.2*\HH}
\draw [thick,<->] (\Lax,0) node[anchor=text] {$\xx$} --(0,0)-- (0,\Lax) node[anchor=text] {$\ii\yy$} ;

\end{tikzpicture}%
%
%
%
%
%
%
%
\begin{tikzpicture}[scale=.75]
	
	
	\def\LL{5.5}
	\def\HH{.6*\LL}
	\def\H{\HH}
	\def\L{.9*\LL}

	\def\tWidth{.025*\LL} 
	\def\xClip{.4*\H}

	\def\HminusD{.25*\LL}
	\def\D{\H-\HminusD}
	\def\mD{-\H+\HminusD}
	\def\hX{-.25*\H}

		\def\cropHeight{\HH+\tWidth}
	\clip (-\tWidth+\hX,\cropHeight) rectangle (\L+\tWidth,-\H-\tWidth);

	\draw[thick]   (0,\H) -- (0,\D) --  (\hX,0)-- (0,\mD)--(0,-\H)   --(\L,-\H)--(\L,\H)--(0,\H);
	\fill[pattern=north east lines, pattern color=black]   (0,\H) -- (0,\D) --  (\hX,0)-- (0,\mD)--(0,-\H)   --(\L,-\H)-- (\L,\H) -- (0,\H)-- (-\tWidth,\H+\tWidth) -- (\L+\tWidth,\H+\tWidth) --(\L+\tWidth,-\H-\tWidth) --(0-\tWidth,-\H-\tWidth) -- (0-\tWidth,\mD)  -- (\hX-\tWidth,0) -- (0-\tWidth,\D)-- (0-\tWidth,\H);
	
	\draw[dashed] (\hX,0)--(\L,0) ;
	\draw[dotted] (-\tWidth+\hX,\mD)--(\L,\mD);
	\draw[dotted] (-\tWidth+\hX,+\D)--(\L,+\D);
	\draw[dashed] (-\tWidth+\hX,\HH-\H)--(\L,\HH-\H) ;
	
	\draw [<->] (.25*\L,0)--(.25*\L,-\H)node[midway,anchor=west] {$H=\zzvar H$};
	\draw [<->] (.65*\L,0)--(.65*\L,\mD)node[midway,anchor=west] {$D=\zzvar D$};
	\draw [<->] (0,.05*\HH-\H)--(\L,.05*\HH-\H)node[midway,anchor=base] {$L<\zzvar L$};

\def\Lax{.2*\HH}
\draw [thick,<->] (\Lax,0) node[anchor=text] {$\xx$} --(0,0)-- (0,\Lax) node[anchor=text] {$\ii\yy$} ;

\end{tikzpicture}%
%
%
%
}%
		\caption{Simplified map setting $\alpha = 1$---map length varies while depths remain fixed.}
		\label{fig:paddleBasic}
	\end{figure}

	The length and depth of the $\zzDomain$ domain are assumed fixed in the model described in \autoref{sec:model}, and
	the expansion factor  $\alpha$ is again introduced to ensure that the right boundary remains stationary. 
	Setting $\alpha = 1$  yields a fully functional wavemaker model, but one in which the domain length $L$ changes with the paddle angle according to 
	$L=\LL +  \mean{\XX}$ (property \eqref{eq:LProps:Sh}).
	This behaviour, illustrated in \autoref{fig:paddleBasic}, leads to a piston-like motion of the far wall, generating unwanted waves that must be suppressed by a numerical beach.

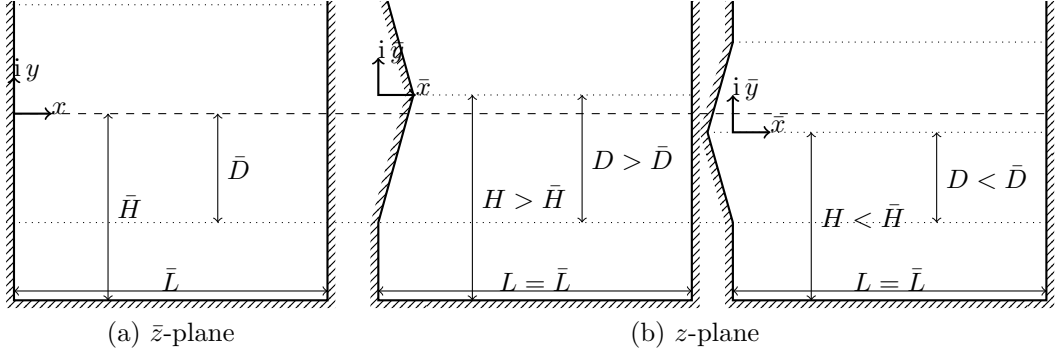
\begin{figure}[H]
		\centering
		\subfloat[$\zz$-plane]{
%
%
%
%
\begin{tikzpicture}[scale=.75]
	
	\def\LL{5.5}
	\def\HH{.6*\LL}
	\def\LLDelta{.25*\HH} 
	\def\tWidth{.025*\LL} 

	\def\HminusD{.25*\LL}
	\def\D{\HH-\HminusD}
	\def\mD{-\HH+\HminusD}
	
		\def\cropHeight{.6*\HH}
	\clip (-\tWidth,\cropHeight) rectangle (\LL+\tWidth,-\HH-\tWidth);

	\draw[thick]   (0,\HH) -- (0,-\HH)   --(\LL,-\HH)--(\LL,\HH);
	\fill[pattern=north east lines, pattern color=black]   (0,\HH) -- (0,-\HH)   --(\LL,-\HH)-- (\LL,\HH) -- (\LL+\tWidth,\HH) --(\LL+\tWidth,-\HH-\tWidth) --(0-\tWidth,-\HH-\tWidth) --  (0-\tWidth,\HH);
	
	\draw[dashed] (0,0)--(\LL,0) ;
		\draw[dotted] (-\tWidth,\mD)--(\LL,\mD);
		\draw[dotted] (-\tWidth,\D)--(\LL,\D);
	
	\draw [<->] (.3*\LL,0)--(.3*\LL,-\HH)node[midway,anchor=west] {$\zzvar H$};

\draw [<->] (.65*\LL,0)--(.65*\LL,\mD)node[midway,anchor=west] {$\zzvar D$};
	\draw [<->] (0,.05*\HH-\HH)--(\LL,.05*\HH-\HH)node[midway,anchor=base] {$\zzvar L$};

\def\Lax{.2*\HH}
\draw [thick,<->] (\Lax,0) node[anchor=text] {$\x$} --(0,0)-- (0,\Lax) node[anchor=text] {$\ii\y$} ;

\end{tikzpicture}%
%
%
%
}%
		\subfloat[$\z$-plane]{
%
%
%
%
\begin{tikzpicture}[scale=.75]
	
	
	\def\L{5.5}
	\def\HH{.6*\L}
	\def\H{1.1*\HH}

	\def\tWidth{.025*\L} 
	\def\xClip{.4*\H}

	\def\HminusD{.25*\L}
	\def\D{\H-\HminusD}
	\def\mD{-\H+\HminusD}
	\def\hX{.17*\H}

		\def\cropHeight{.6*\HH}
	\clip (-\tWidth-\hX,\cropHeight - \H+\HH) rectangle (\L+\tWidth,-\H-\tWidth);

	\draw[thick]   (0,\H) -- (0,\D) --  (\hX,0)-- (0,\mD)--(0,-\H)   --(\L,-\H)--(\L,\H);
	\fill[pattern=north east lines, pattern color=black]   (0,\H) -- (0,\D) --  (\hX,0)-- (0,\mD)--(0,-\H)   --(\L,-\H)-- (\L,\H) -- (\L+\tWidth,\H) --(\L+\tWidth,-\H-\tWidth) --(0-\tWidth,-\H-\tWidth) -- (0-\tWidth,\mD)  -- (\hX-\tWidth,0) -- (0-\tWidth,\D)-- (0-\tWidth,\H);
	
	\draw[dotted] (\hX,0)--(\L,0) ;
	\draw[dotted] (-\tWidth-\hX,\mD)--(\L,\mD);
	\draw[dotted] (0,+\D)--(\L,+\D);
	\draw[dashed] (-\tWidth-\hX,\HH-\H)--(\L,\HH-\H) ;

	\draw [<->] (.3*\L,0)--(.3*\L,-\H)node[midway,anchor=west] {$H>\zzvar H$};
\draw [<->] (.65*\L,0)--(.65*\L,\mD)node[midway,anchor=west] {$D>\zzvar D$};
	\draw [<->] (0,.05*\HH-\H)--(\L,.05*\HH-\H)node[midway,anchor=base] {$L=\zzvar L$};

	\def\Lax{.2*\HH}
	\draw [thick,<->] (\Lax,0) node[anchor=text] {$\xx$} --(0,0)-- (0,\Lax) node[anchor=text] {$\ii\yy$} ;

\end{tikzpicture}%
%
%
%
%
%
%
%
\begin{tikzpicture}[scale=.75]
	
	
	\def\L{5.5}
	\def\HH{.6*\L}
	\def\H{.9*\HH}

	\def\tWidth{.025*\L} 
	\def\xClip{.4*\H}

	\def\HminusD{.25*\L}
	\def\D{\H-\HminusD}
	\def\mD{-\H+\HminusD}
	\def\hX{-.15*\H}

	\def\cropHeight{.6*\HH}
	\clip (-\tWidth+\hX,\cropHeight- \H+\HH) rectangle (\L+\tWidth,-\H-\tWidth);

	\draw[thick]   (0,\H) -- (0,\D) --  (\hX,0)-- (0,\mD)--(0,-\H)   --(\L,-\H)--(\L,\H);
	\fill[pattern=north east lines, pattern color=black]   (0,\H) -- (0,\D) --  (\hX,0)-- (0,\mD)--(0,-\H)   --(\L,-\H)-- (\L,\H) -- (\L+\tWidth,\H) --(\L+\tWidth,-\H-\tWidth) --(0-\tWidth,-\H-\tWidth) -- (0-\tWidth,\mD)  -- (\hX-\tWidth,0) -- (0-\tWidth,\D)-- (0-\tWidth,\H);
	
	\draw[dotted] (\hX,0)--(\L,0) ;
	\draw[dotted] (-\tWidth+\hX,\mD)--(\L,\mD);
	\draw[dotted] (0,+\D)--(\L,+\D);
	\draw[dashed] (-\tWidth+\hX,\HH-\H)--(\L,\HH-\H) ;
	
	\draw [<->] (.25*\L,0)--(.25*\L,-\H)node[midway,anchor=west] {$H<\zzvar H$};
	\draw [<->] (.65*\L,0)--(.65*\L,\mD)node[midway,anchor=west] {$D<\zzvar D$};
	\draw [<->] (0,.05*\HH-\H)--(\L,.05*\HH-\H)node[midway,anchor=base] {$L=\zzvar L$};

	\def\Lax{.2*\HH}
	\draw [thick,<->] (\Lax,0) node[anchor=text] {$\xx$} --(0,0)-- (0,\Lax) node[anchor=text] {$\ii\yy$} ;

\end{tikzpicture}%
%
%
%
}%
		\caption{Preferred map, setting $\alpha$ from \eqref{eq:alphaFlap} and varying $D$ to match the hinge position.}
		\label{fig:paddleSteady}
	\end{figure}

	With the  expansion in $\alpha$, a fixed far wall  is achieved by setting
	\begin{equation}
		\alpha = 1-\frac{\mean{\XX}}{\LL}.
		\label{eq:alphaFlap}
	\end{equation}
	A resulting complication is that the heights $H$ and $\Wbl$ now vary as function of $\XX$, respectively equalling $\HH$ and $\DD$ when $\XX(\yy)=0$, as illustrated in \autoref{fig:paddleSteady}.
	Evaluating \eqref{eq:ffPadleNull} at $\zz=-\ii \HH$, we obtain
	\begin{equation}
		H = \alpha \HH \quad\text{and}\quad D = \DD + (1-\alpha)\HH.
		\label{eq:HD}
	\end{equation}
Thus, one must jointly iterate on the flap displacement $\XX$ and the updated flap height $\Wbl$.
However, both $\XX$ and $\Wbl$ can be handled within the same iteration loop, adding no extra computational overhead. A code example is provided in \autoref{list:iterations}, \appendixref{sec:listings}.

\begin{figure}[H]
	\centering
		\includegraphics[width=.33\columnwidth]{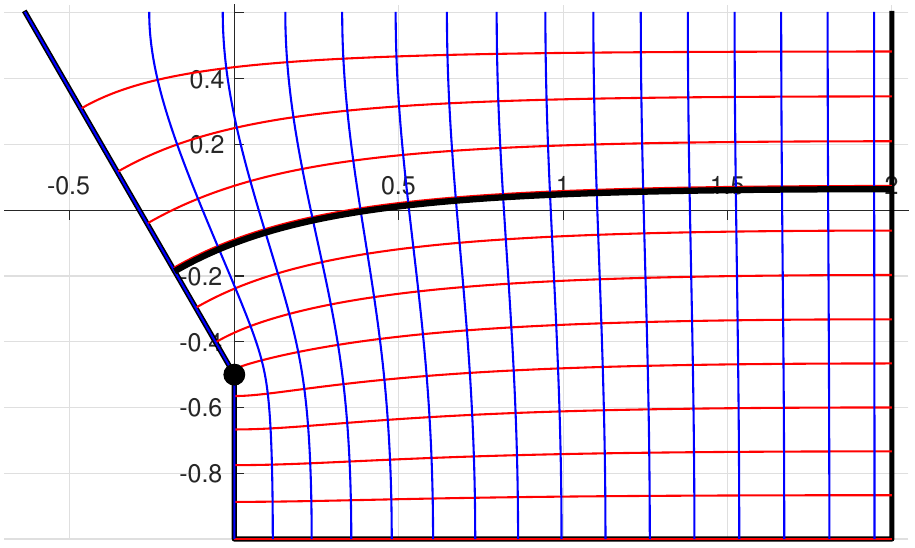}%
		\includegraphics[width=.33\columnwidth]{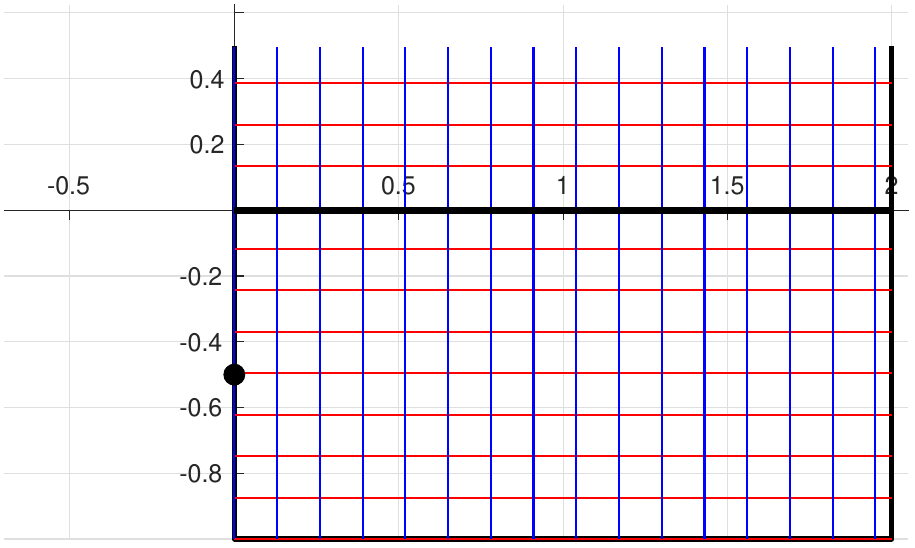}%
		\includegraphics[width=.33\columnwidth]{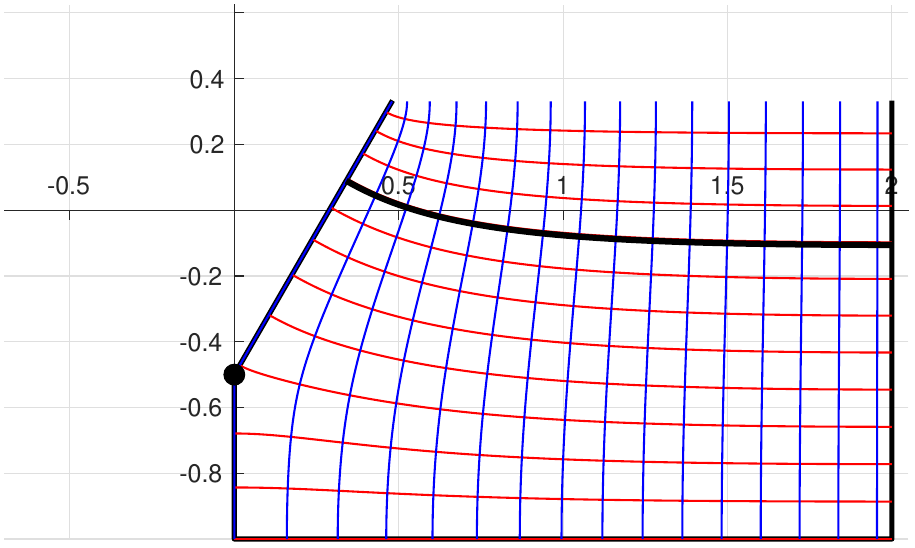}%
	\caption{Paddle map \eqref{eq:ffPadleNull}--\eqref{eq:ffPaddleShift} for $\theta=-30\degree$, $30\degree$ and $30\degree$, respectively, with $\h = 1.0$, $\Delta = 0.5$, $\wbl = 0.5$ and $\L=2.0$. 
Black isocline corresponds to $\yy=0$. }
	\label{fig:paddleMapSim}
\end{figure}

	Finally, the map must be repositioned such that the waterline touches the paddle face as opposed to the `tip'  created by the mirroring seam:
	\begin{equation}
		\zmap(\zz,t) = \ffNull(\zz-\ii\Delta) + \ii(H-h).
	\label{eq:ffPaddleShift}
	\end{equation}
Here,  $\Delta$ is the freeboard height (vertical distance to the mirroring seam) when  $\theta=0$, and $\h=\hh$ the mean water depth when  $\theta=0$.
	The final mapping is the one sketched in \autoref{fig:fSketch:z}, which also introduces the hinge depth $\wbl$.
	As expressed in \eqref{eq:alphaFlap} and \eqref{eq:HD},  the basin volume changes slightly with paddle angle,  causing the mean water level to rise slightly above ($\theta>0$), or fall slightly below ($\theta<0$), the still-water line $\z=\x$.
	\autoref{fig:paddleMapSim} illustrates  the mapping in practice, highlighting the curvature of the reference line $\zz=\xx$ and demonstrating that the basin length and depth remain fixed.

\subsection{The background velocity field $\WW_\zz^*$}
\label{sec:map:WW}
The background velocity potential is pre-computed according to the wall and floor conditions in \eqref{eq:conditionsWW}, which depend solely on the mapping function $\zmap$.
Neither $\zmap_\zz$ nor $\zmap_t$ are pure real along the inclined paddle surface, which rules out the trick used with the piston wavemaker, where setting $\WW_\zz =  \zmap_\zz \zmap_t$ satisfies  \eqref{eq:ReWWzz}. 
Instead, the projection kernels from \eqref{eq:mapSC} are again employed, setting
\begin{equation}
 \WW_\zz =  \SS{\LL}{\muw_1}(-\ii\zz)  - \mean{\muw_1}\frac{\zz + \ii \hh}{\LL}.
	\label{eq:WWFlap}%
\end{equation}
Similar to earlier, the last term is a pulsation that  removes the zero mode when evaluated at the far wall, ensuring that  $\muw_2=0$ along $\zz=\xx\^w_2+\ii \yy$.
The mirroring inherent in the projection operator also satisfies condition \eqref{eq:ImWWzz}.
Unlike \eqref{eq:ffPadleNull}, \eqref{eq:WWFlap} is an explicit expression.

	\subsection{Flap motion}

\newcommand{\zmapTh}{\tilde f}
\newcommand{\muwTh}{\tilde \mu\^w}
\newcommand{\WWTh}{\tilde W}
The previous sections describe instantaneous mappings at prescribed wavemaker positions.
The range of these  span the paddle stroke limit and are independent of the paddle's motion in time. 
Time dependencies are then captured using the chain rule;
assuming a position-determined mapping function 
$\zmapTh(\zz,\theta)$
defined
\begin{align}
\zmap(\zz,t) = \zmapTh[\zz,\theta(t)],
\label{eq:zmapTh}
\end{align}
	along with a position-dependent background  potential 
$\WWTh(\zz,\theta)$
that equals \eqref{eq:WWFlap} with
\begin{equation}
	\muwTh_1(\yy,\theta) =  \Re(\zmapTh_\zz \zmapTh_\theta^*)\big|_{ \xx=\xx\^w_1}
\end{equation} 
replacing $\muw_1$,
the following properties result:
\begin{align}
	\zmap_\zz &= \zmapTh_\zz,&
	\zmap_t &= \theta_t \zmapTh_\theta,&
	\WW_\zz &=  \theta_t  \WWTh_\zz,&
	\WW &=  \theta_t  \WWTh,&
	\WW_t &=  \theta_{tt}  \WWTh +  \theta_t^2  \WWTh_\theta.
	\label{eq:signalDerivatives}
\end{align}
Discrete differentiation 
\begin{equation}
	\zmapTh_\theta(\zz,\theta) \approx \frac{\zmapTh(\zz,\theta+\Delta\theta/2)-\zmapTh(\zz,\theta-\Delta\theta/2)}{\Delta\theta}
\label{eq:discreteDifferentiation}%
\end{equation} 
is used for the map position derivative, and similar for $\WWTh_\theta$.
One accordingly pre-computes the positional functions $\zmapTh$, $\zmapTh_\zz$, $\zmapTh_\theta$, $\WWTh$, $\WWTh_\zz$ and  $\WWTh_\theta$ prior to simulation, along with the wavemaker signals 
$\theta$, $\theta_t$ and $\theta_{tt}$. 
These are then interpolated during simulation, with the expressions \eqref{eq:zmapTh} and \eqref{eq:signalDerivatives} entering the model in \autoref{sec:model:core}.
\\

\begin{figure}[H]
	\centering
	\subfloat[Surface elevation in space and time.]{\includegraphics[width=.5\columnwidth]{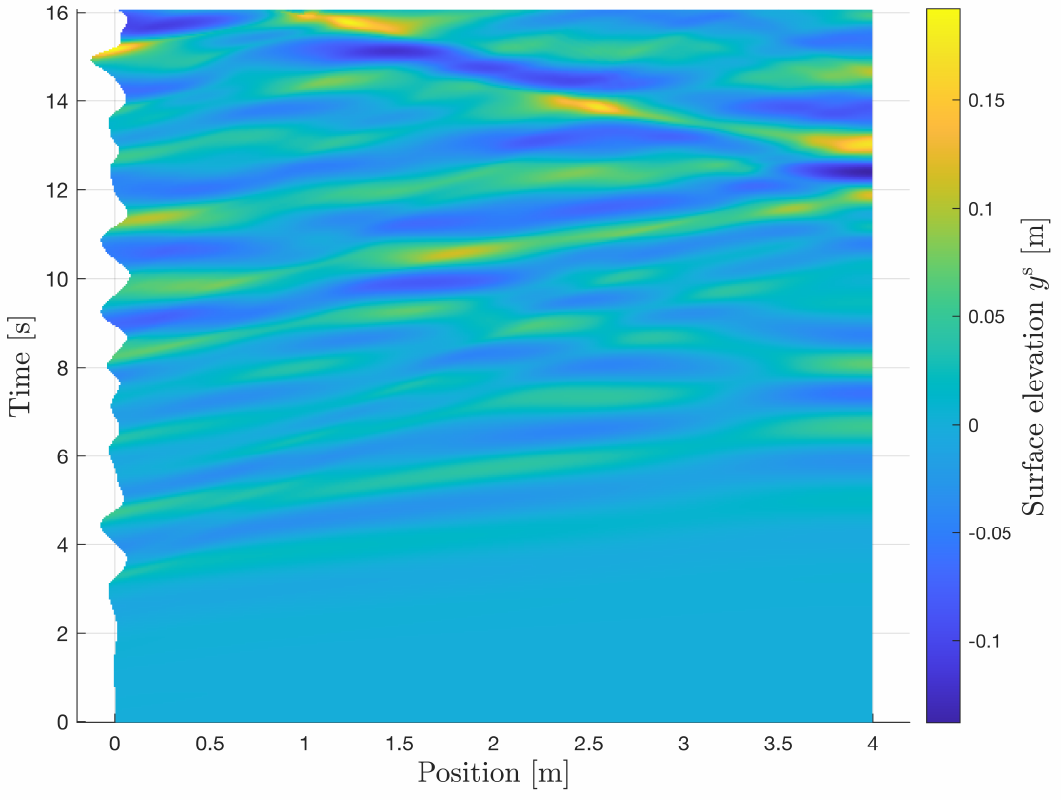}\label{fig:exampeSim:imsc}}\\
	\subfloat[Snapshot at largest negative paddle stroke; map isoclines (left), potential field $\phi$ and vectors of velocity field $\w_z^*+\W_z^*$ (right).]{\includegraphics[width=.5\columnwidth]{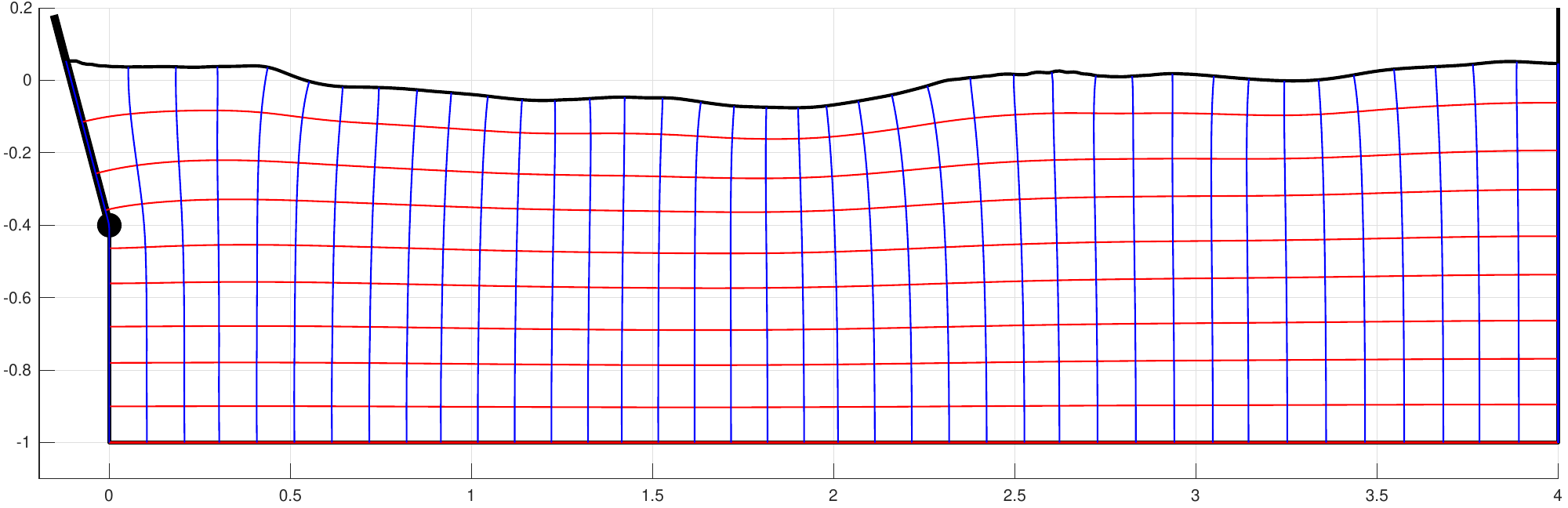}
\includegraphics[width=.5\columnwidth]{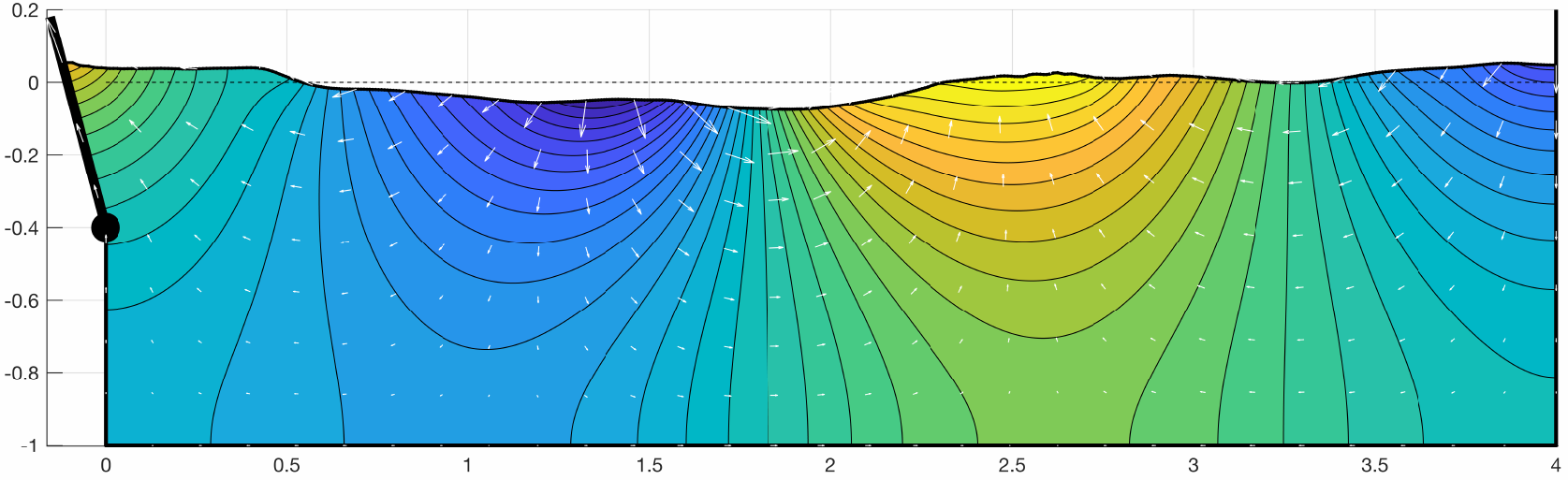}\label{fig:exampeSim:neg}}\\
	\subfloat[Snapshot at largest positive paddle stroke.]{\includegraphics[width=.5\columnwidth]{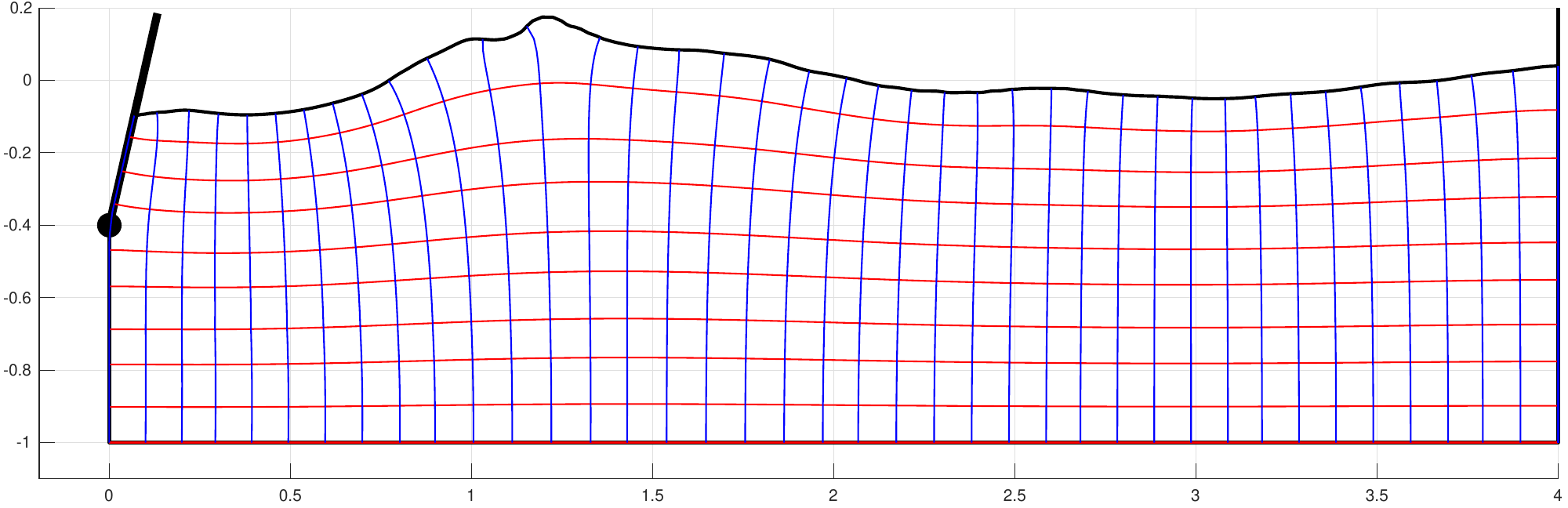}
\includegraphics[width=.5\columnwidth]{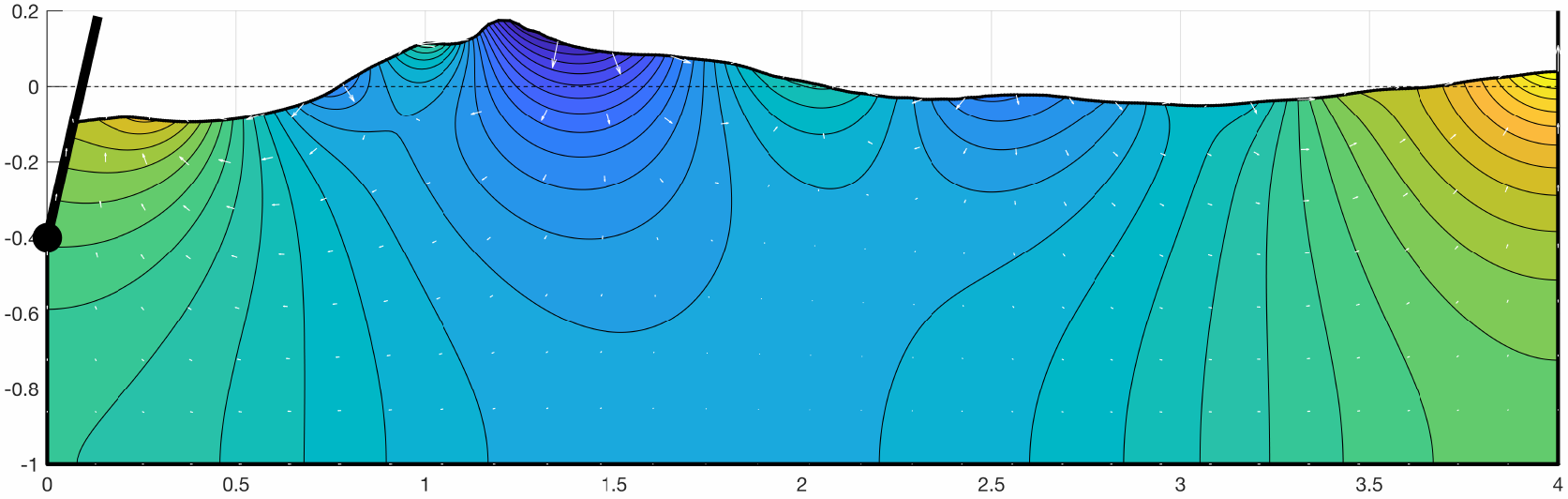}\label{fig:exampeSim:pos}}%
	\caption{Example simulation with irregular flap wavemaker motion in a small basin without a beach. Excessive parameters for illustration: JONSWAP wave spectrum with 1.5\,s peak period, 0.15\,m significant wave height and shape factor 3.0; $h=1.0$\,m, $d=0.4$\,m, $L=4.0$\,m $\kd=0.5\,\kMax$, $\rDamping=0.5$.}
	\label{fig:exampeSim}
\end{figure}

To conclude this section, an example simulation is presented in \autoref{fig:exampeSim} to demonstrate the capabilities of the model.
 For visual emphasis, exaggerated parameters---including a relatively strong modal damping---have been used.
The simulated basin is short and without any  beach, so waves are reflected off  the back wall and interact with incoming waves.
Over time, the resulting sloshing accumulates and eventually causes the simulation to crash.
The left edge of \autoref{fig:exampeSim:imsc} reveals the altering horizontal position of the water surface at the wavemaker.

	\section{Numerical validation}
\label{sec:validation}

\begin{figure}[H]
	\centering
	\subfloat[Piston at depths $\yyy=-\h/3$, $-\h/2$ and$-\h$ (overlapping curves).]{%
		\includegraphics[width=.5\columnwidth]{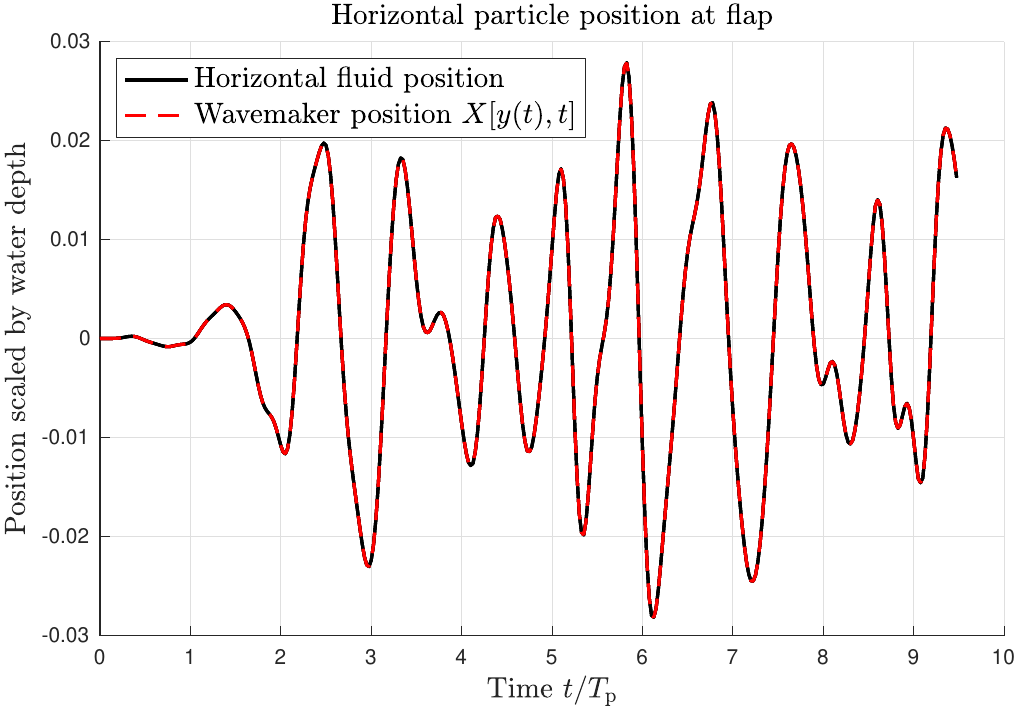}%
		\includegraphics[width=.5\columnwidth]{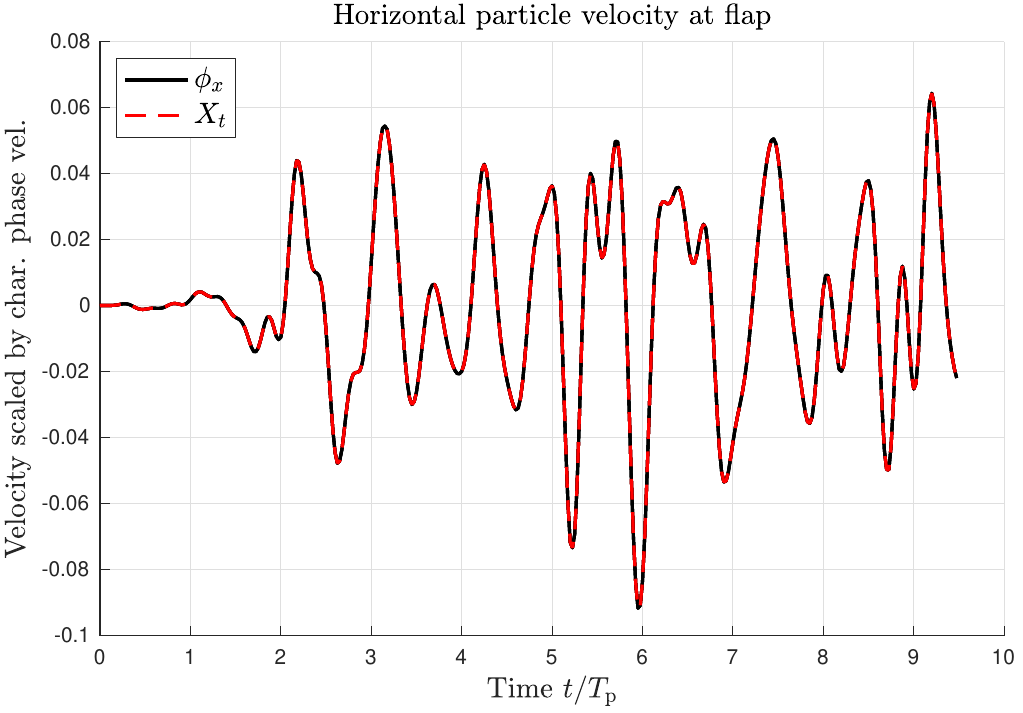}}\\
	\subfloat[Flap at depths $\yyy=-2\wbl/3$, $-\wbl/3$ and $-4\wbl/3$. ]{%
		\includegraphics[width=.5\columnwidth]{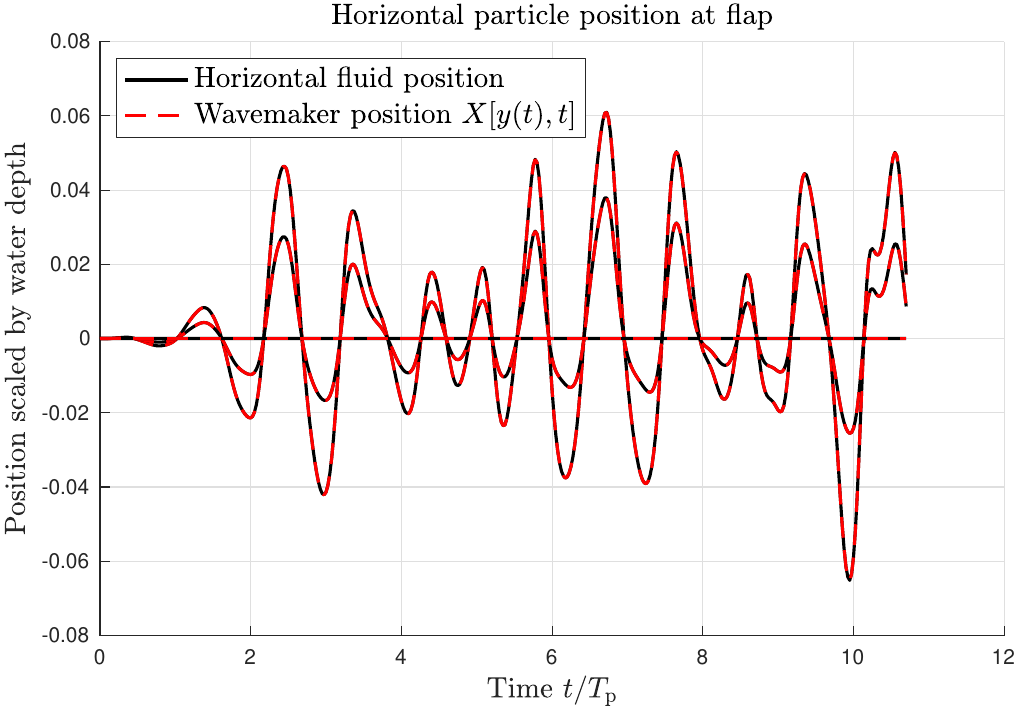}%
		\includegraphics[width=.5\columnwidth]{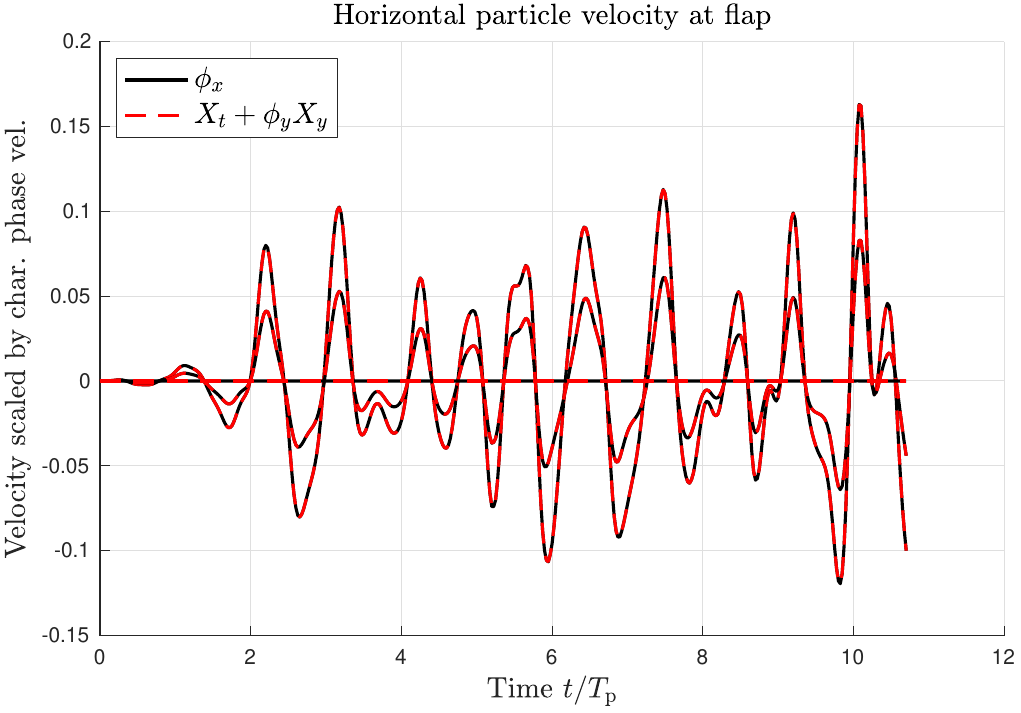}}%
	\caption{Validation of the kinematic wavemaker boundary condition.}
	\label{fig:BC_wm}
\end{figure}

We begin by validating the kinematic wavemaker boundary condition, which, by design, should be exactly satisfied.
\autoref{fig:BC_wm} confirms this, showing examples from both the piston mapping in \autoref{sec:map:piston}, and the flap mapping in \autoref{sec:map:flap}. 
The figure presents fluid particle positions and velocities over time at various depths, demonstrating that the motion closely follows the wavemaker boundary.
The  evaluated  depths are fixed in the $\zzz$-plane, which vary with time in the $\z$-plane. Wavemaker signals are irregular and correspond to the example presented in the  previous section.
\\

\begin{figure}[H]
	\centering
	\subfloat[Intermediate water depth flap; $T=1.0$\,s, $\h=1.0$\,m, $\wbl=0.5$\,m, $15\degree$ flap stroke.]{\parbox{.5\columnwidth}{\includegraphics[width=.5\columnwidth]{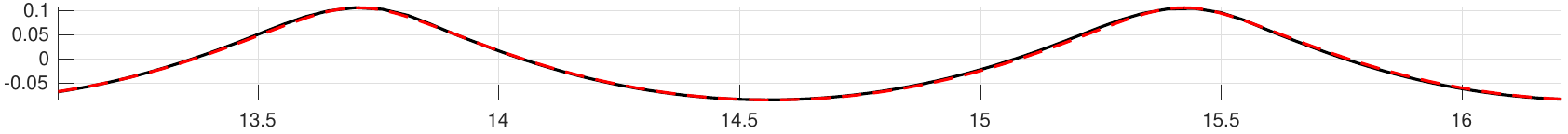}\\
		\includegraphics[width=.5\columnwidth]{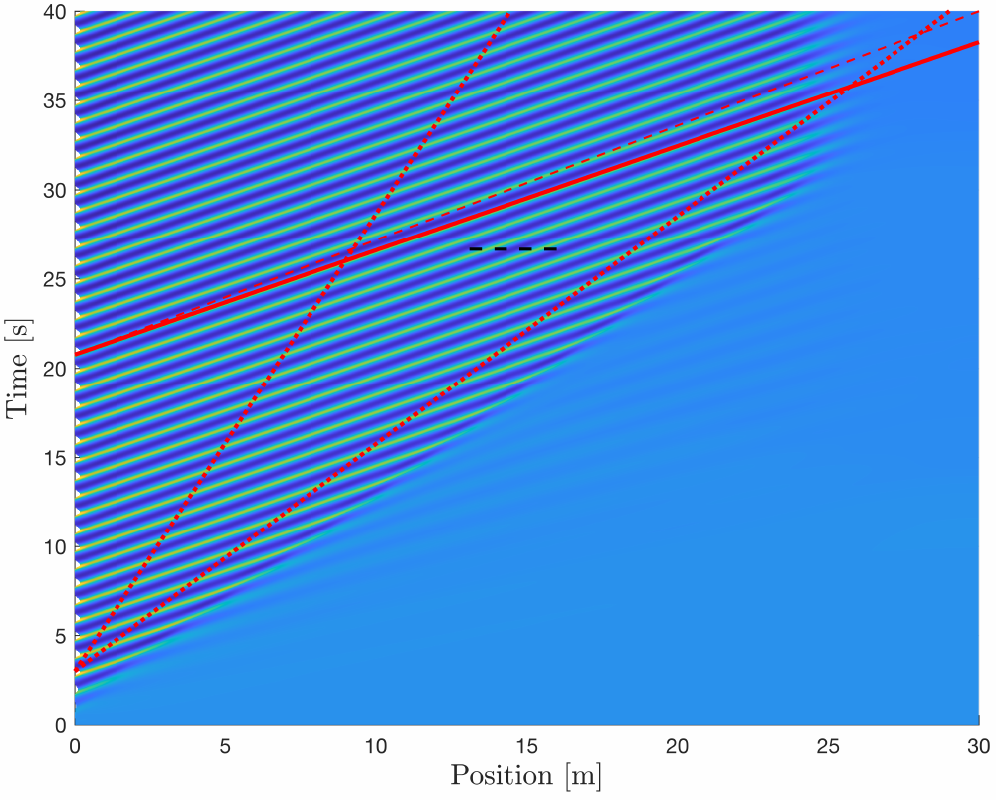}%
		}}%
	\subfloat[Shallow water piston, $T=1.0$\,s, $\h=0.1$\,m, 2.0\,m piston stroke.]{\parbox{.5\columnwidth}{\includegraphics[width=.5\columnwidth]{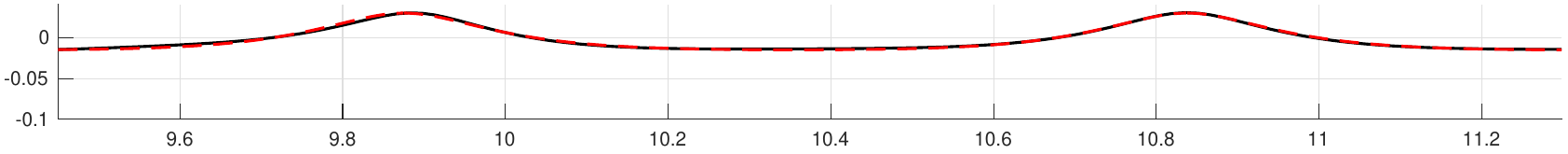}\\
		\includegraphics[width=.5\columnwidth]{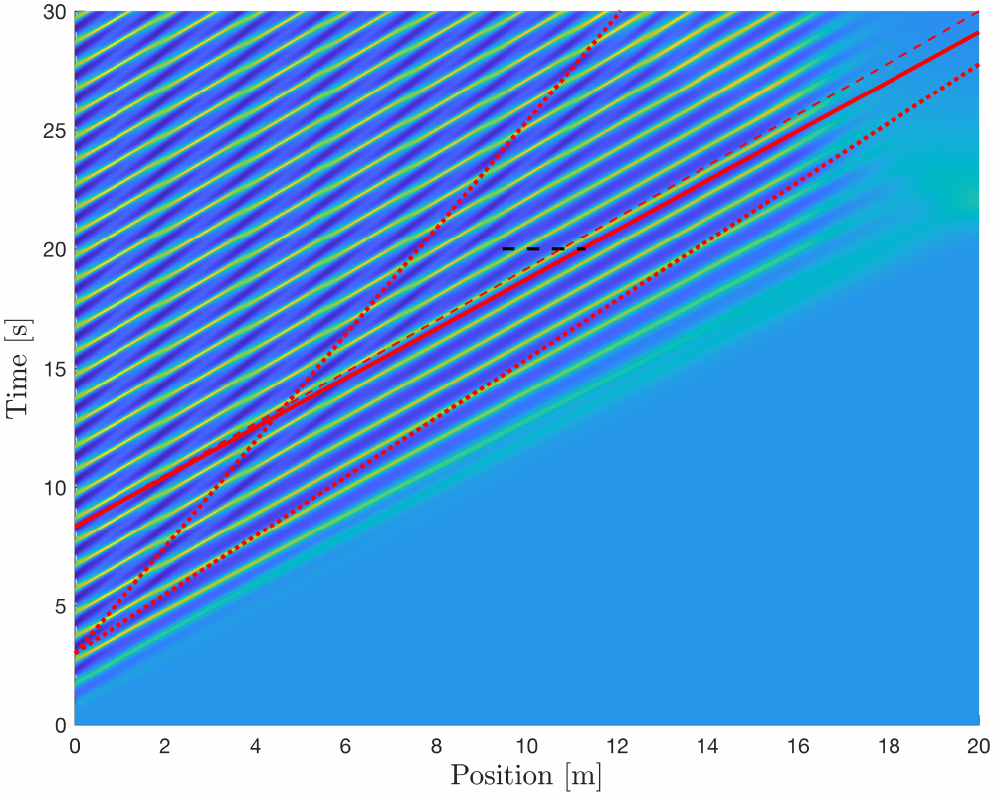}%
		}}%
	\caption{Bottom panels: test on phase and group velocities, compared to linear theory (dashed and dotted lines), and the exact solution from SSGW using Sokes' second definition of phase velocity (solid line).
	Top panels: surface elevation profile compared to SSGW solution. Evaluation made between the principal and second-order spurious wave train fronts in the area indicated with a black dashed line in lower panels.}
	\label{fig:cp}
\end{figure}

The phase velocity of regular waves is considered in  \autoref{fig:cp},
comparing simulated results to linear theory and Stokes wave solutions from the SSGW model of \citet{clamond2018accurate}.
Precise agreement with the nonlinear phase velocity is observed.
Assessing phase velocity in a wave basin is more challenging than in a periodic domain due to spurious waves generated by the wavemaker and the inherent instability at the front of the wave train. 
Additional complications arise from the return flow and the set-down effect, both of which become increasingly prominent in shallower water. 
The simulations intrinsically include a uniform return flow and so Stokes’ second definition of wave celerity is used when comparing with the SSGW predictions.
\autoref{fig:cp} also includes the linear group velocities of the principal and second-order spurious waves,
along with surface elevation profiles are compared within a relatively uncontaminated portion of the wave train,
both showing good agreement with the observed wave field.
\\

\begin{figure}[H]
	\subfloat[Flap wavemaker case 70040 from \autoref{fig:starMap:flap} with flap stroke angles $2.5\degree$, $5.0\degree$ and $7.5\degree$.]{
\parbox{.5\columnwidth}{\centering\includegraphics[width=.25\columnwidth]{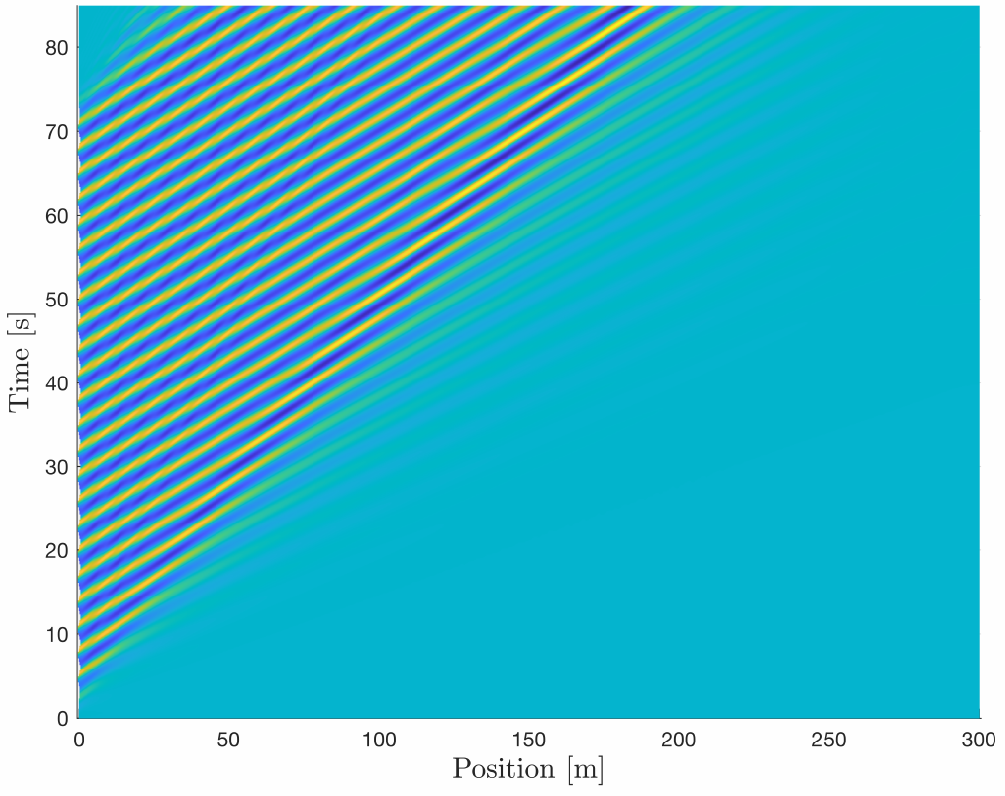}\\
\includegraphics[width=.5\columnwidth]{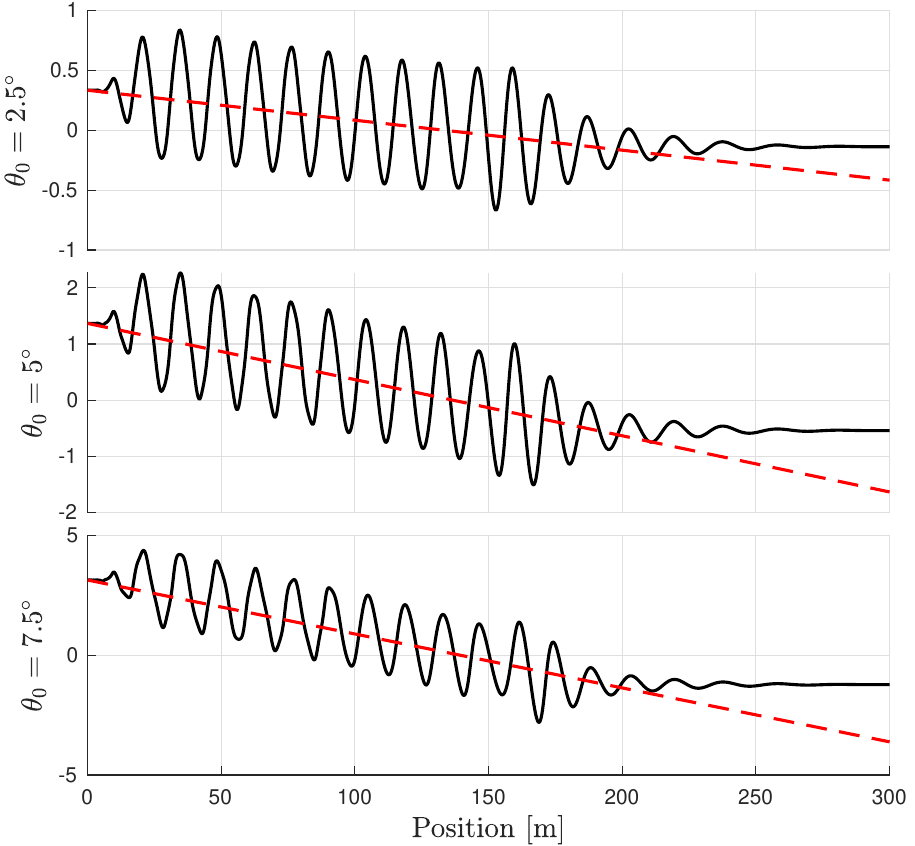}
}}%
	\subfloat[Shallow water piston  wavemaker from \autoref{fig:starMap:piston} with paddle strokes  $2.5$\,m, $5.0$\,m and $7.5$\,m.]{
\parbox{.5\columnwidth}{\centering\includegraphics[width=.25\columnwidth]{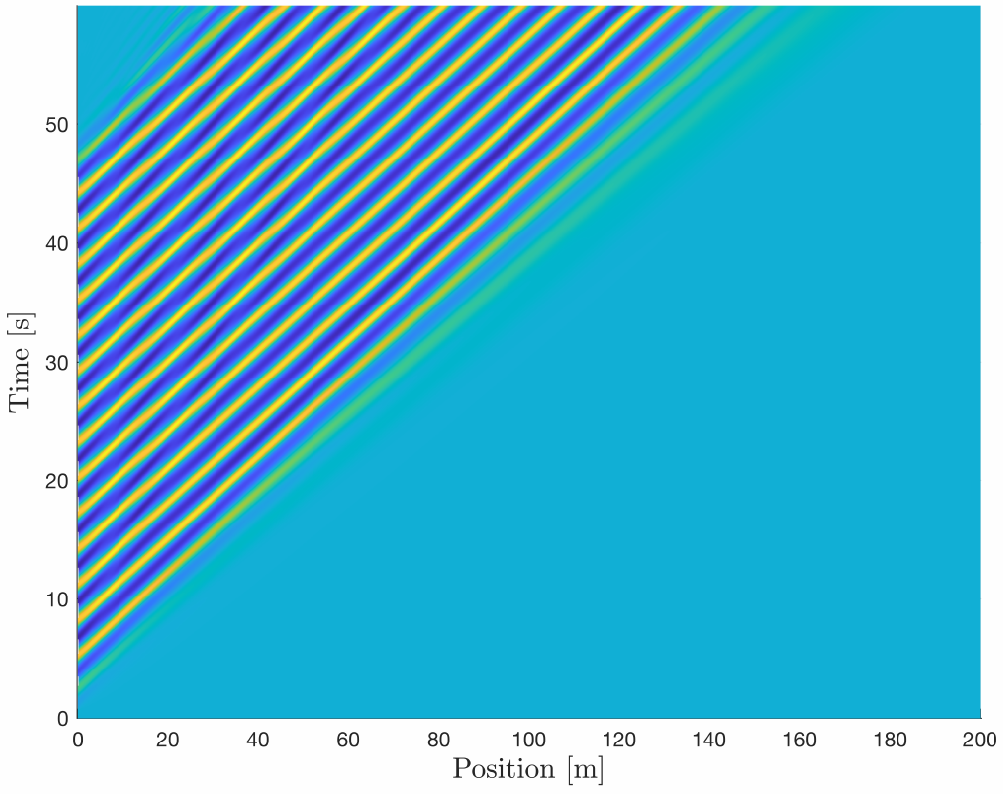}\\
\includegraphics[width=.5\columnwidth]{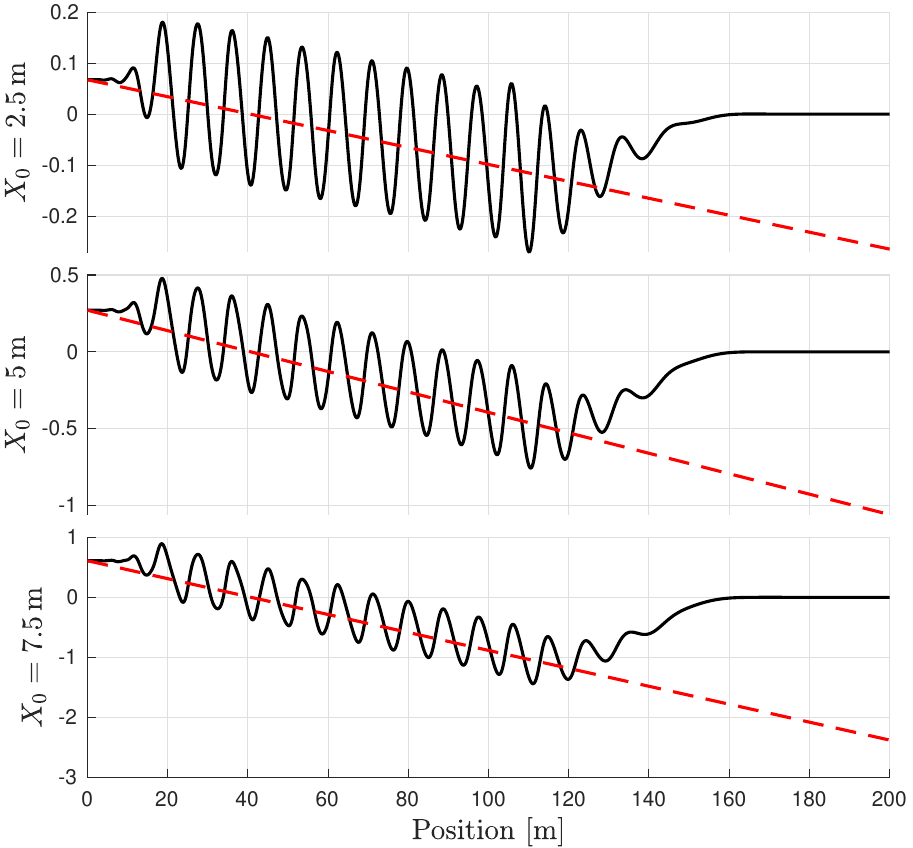}
}}%
	\caption{The return current set up by stokes drift and mass conservation---plotted are the surface potential $\phi\S(\x,t)$ (solid black line) and the return current potential $U_0\,\x$ (dashed red line) as predicted by the average Stokes drift or second order wavemaker theory.}
	\label{fig:U0}
\end{figure}

Finally, we verify that the model  captures basin return flow correctly. 
The author showed in \citet{AHA2025_multihinge} that the return flow is inherent in  second-order wavemaker theory \citep[e.g.,][]{schaffer_1996}, appearing as the zero-mode limit of the second-order free spurious wave component.
Through mass conservation, this further equals the mean Stokes drift across the water column, which can be evaluated analytically;
given that the wavemaker generates a  principal wave field of amplitude components $a_n$, angular frequencies $\omega_n$ and progressive wavenumbers $k_{0n}$, the return flow becomes
\begin{equation}
U_0 = -\mean{ U_\mathrm{Stokes} }
=-\overline{\frac{1}{h} \int_0^{\y\S} \!\!\phi_x\,\dd \y}
=-\frac g{4h}\sum_n \frac{k_{0n} |a_n|^2 }{\omega_n }  + O\big(|a_n|^3\big).
\label{eq:U0}
\end{equation}
\autoref{fig:U0}  demonstrates, again by example,  the  return flow current in intermediate and shallow water depths.
To simplify the presentation, only regular waves are considered, using the same flap and piston motions as before.
The comparison is carried out by first generating a wave train and then returning the wavemaker to rest.
Once stationary, the background velocity field $\WW_\zz^*$ is zero everywhere, allowing us to compare the surface potential $\phi\S(\x,t)$ directly to  the return flow potential $U_0 \x$.
 The surface potential is seen to  exhibit a mean gradient that matches the return current velocity  $U_0$.
Similar to the  set-down effect, this current is established from the onset of wave generation, running underneath the wave train as it progresses.
The return flow manifests as a uniform current in potential theory, but will in reality attain a sheared profile.

\section{Experimental validation}
\label{sec:exp}

In this section, we evaluate the model’s ability to replicate experimental results by comparing simulations to measurements conducted in the towing tank laboratory at SINTEF Ocean in Trondheim, Norway
The water depth is $h=5.415$ meters and the hinge depth  is $\wbl=3.015$  meters.
A harp, consisting of 23 wave gauges, spaced 15 centimetres apart, was positioned approximately 90 meters down the flume.
An additional transverse row of wave gauges, aligned perpendicularly to the harp, is included to monitor sloshing, and permanent wave gauges are mounted to the wavemaker flaps as part of the active wave absorption system.
The number of tests, tabulated in \autoref{tab:tests}, is moderate, intended  validation and to assess the model's predictive capabilities and limitations. 
Further experimental investigations are anticipated in near future.

\begin{table}[H]
\centering\small
\subfloat[Regular wave tests.]{
\begin{tabular}{r|cccccccc}
Test no.& 70010&70020&70030&70040&70050&70060&70070&70080\\\hline
Period $T$ [s]  &1.5   &2.0  &2.5  &3.0  &3.0   &3.5  &4.0  &4.5 \\
Wave height $H$ [m]& 0.15&0.25&0.40&0.40&0.20&0.40&0.30&0.25
\end{tabular}
}\\
\subfloat[Irregular wave tests, all being JONSWAP spectra with shape factor 3.0. Last test number digit   indicates  calibration increment.]{
\begin{tabular}{r|cccccc}
Test no.      & 8012--&8006--&8010--&8008--&8011--&8100--\\\hline
Peak period $T\_p$ [s]  &1.5      &1.5      &2.0    &2.0      &2.5    &3.0   \\
Significant wave height $H\_s$ [m]& 0.15    &0.17   &0.20  &0.30    &0.25 &0.10 
\end{tabular}
}
\caption{Experimental towing tank tests.}
\label{tab:tests}
\end{table}

\subsection{Regular waves and spurious waves}
\label{sec:exp:recular}
Simulations are conducted by plugging signal paddle position signals recoded during experiments into the numerical simulator and extracting equivalent harp measurement signals therefrom. 
The resolution is set to 100 points per wavelength and passive modal damping parameters $\kd=0.5\,\kMax$ and $\rDamping=0.01.$ are adopted.

We begin by examining the wavemaker characteristics in terms of regular wave amplitudes and the emergence of higher-order spurious waves.
To this end, surface elevation signals are decomposed through Fourier transformation.
These are spatially fitted to their respective linear wave components via linear regression---an estimation method commonly used to estimate wave reflection \citep{mansard1980}, here applied to higher-order wave components. 
The fitting is carried out within a sliding time window spanning one wavemaker period, giving the amplitude evolution in time in the vicinity of the wave gauges.

\begin{figure}[H]
		\subfloat[70050; $T=3.0$\,s, $H=0.20$\,m.]{\includegraphics[width=.5\columnwidth]{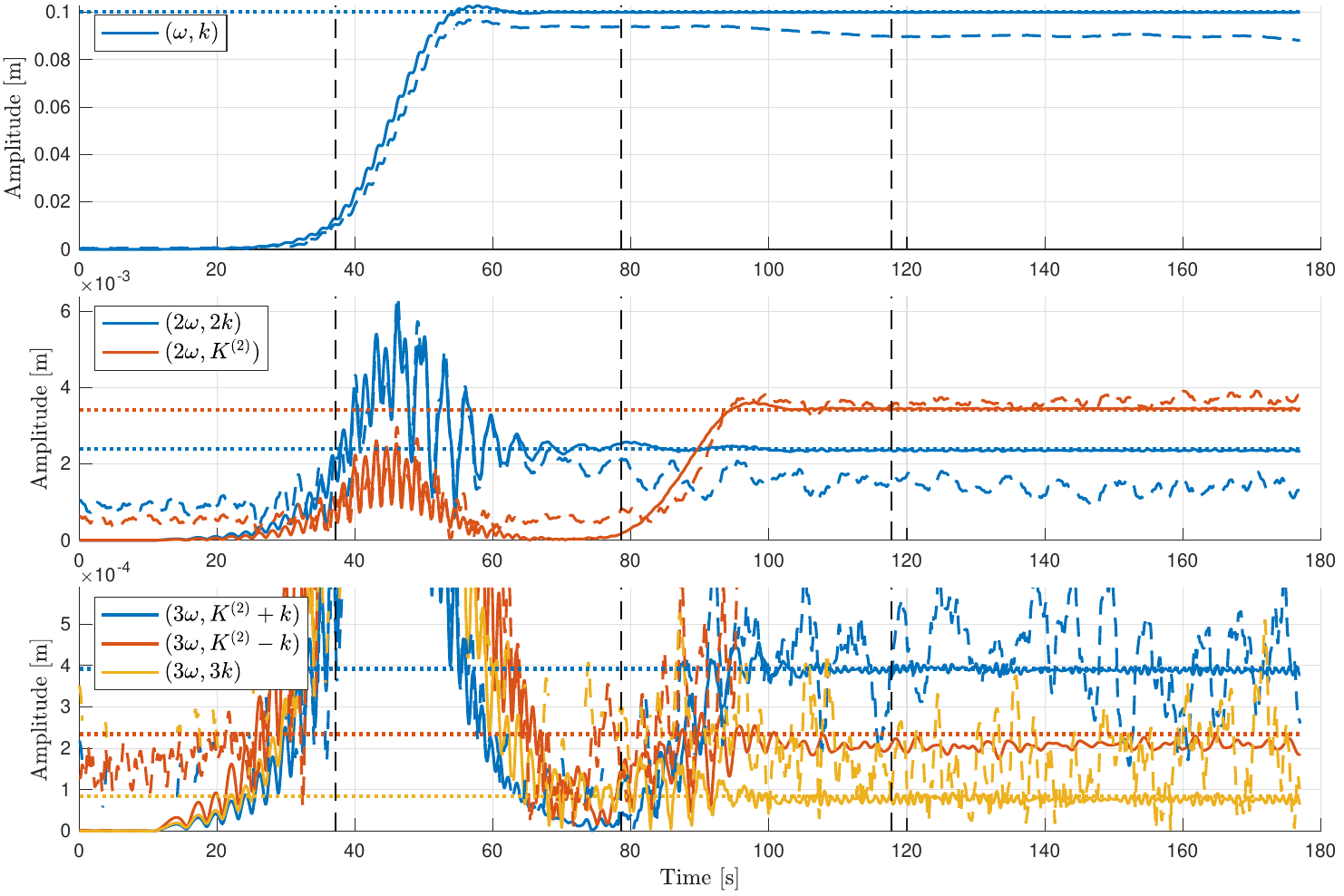}}%
		\subfloat[70060; $T=3.5$\,s, $H=0.40$\,m.]{\includegraphics[width=.5\columnwidth]{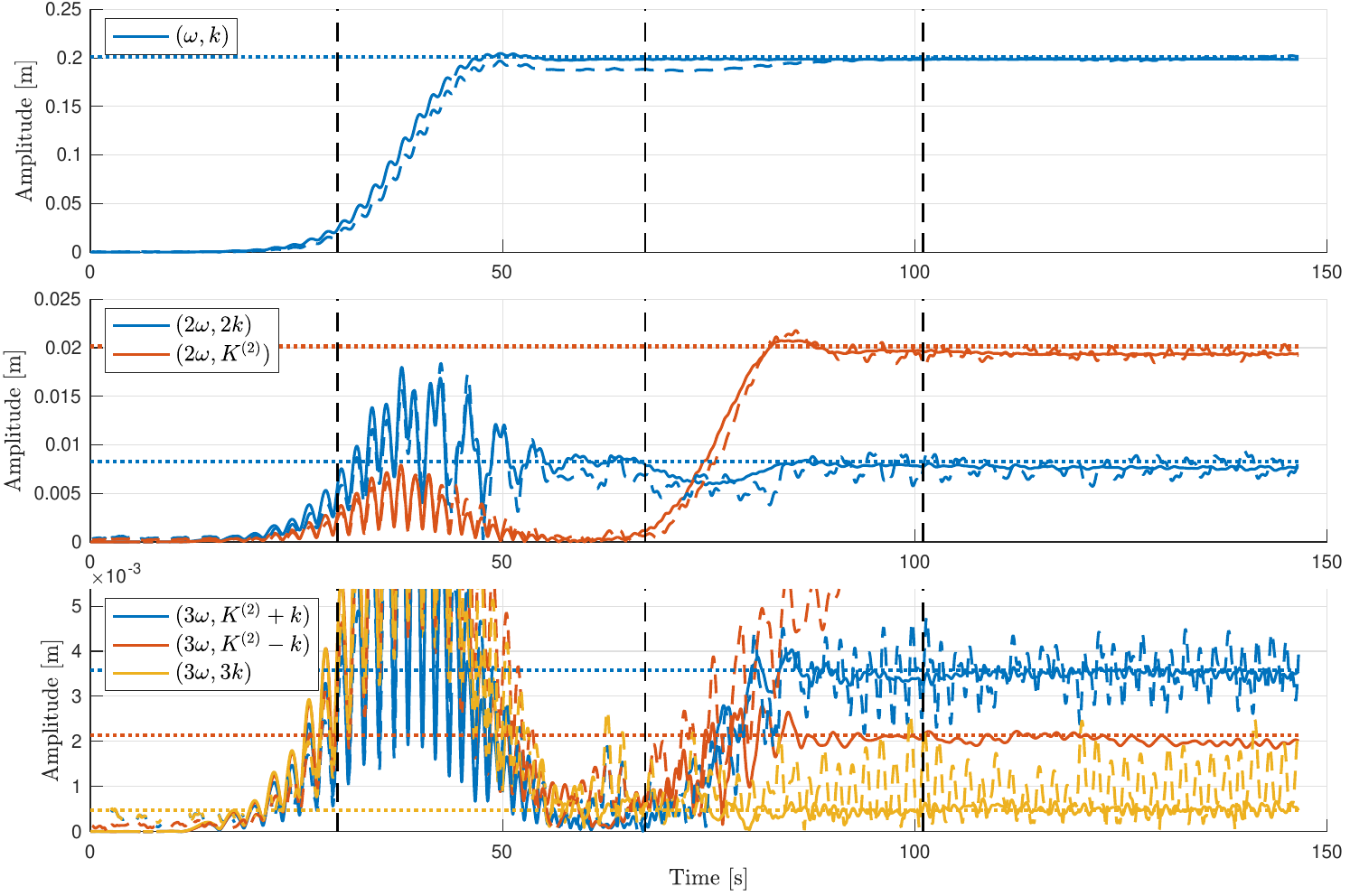}}\\
		\subfloat[70070; $T=4.0$\,s, $H=0.30$\,m.]{\includegraphics[width=.5\columnwidth]{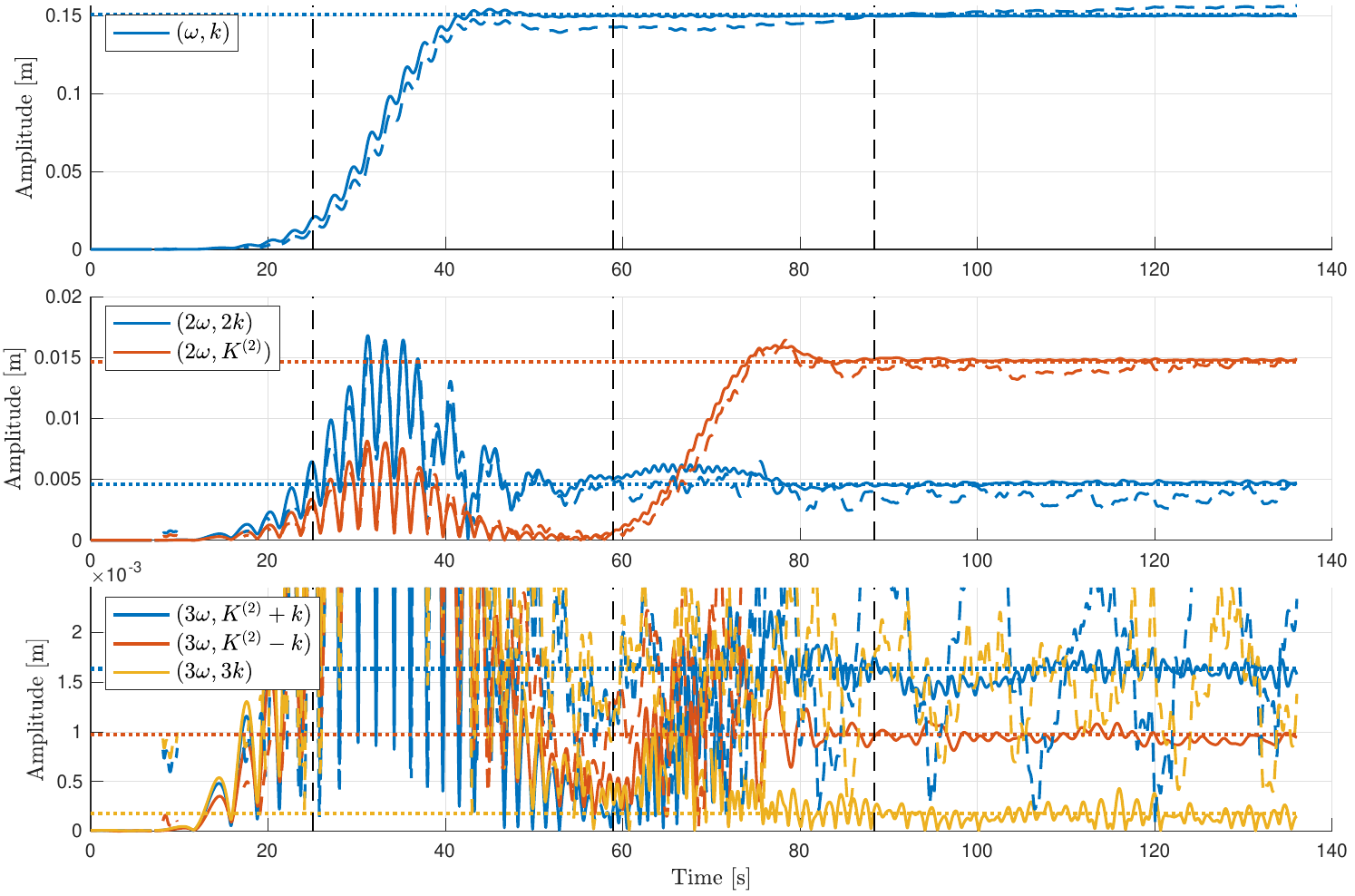}}%
		\subfloat[70080; $T=4.5$\,s, $H=0.25$\,m.]{\includegraphics[width=.5\columnwidth]{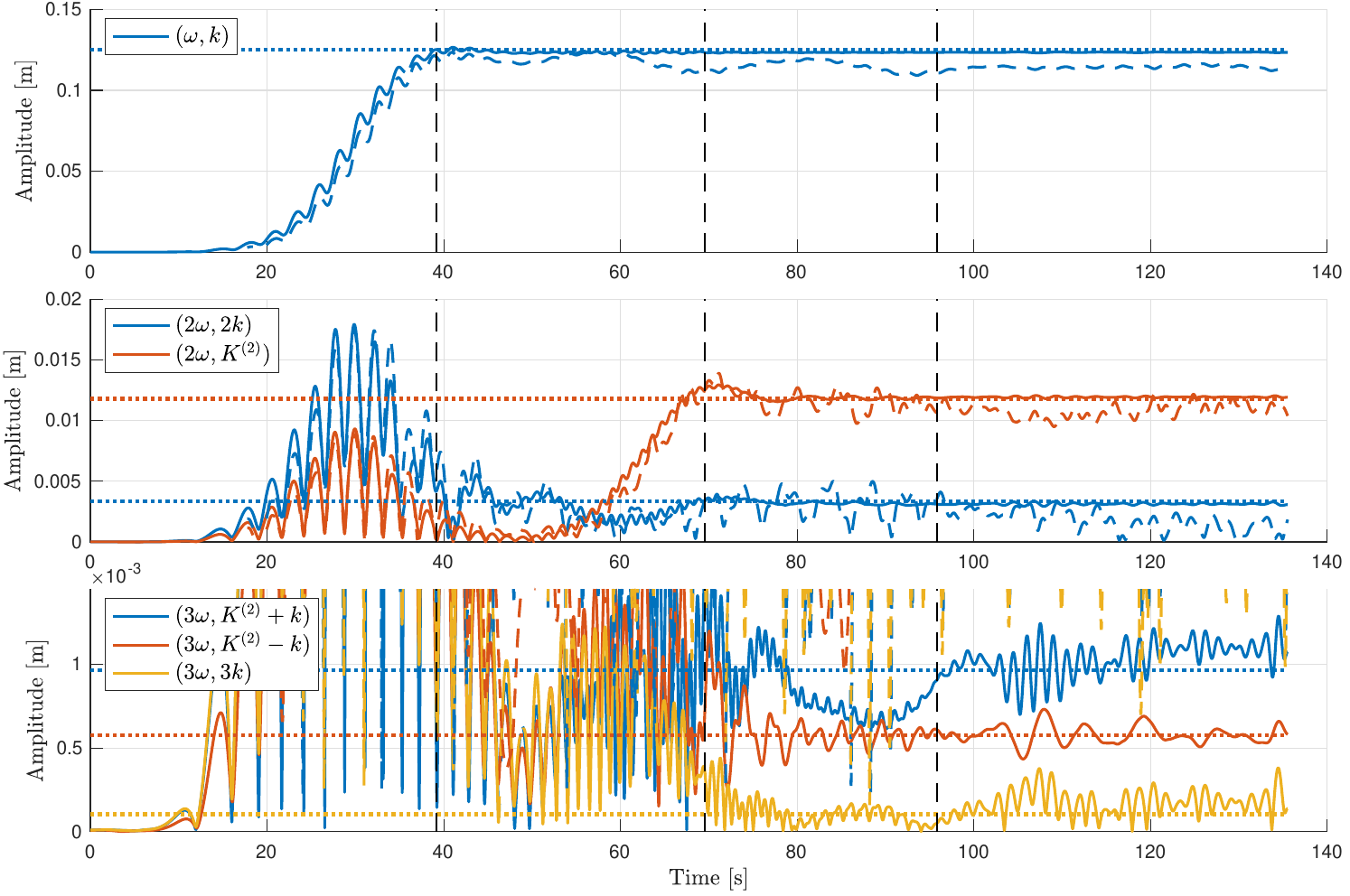}}%
		\caption{Principal and higher-order wave amplitude components of simulation (solid),  experiment (dashed) and wavemaker theory (dotted). Dashed vertical black lines indicate the arrival of free wave components, estimating using linear group velocities.}
		\label{fig:compExp}
	\end{figure}

\autoref{fig:compExp} presents the amplitudes compared up to third order.
Predictions from wavemaker theory are also included.
While conventional wavemaker theory \citep{schaffer_1996} is limited to second order, third-order predictions are here included, using the author's own, as yet unpublished, extensive wavemaker theory. 
Second-order free spurious wavenumbers---those satisfy the dispersion relation at frequency $2\omega$---are denoted $K^{(2)}$ in the figure captions.
Similarly, third-order free wavenumbers are denoted $K^{(3)}$.
Pairs of the form $(2\omega, K^{(2)} \pm k)$ correspond to third-order bound subharmonics and superharmonics, which can be estimated by applying Sch{\"a}ffer's theory to a bichromatic wave. 
The arrival times of free waves, estimated from group velocities, are indicated by dashed vertical black lines, such that the ramp of the principal wave appears after the first line, the ramp of the second-order spurious wave appears after the second line, and so on.

Experimental amplitudes are seen to match the theoretical predictions  well, although  third-order components exhibit notable noise and are exceptionally  sensitive to measurement inaccuracies. 
Simulations, containing fewer sources of disturbance, show a better agreement with theoretical estimates. 
The cosine ramping of the wavemaker signal appears in the analysis as patches of noise preceding a stable amplitude estimate.
A slight long-term variation is observed in the measured amplitudes, absent in simulations, which  is likely attributable to three-multidimensional effects, build-up of transverse modes, and reflections from the beach.
\\

	\begin{figure}[H]
		\centering
		\subfloat[70010; $T=1.5$\,s, $H=0.15$\,m.]{\includegraphics[width=.5\columnwidth]{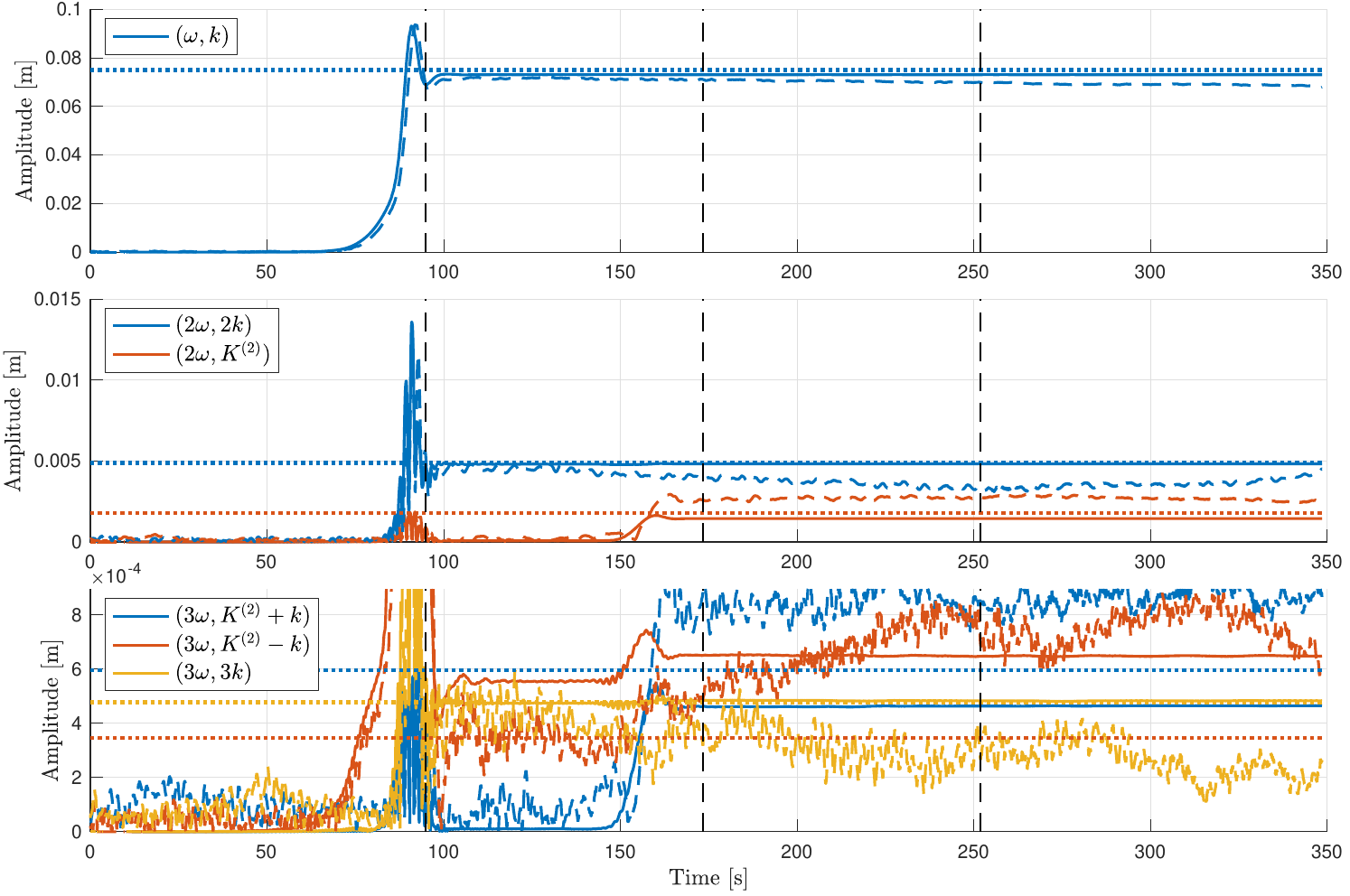}}%
		\subfloat[70020; $T=2.0$\,s, $H=0.25$\,m.]{\includegraphics[width=.5\columnwidth]{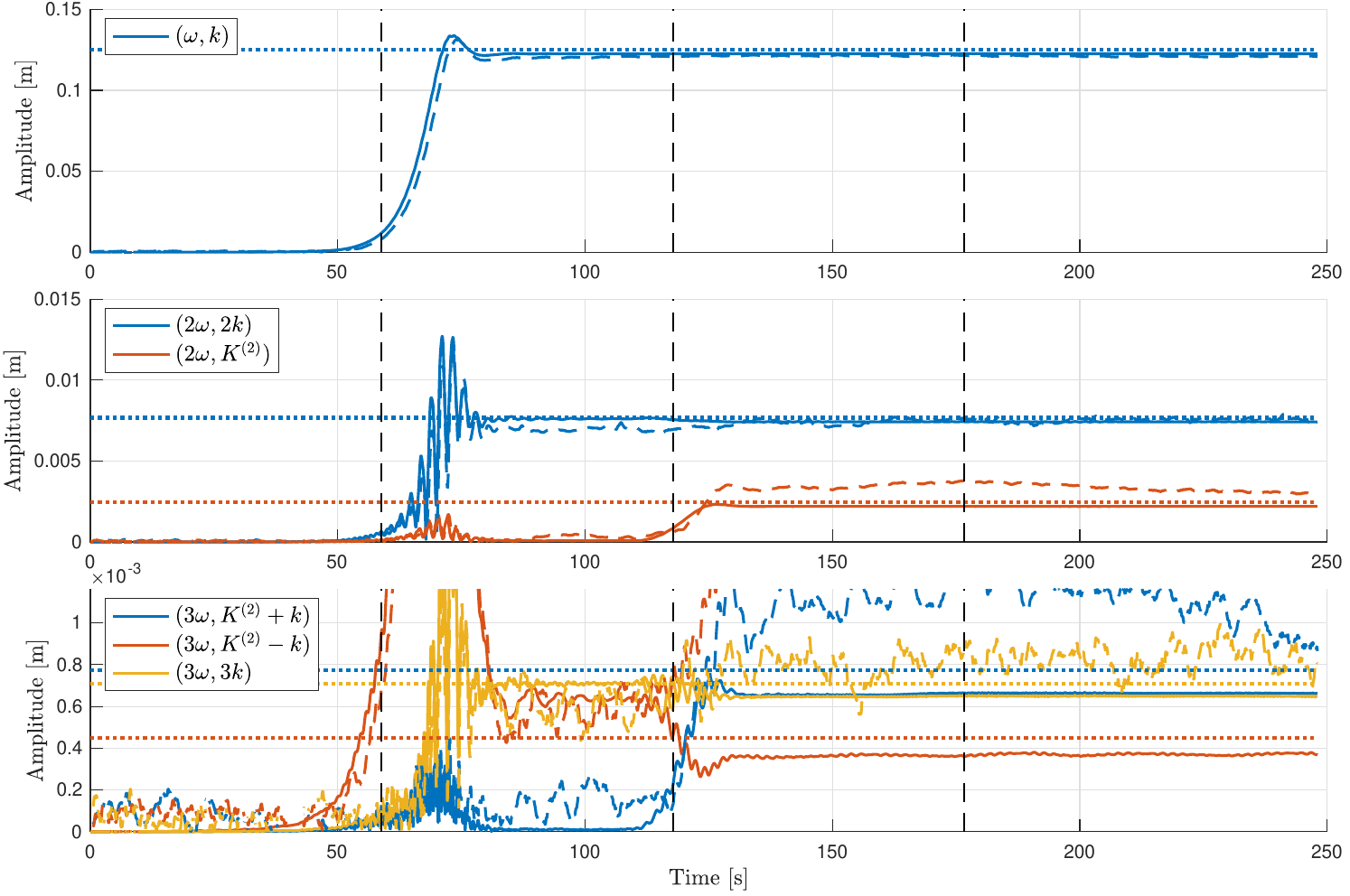}}\\
		\subfloat[70030; $T=2.5$\,s, $H=0.40$\,m.]{\includegraphics[width=.5\columnwidth]{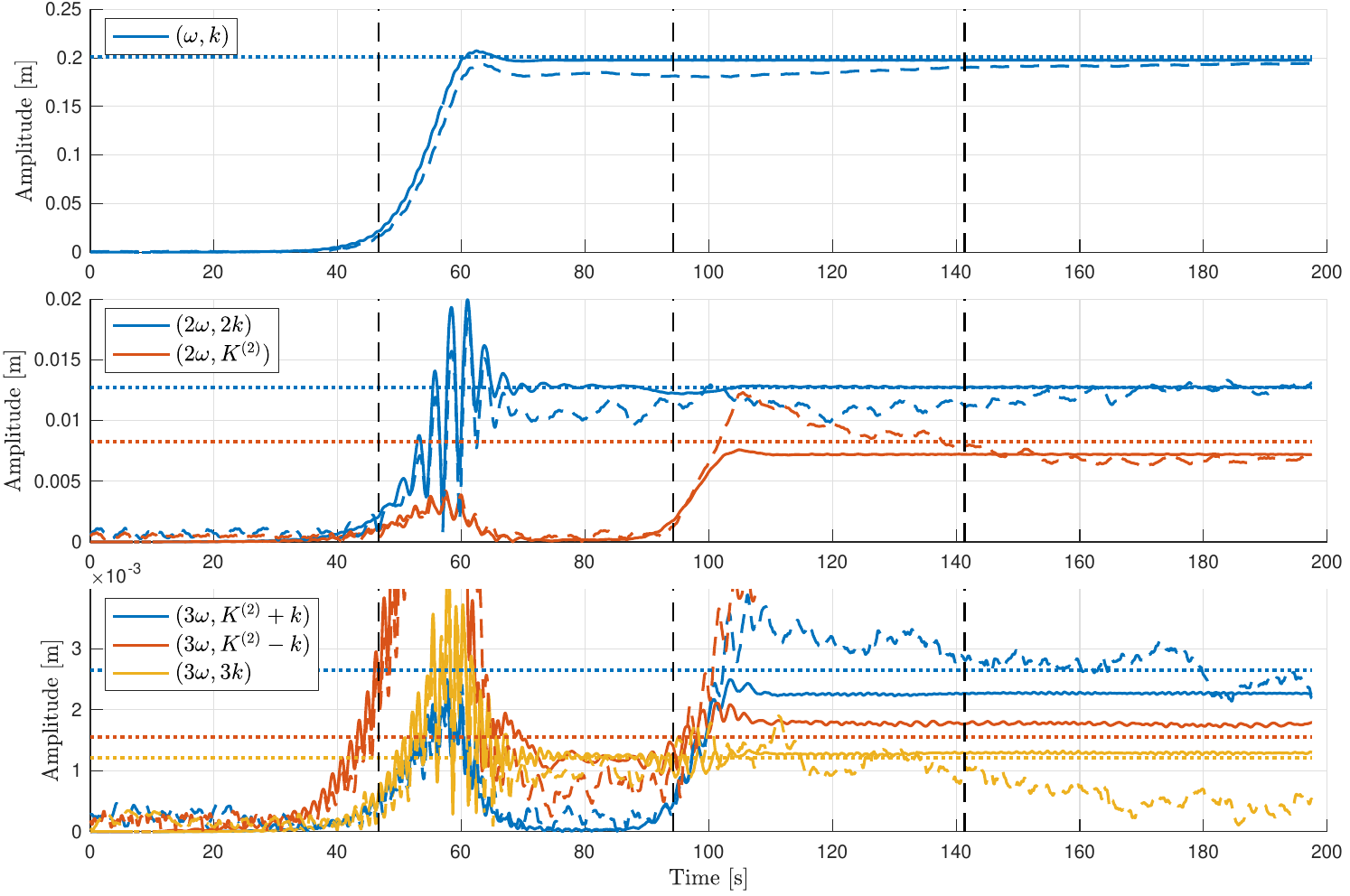}}%
		\subfloat[70040; $T=3.0$\,s, $H=0.40$\,m.]{\includegraphics[width=.5\columnwidth]{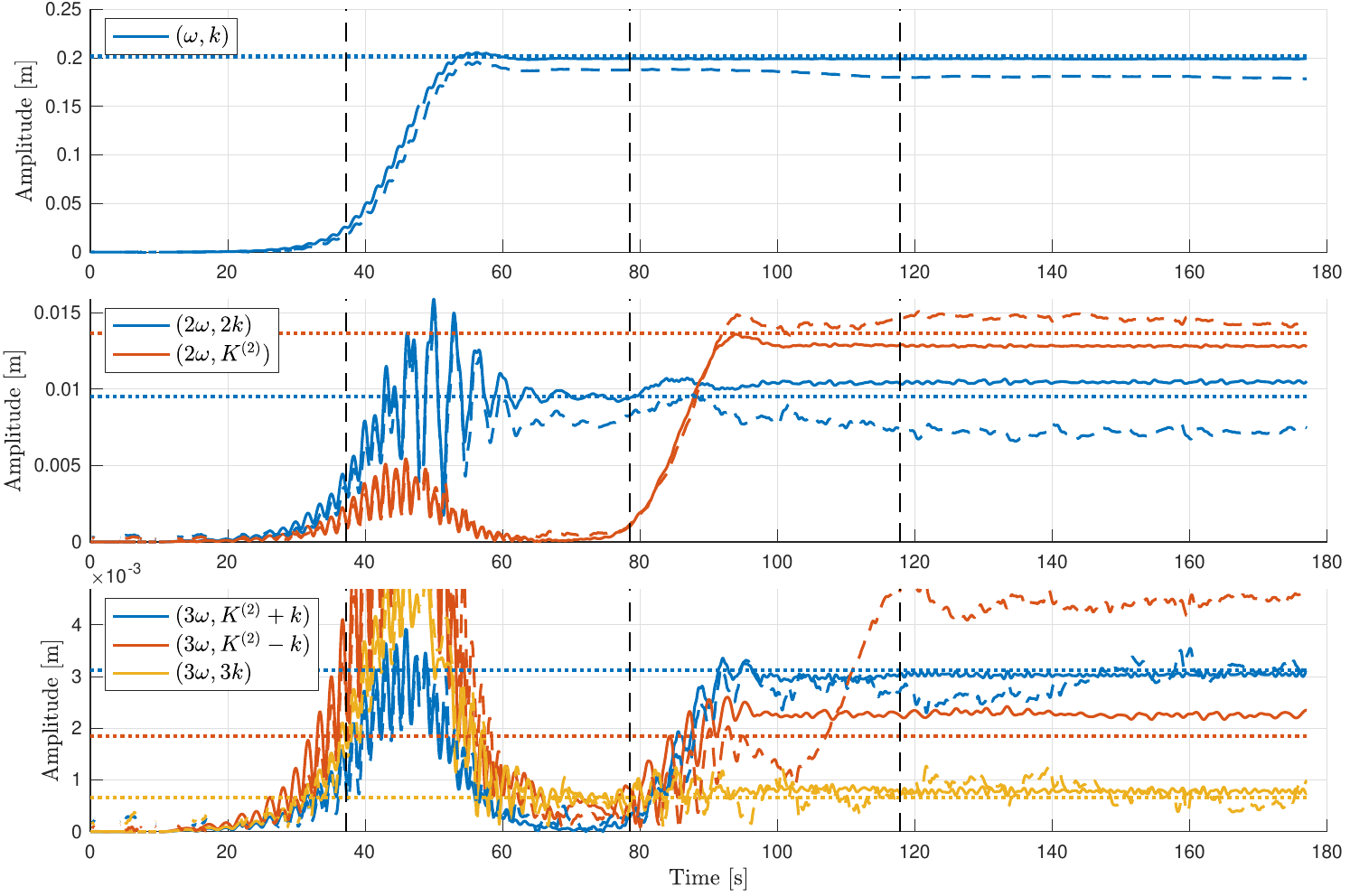}}
	\end{figure}

\begin{figure}[H]
		\centering
		\subfloat[Case 70040; $T=3.0$\,s, $H=0.40$\,m, $h=5.415$\,m .]{%
\includegraphics[width=.5\columnwidth]{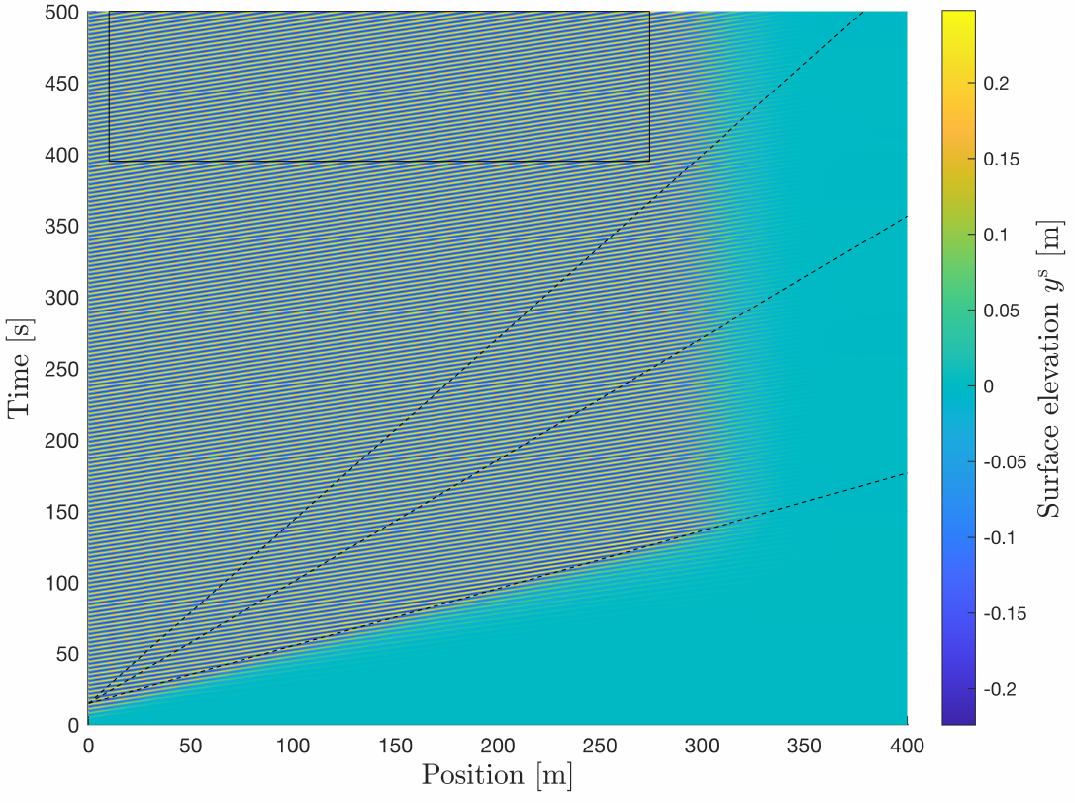}%
\includegraphics[width=.5\columnwidth]{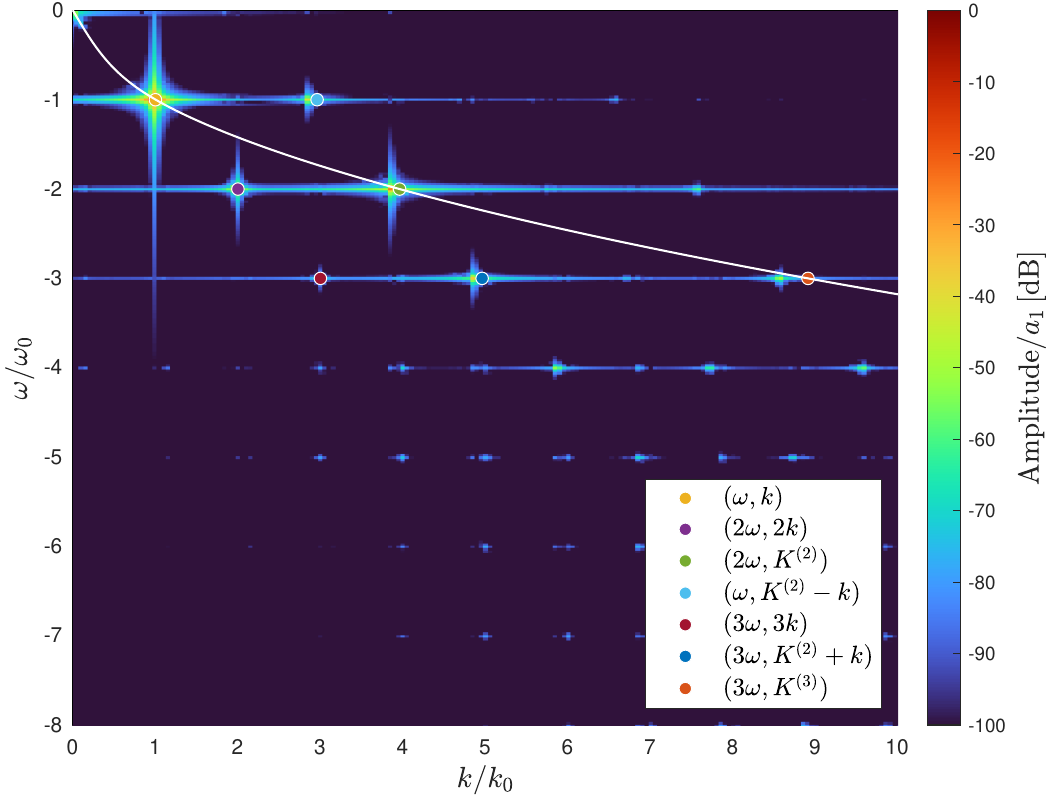}%
\label{fig:starMap:flap}
}\\
		\subfloat[Shallow water piston, $T=3.0$\,s, $h=1.0$\,m, 7.5\,m stroke. ]{
\includegraphics[width=.5\columnwidth]{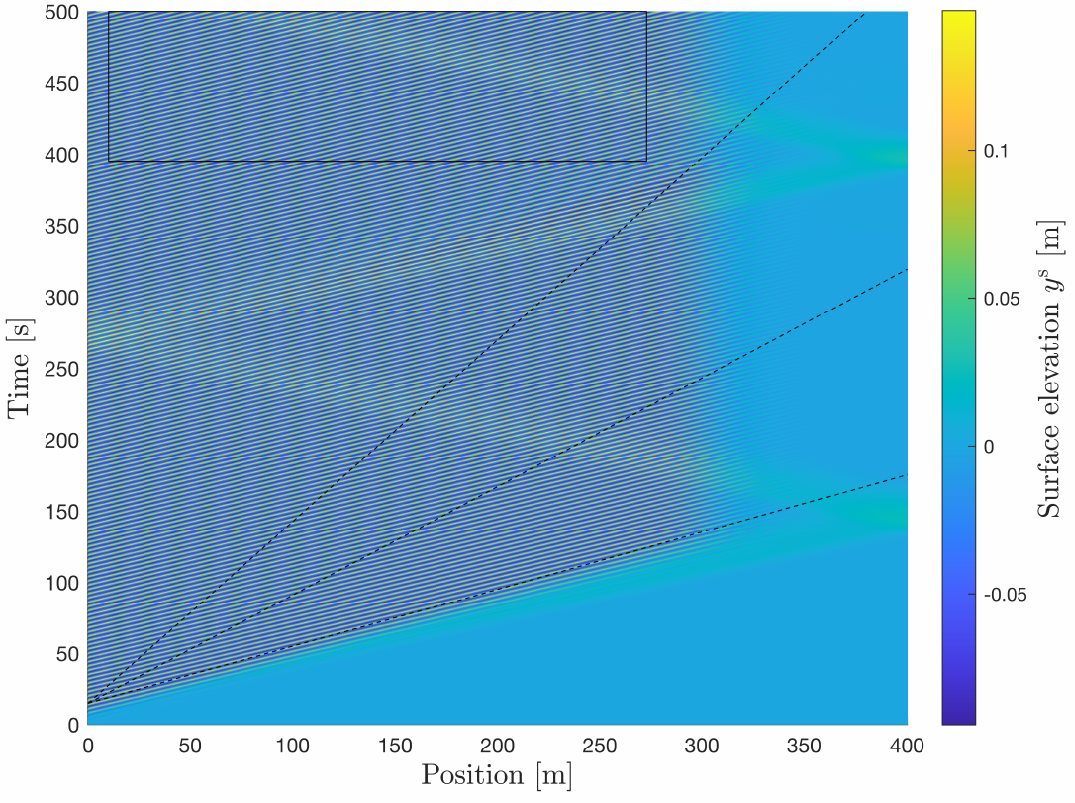}%
\includegraphics[width=.5\columnwidth]{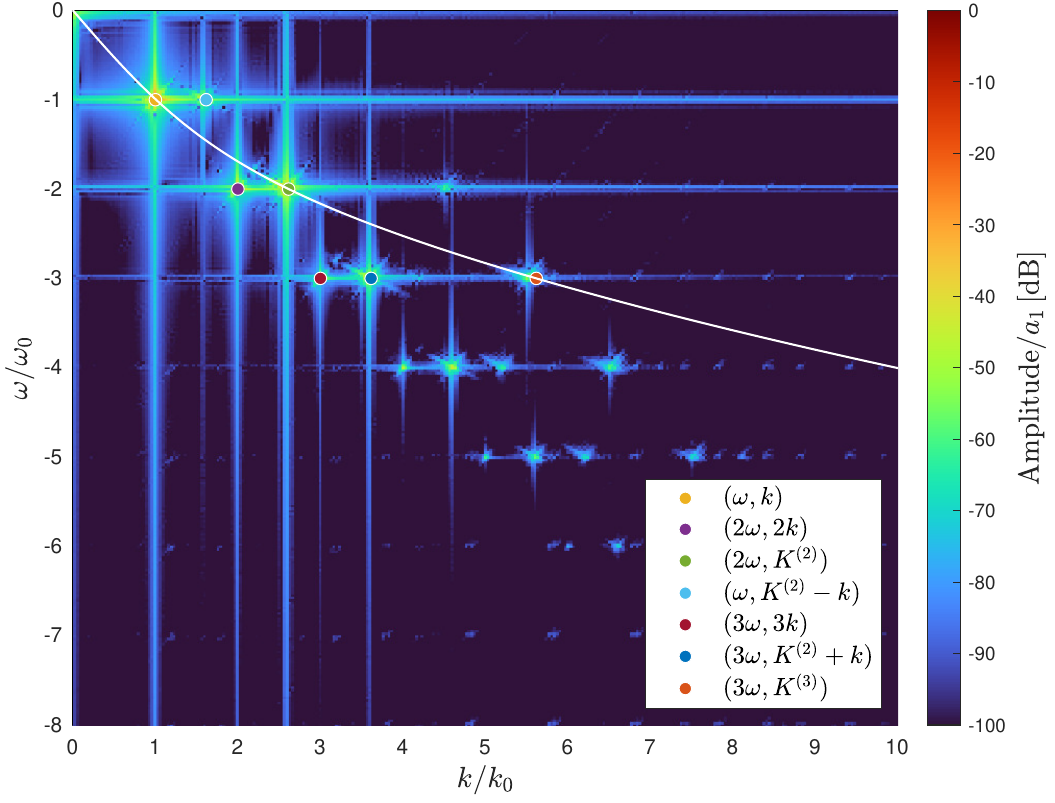}%
\label{fig:starMap:piston}
}%
		\caption{The spectral content of generated wave trains; modal amplitude contained within the spatio-temporal rectangular window shown in the left panels. Dashed lines indicate the group velocities of principal, secondary and tertiary free waves components.
		The white line in the right panel shows the dispersion relation, and round markers indicate the main harmonics according to perturbation theory. 
}
		\label{fig:starMap}
	\end{figure}

The full spectral content is presented next. 
This is available  only from simulated data, where the spatial resolution is complete, and is shown for an intermediate- and a shallow-water case in  \autoref{fig:starMap}.
Displayed are spatio-temporal Fourier decompositions contained within windows encompassing free wave components up to third order. 
Being more nonlinear, the shallow-water case exhibits a greater concentration of high-order modal energy than the intermediate-water case.
Spectral leakage is a prominent feature in spectral plots, and to improve clarity, a Kaiser window has been applied.
Because windowing affects amplitude accuracy, the spectra should be interpreted qualitatively rather than quantitatively.

Included in \autoref{fig:starMap} is the dispersion relation, displayed as a white line, along with markers indicating the composition of free and bound wave components as predicted by the third-order dispersion relation. 
We see, particularly in the intermediate-depth case, that peaks of concentrated modal energy are slightly shifted relative to these markers.
The shift becomes more pronounced  with increasing wave steepness and corresponds to a discrepancy in phase velocity on the order of the surface Stokes drift.
Although somewhat mysterious, this observation aligns well with tertiary phase velocity modulations, originally postulated by \citet{longuet1962_phaseVelocityShift}. 
These are  induced by the principal wave train, which smaller free waves conceptually `ride' in an advective capacity.
To rule out the possibility of  a model-induced artifact, the same observation was independently confirmed in additional simulations using the HOS method, employing a linearised wavemaker boundary condition.
\\

\begin{figure}[H]
		\centering
		\subfloat[Case 70040.]{\includegraphics[width=.5\columnwidth]{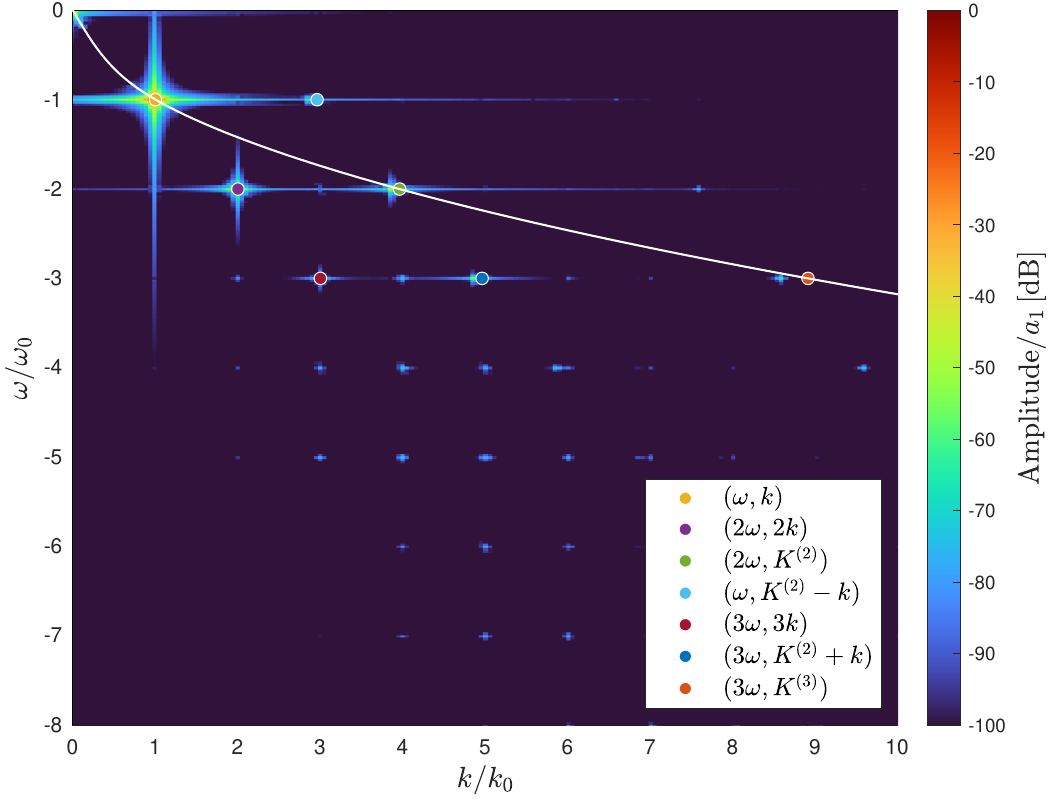}\label{fig:starMapCorr:flap}}%
		\subfloat[Shallow water piston.]{\includegraphics[width=.5\columnwidth]{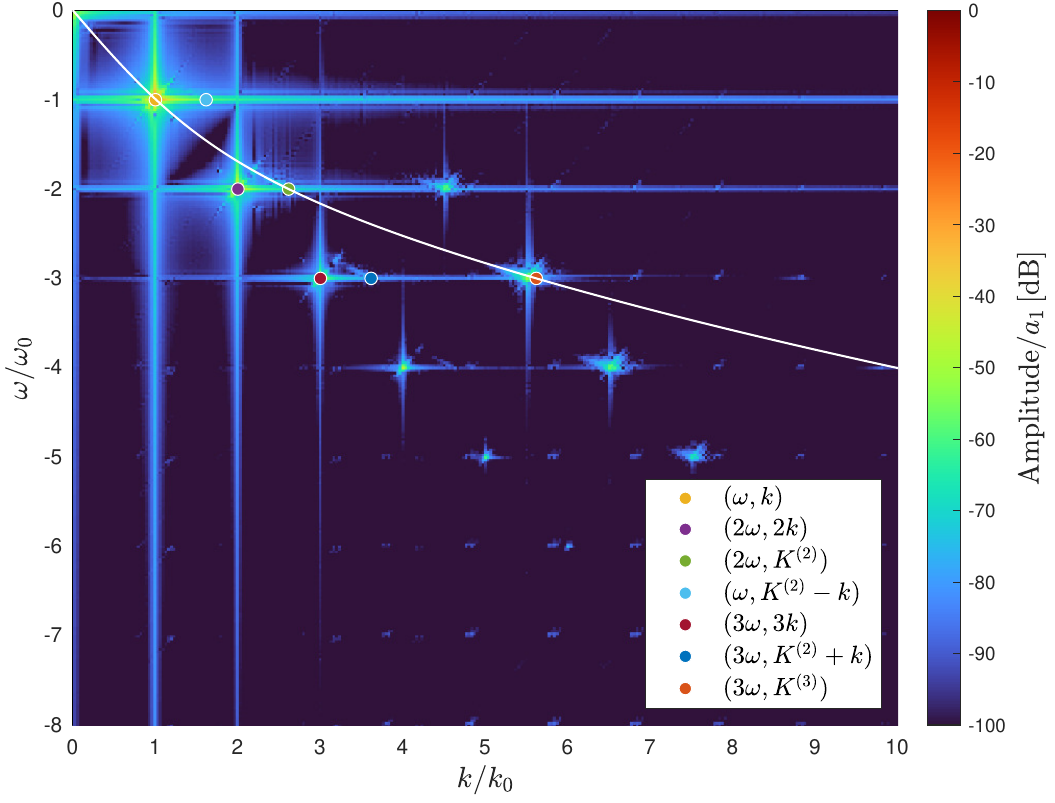}\label{fig:starMapCorr:piston}}%
		\caption{Cases corresponding to \autoref{fig:starMap} but with Sch{\"a}ffer's second order wave correction.}
		\label{fig:starMapCorr}
	\end{figure}

The wavemaker signal correction  proposed by \citet{schaffer_1996} is now widely implemented in wave generation systems to suppress second-order spurious waves. 
Its effectiveness  is evident in \autoref{fig:starMapCorr}, which repeats  the previous simulations with the corrected paddle motion applied. 
In the intermediate water depth case, both second- and third-order spurious wave amplitudes are significantly reduced.
Only the second-order spurious wave is reduced in the shallow-water case, the third-order free wave remaining strong or even magnified.
\\

\subsection{Time series of irregular waves}
\label{sec:exp:irr}

\begin{figure}[H]
	\centering
	\subfloat[80121; $T\_p=1.5$\,s, $H\_s=0.15$\,m.]{\includegraphics[width=.95\columnwidth]{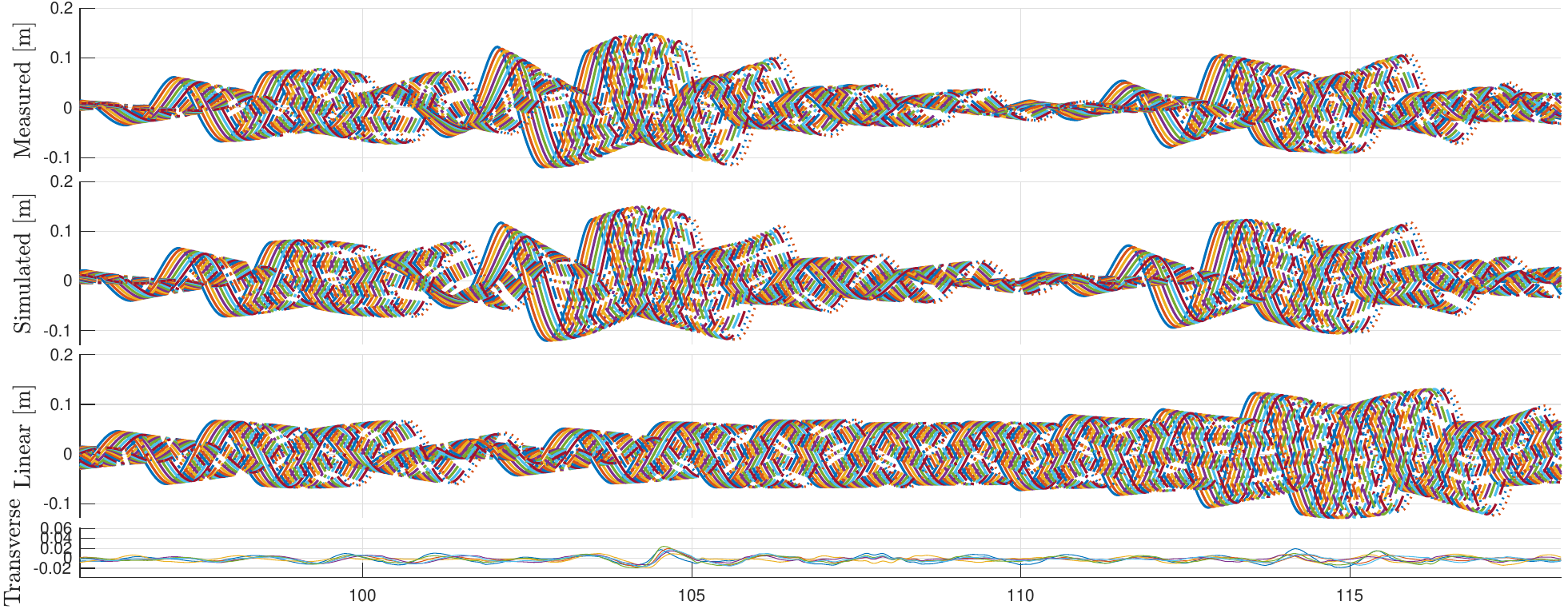}}\\
	\subfloat[80103; $T\_p=2.0$\,s, $H\_s=0.20$\,m.]{\includegraphics[width=.95\columnwidth]{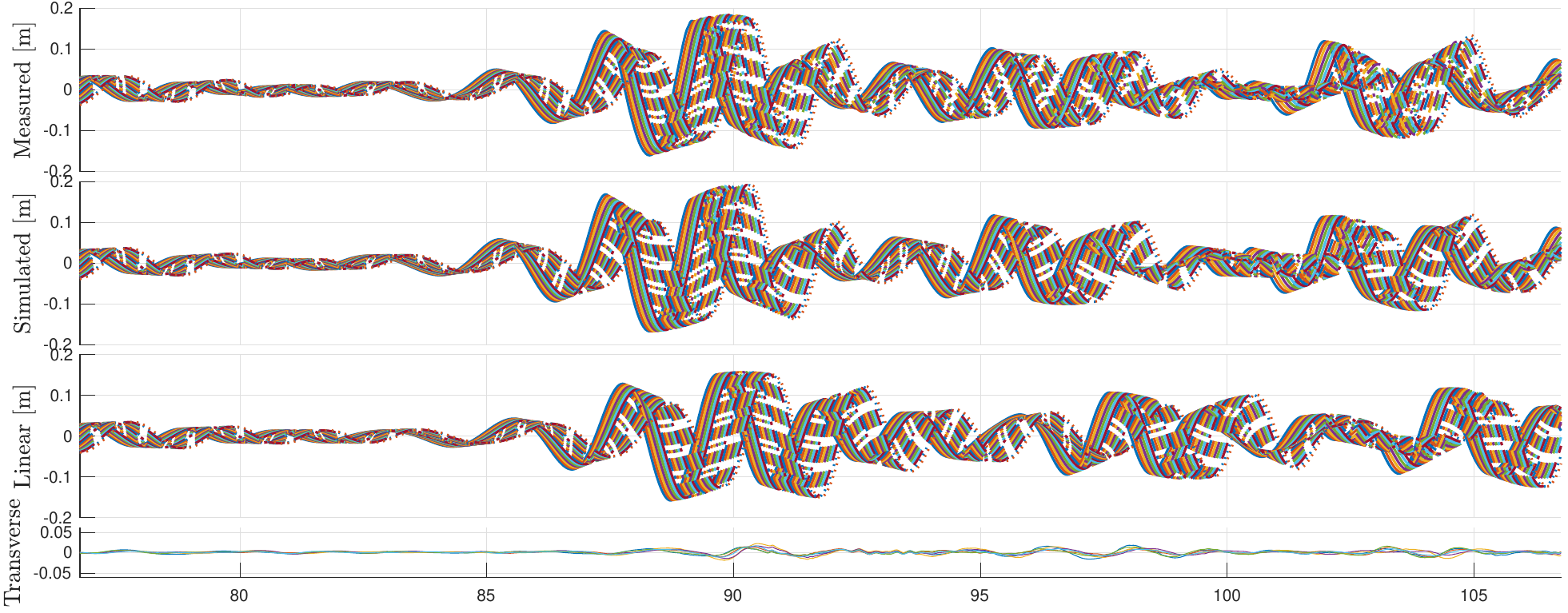}\label{fig:expIrr:refcase}}\\
	\subfloat[80112; $T\_p=2.5$\,s, $H\_s=0.25$\,m.]{\includegraphics[width=.95\columnwidth]{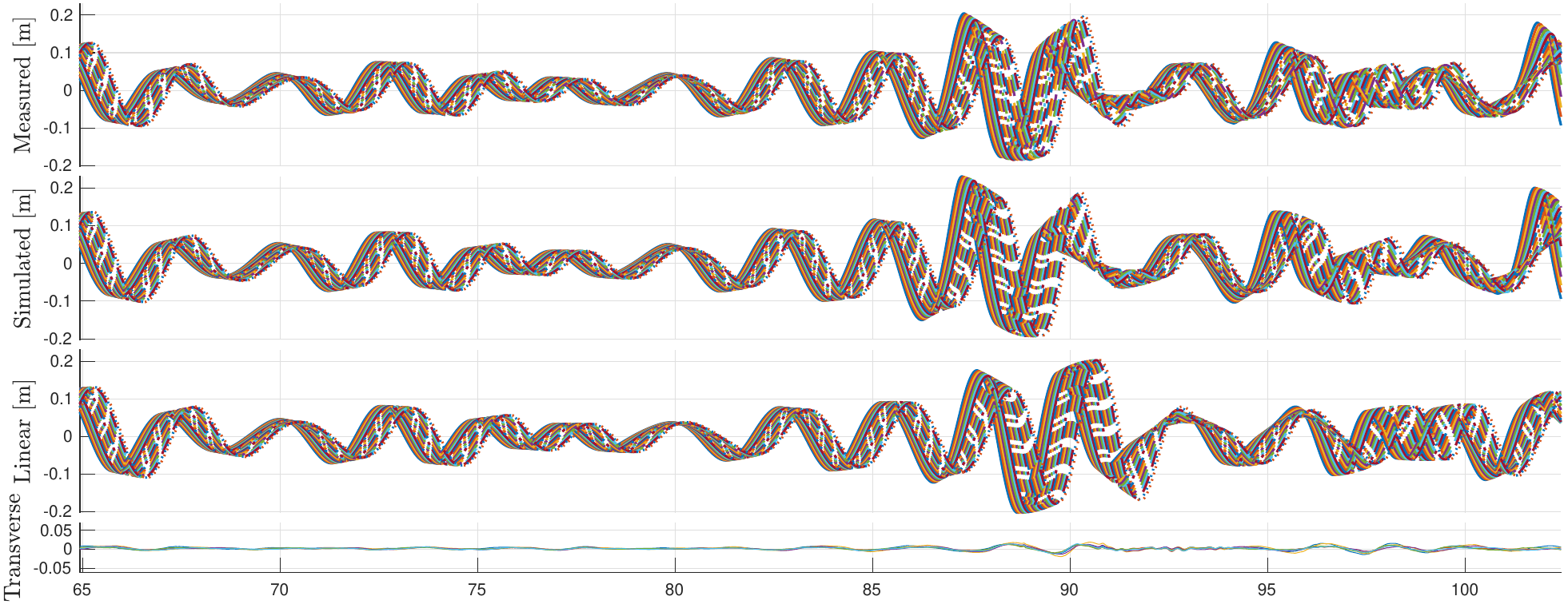}}
	\caption{Irregular wave time signals measured with wave gauge harp, 90 meters downstream.
					Top panel: experiments. Second from top: simulations. Third panel: signal predictions from linear wavemaker theory. Bottom panel: Difference between transverse wave gauge signal and middle gauge reference. Passive modal damping coefficient $\kd=0.5\,\kMax$, $\rDamping=0.025$.}
	\label{fig:expIrr}
\end{figure}

\begin{figure}[H]
	\centering
	\subfloat[During passage of wave train front.]{\includegraphics[width=1\columnwidth]{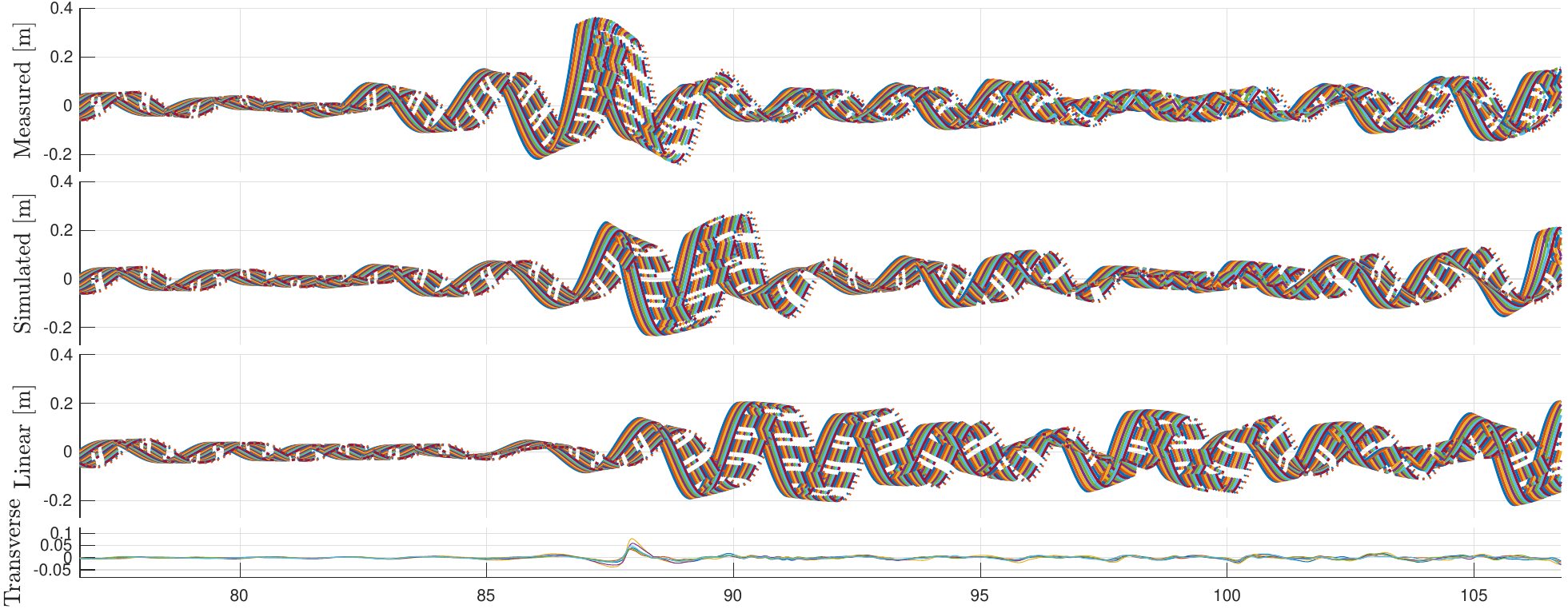}\label{fig:expIrr:steep:front}}\\
	\subfloat[Later in signal.]{\includegraphics[width=1\columnwidth]{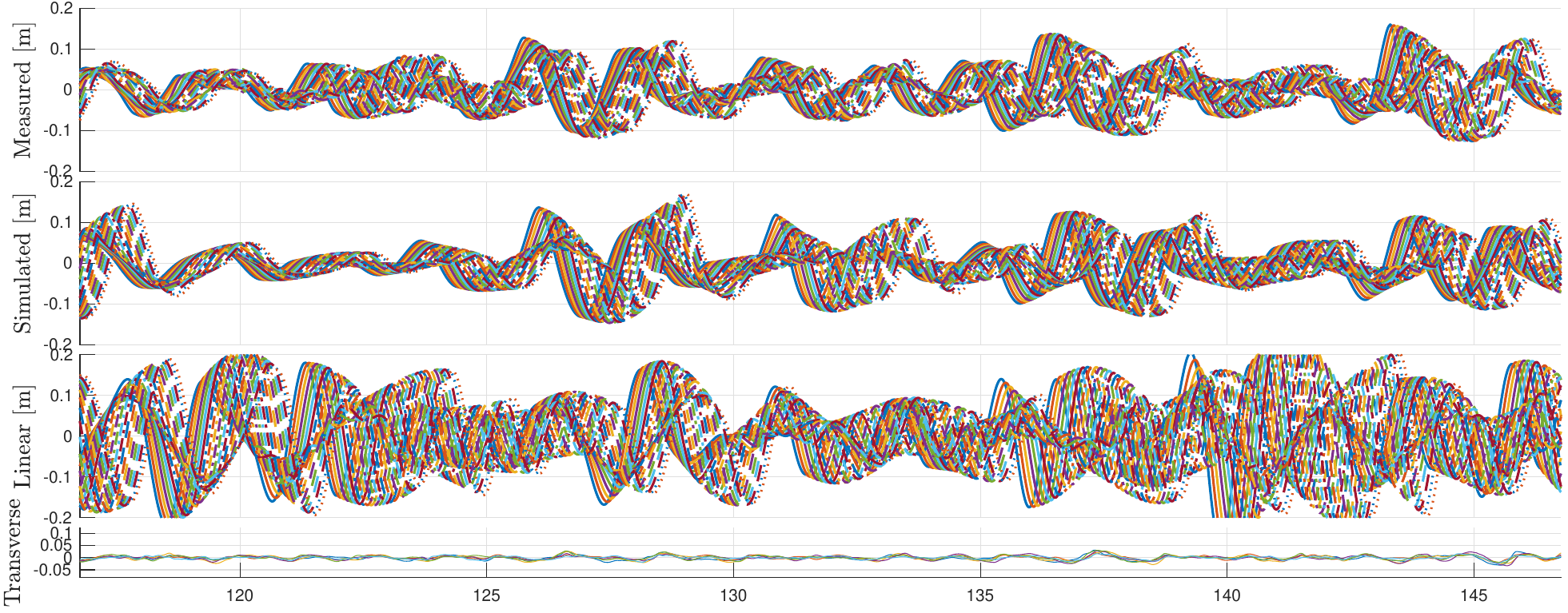}\label{fig:expIrr:steep:interior}}%
	\caption{Steeper case 80084; $T\_p=2.0$\,s, $H\_s=0.30$\,m and stabilising damping coefficients $\kd=0.25\,\kMax$, $\rDamping=0.25$, sampled at different times.}
	\label{fig:expIrr:steep}
\end{figure}

A direct comparison of irregular wave time series is presented in \autoref{fig:expIrr} and \ref{fig:expIrr:steep}.
All the wave gauge signals of the harp are included to illustrate the spatial variation of the wave field.
Each figure also shows predictions from linear wavemaker theory (the Bi{\'e}sel transfer function) to provide a reference for assessing the varying degree of nonlinearity in the different cases.
The waves shown in \autoref{fig:expIrr} are relatively gentle and can be simulated with passive modal damping parameters $\kd=0.5\,\kMax$, $\rDamping=0.025$.
In contrast, the steeper waves in \autoref{fig:expIrr:steep} exhibit slight wave front braking and require a broader damping range to maintain stability: $\kd=0.25\,\kMax$, $\rDamping=0.25$. 
The latter damping modifies the wave train front (comparing \autoref{fig:expIrr:steep:front} to \autoref{fig:expIrr:refcase}), though the interior of the train remains largely unaffected (\autoref{fig:expIrr:steep:interior}).
Minor spectral phase shifts are  visible in other signals; these appear insensitive to simulation parameters,  and their precise cause remains unknown.
However, it should be emphasised that the measurements are sampled 90 meters down the flume, where even small discrepancies can become significantly amplified.
Three-dimensional effects and the excitation of transverse sloshing modes are likely sources of discrepancy.
To monitor such directional components, signals from the transverse row of wave gauges are included in the bottom panel of the figure for reference.

\subsection{Energy spectra and wave calibration}
\label{sec:exp:spec}
We now shift our focus to wave spectra (often called power or variance spectra) in the perspective of numerical wave calibration.
Ideally, numerical wave tanks like the one presented here can serve as tools for calibrating wavemaker signals to a target wave spectrum, thereby reducing the need for time-consuming calibration tests in physical basins.
To evaluate this potential, wavemaker signals generated during the iterative steps of a calibration procedure are simulated, and the resulting wave spectra are compared to those observed experimentally.

\begin{figure}[H]
	\centering
	\includegraphics[width=.6\columnwidth]{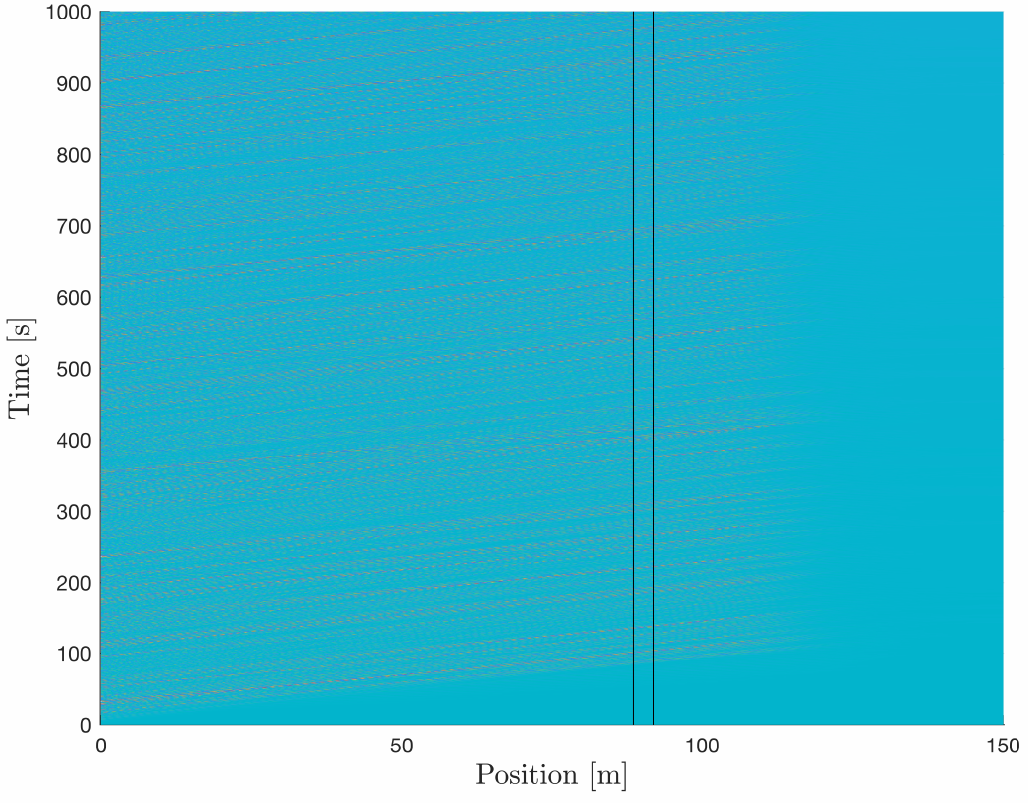}%
	\caption{Full test simulation of case 80060; $T\_p=1.5$\,s, $H\_s=0.17$\,m.}
	\label{fig:specWaveField}
\end{figure}

The time spent on one test iteration varies, typically ranging from 15 to 40 minutes. Each simulation produces a substantial amount of data, as illustrated in \autoref{fig:specWaveField}.  
To ensure consistent spatial resolution, a fixed number of grid points per characteristic wavelength $2\pi/k\_p$ is used,  $k\_p$ satisfying the linear dispersion relation for a peak frequency $2\pi/T\_p$.
This implies that computational cost increases with decreasing peak period.  
Cases presented here are simulated on a modern laptop with 50 points per wavelength, achieving speeds faster than real time but for the shortest peak period $T\_p=1.5$\,s.
Computational efficiency is discussed further in \autoref{sec:closingRemarks}.
\\

\begin{figure}[H]
	\centering
	\subfloat[80060--3; $T\_p=1.5$\,s, $H\_s=0.17$\,m.]{\includegraphics[width=.75\columnwidth]{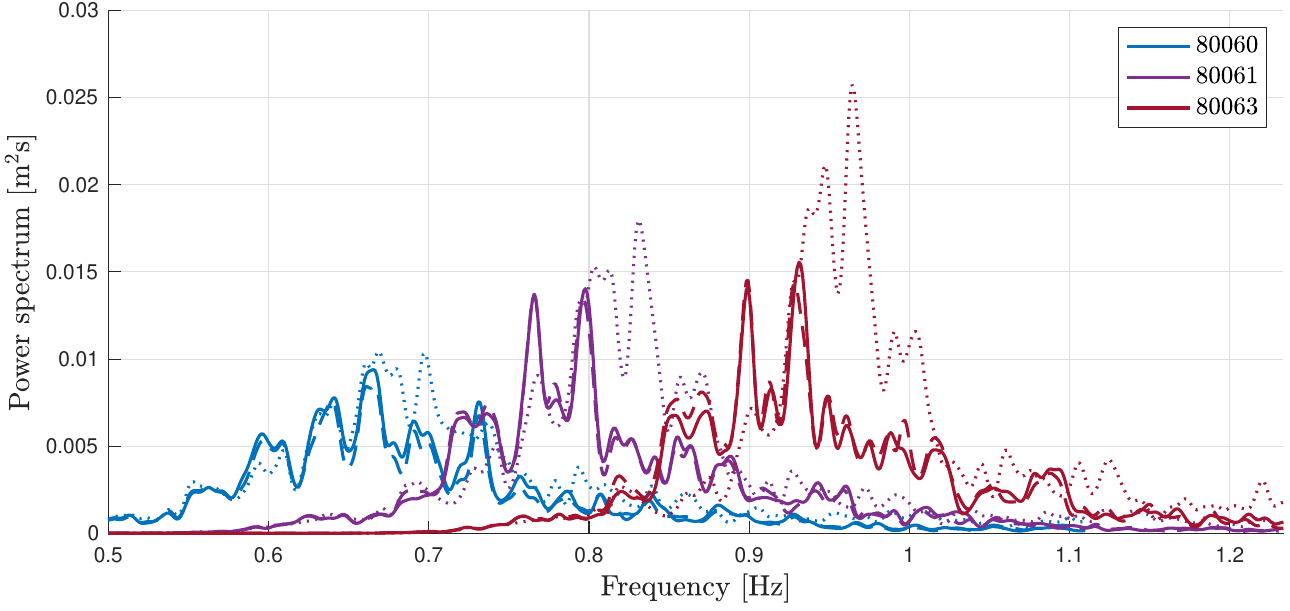}\label{fig:spec:Tp15}}\\
	\subfloat[80100--3; $T\_p=2.0$\,s, $H\_s=0.20$\,m.]{\includegraphics[width=.75\columnwidth]{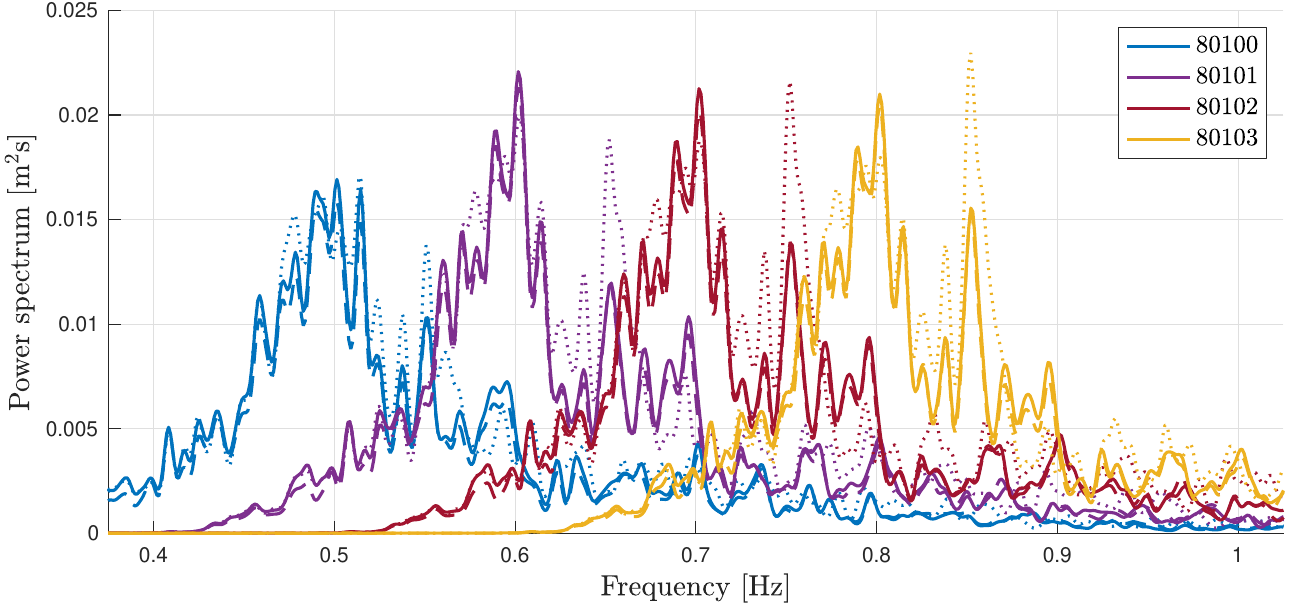}\label{fig:spec:Tp20}}\\
	\subfloat[81000--4; $T\_p=3.0$\,s, $H\_s=0.10$\,m.]{\includegraphics[width=.75\columnwidth]{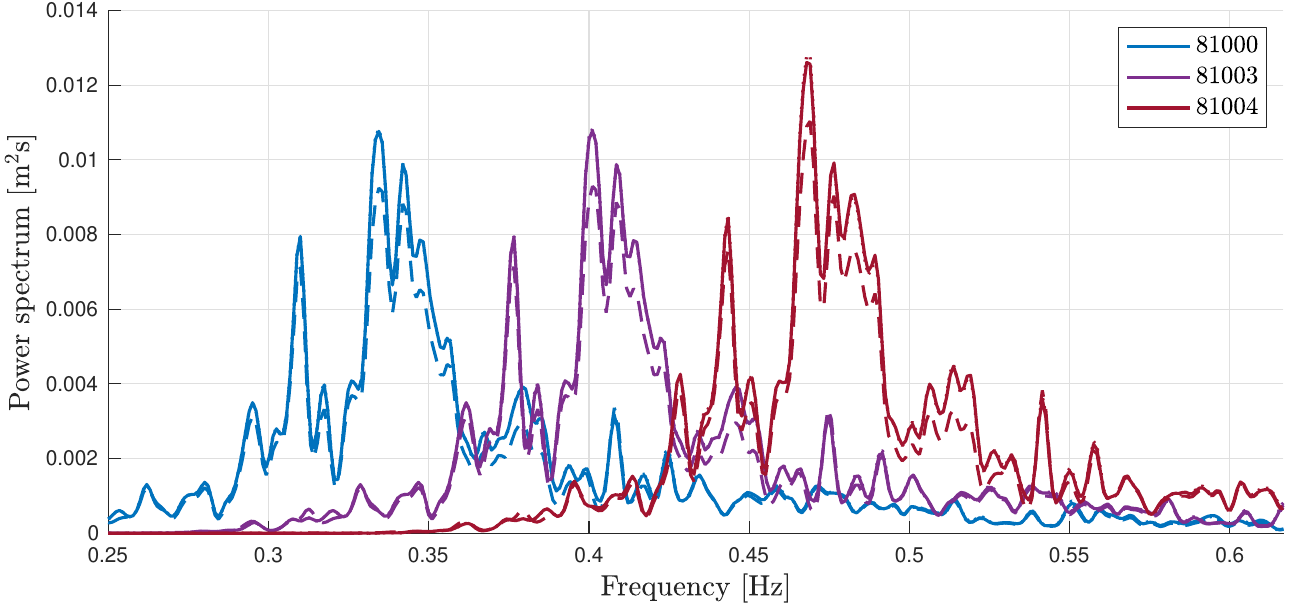}\label{fig:spec:Tp30}}%
	\caption{Power spectral density as measured during calibration procedure. 
Solid: simulated; dashed: experimental (partially overlapping); dotted: linear wavemaker theory. 
Curve frequencies are shifted $0.2/T\_p$ between each test for visibility. 
Wavemaker signals are identical in experiment, simulation and linear theory. Spectra subjected to Gaussian smoothing. Non-intrusive damping $\kd=0.5\,\kMax$, $\rDamping=0.025$.
}
	\label{fig:spec}
\end{figure}

Calibration test spectra are shown in \autoref{fig:spec} and are based on the mean spectrum across all harp wave gauges. 
The final digit of each test number indicates the corresponding iteration number of the calibration process.  
The applied calibration algorithm updates the wavemaker signal using a linear transfer function proportional to the observed discrepancies, also imposing constraints on response coherence and crest statistics.
Simulated wavemaker motions are identical to those applied during the 
Identical wavemaker motions are applied in simulations (solid lines) and experiments (dashed lines), calibrated based on the latter. 
For comparison, the power spectra resulting from linear wavemaker theory are shown with dotted lines. 
These are computed from the same wavemaker signals and are insensitive to modal phase.
All spectra are smoothed using a Gaussian filter of fixed width, chosen to suppress noise while preserving the stochastic frequency scattering that characterises the phase-resolved signals.

For the two shorter peak periods, $T\_p = 1.5$ and $2.0$\,s,  simulated and experimental signals match well but for some slight energy loss in the latter.
The linear signals, which do not account for the nonlinear evolution of the wave field, over-predict the power content at high frequencies and under-predict it in lower ones---a deviation which increasers with wave steepnesses.
 
A non-steep calibration case is presented in the final panel, \autoref{fig:spec:Tp30}.
In this case, the simulated wave spectra closely match the predictions of linear theory, whereas the measured power content is lower.  
This reduction is presumably the same energy loss observed in the previous cases, but here more apparent due to the lower wave amplitudes.
Physical dissipation should be minor in this period range, so a more plausible explanation for the energy discrepancy involves three-dimensional effects, the gaps between wavemaker paddles, or measuring inaccuracies; 
identifying the exact cause is beyond the scope of the present work, but we note that such discrepancies could potentially be corrected for using a simple transfer function.
\\

\begin{figure}[H]
	\centering
	\includegraphics[width=.75\columnwidth]{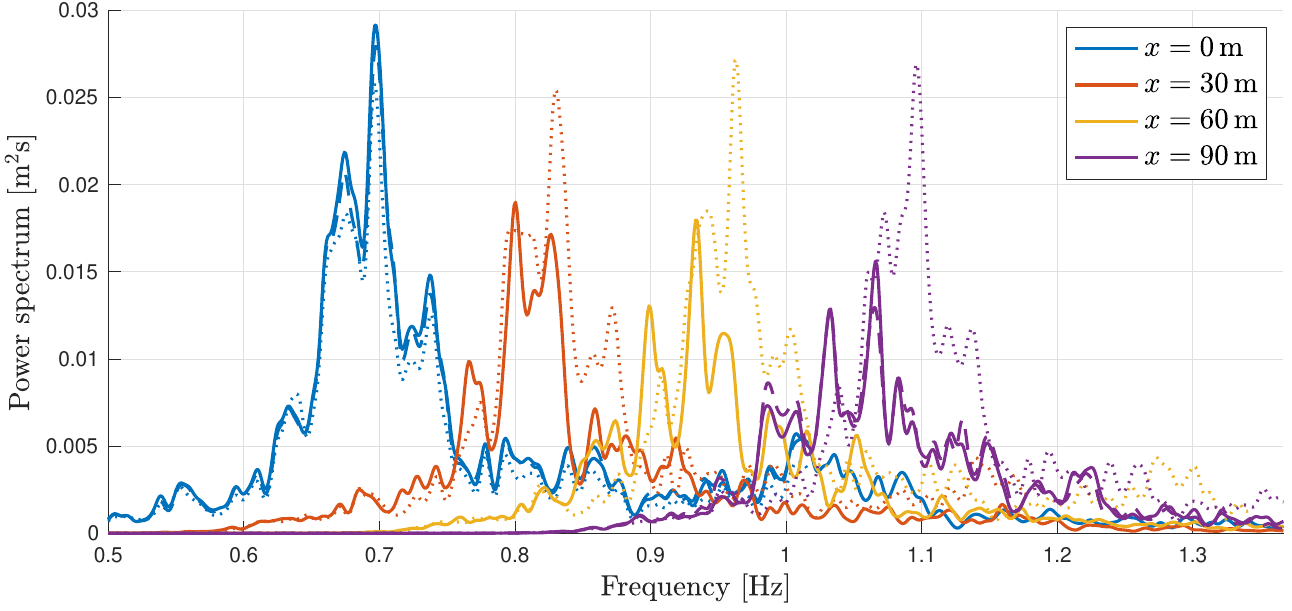}
	\caption{	Spatial variation of wave spectrum in test 80063 ($T\_p=1.5$\,s, $H\_s=0.17$\,m). 
  Solid: simulation; dashed: experimental;  dotted: linear wavemaker theory. (Variation in the latter is due to time window cropping.)
The mean signal from the wave gauges mounted on the paddle flaps represents the experimental measurement at $x=0$\,m  (closely overlapping simulated spectrum). No measurements are available for $x=30$ and $60$\,m.}
	\label{fig:specSpatial}
\end{figure}

The two main challenges of numerical  wave calibration are the precise wavemaker boundary motion  and the spatial evolution of the wave field, predominantly  due to tertiary modal interactions.
The latter is demonstrated in \autoref{fig:specSpatial}, displaying the calibrated test from \autoref{fig:spec:Tp15} at four spatial locations.  
The first measured location is at the wavemaker itself, using the mean signal from the wave gauges that are permanently mounted on the wavemaker paddles.  
In simulations, these signals are represented by the surface elevation at the moving boundary ($\zzz=0$).
Only the progressive mode is included in the approximation from linear theory, neglecting the near-field.
All three spectra are seen to match reasonably well at the wavemaker.

No experimental measurement is available at the intermediate locations until we reach the wave gauge harp 90 meters downstream.
At this location, the spectrum has evolved considerably, yet there is a remarkable agreement between simulation and experiment.
In contrast, linear theory predicts no spectral evolution but for the slight differences produced when cropping the time signal to avoid ramp effects. 
\\

\begin{figure}[H]
	\centering
	\includegraphics[width=.75\columnwidth]{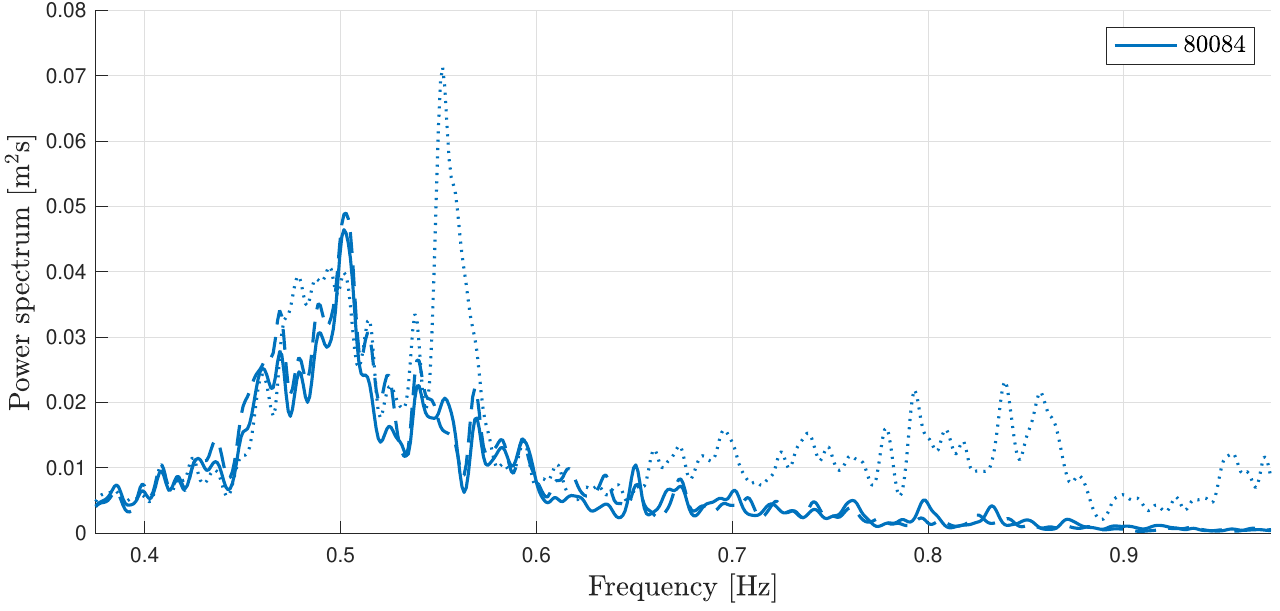}
	\caption{	Power spectrum of steeper case 80084; $T\_p=2.0$\,s,  $H\_s=0.30$\,m and stabilising damping coefficients $\kd=0.25\,\kMax$, $\rDamping=0.25$ (see~\autoref{fig:expIrr:steep}). Solid: simulation; dashed: experimental;  dotted: linear wavemaker theory.}
	\label{fig:specSteep}
\end{figure}

Finally, as with the phase-resolved comparison, the wave spectrum of the steeper wave, test 80084, is shown in \autoref{fig:specSteep}, again simulated with stabilising modal damping $\kd=0.25\,\kMax$, $\rDamping=0.25$.  
The time series of \autoref{fig:expIrr:steep} indicated that the damping has no notable effect on the energy distribution of the wave train interior.
Indeed, the  spectrum in \autoref{fig:specSteep} demonstrates strong agreement between simulation and experiment across a wide  frequency range.

\subsection{Statistical distribution}
\label{sec:exp:stat}

\begin{figure}[H]
	\centering
	\subfloat[80063; $T\_p=1.5$\,s, $H\_s=0.17$\,m (see \autoref{fig:spec:Tp15}).]{\includegraphics[width=.5\columnwidth]{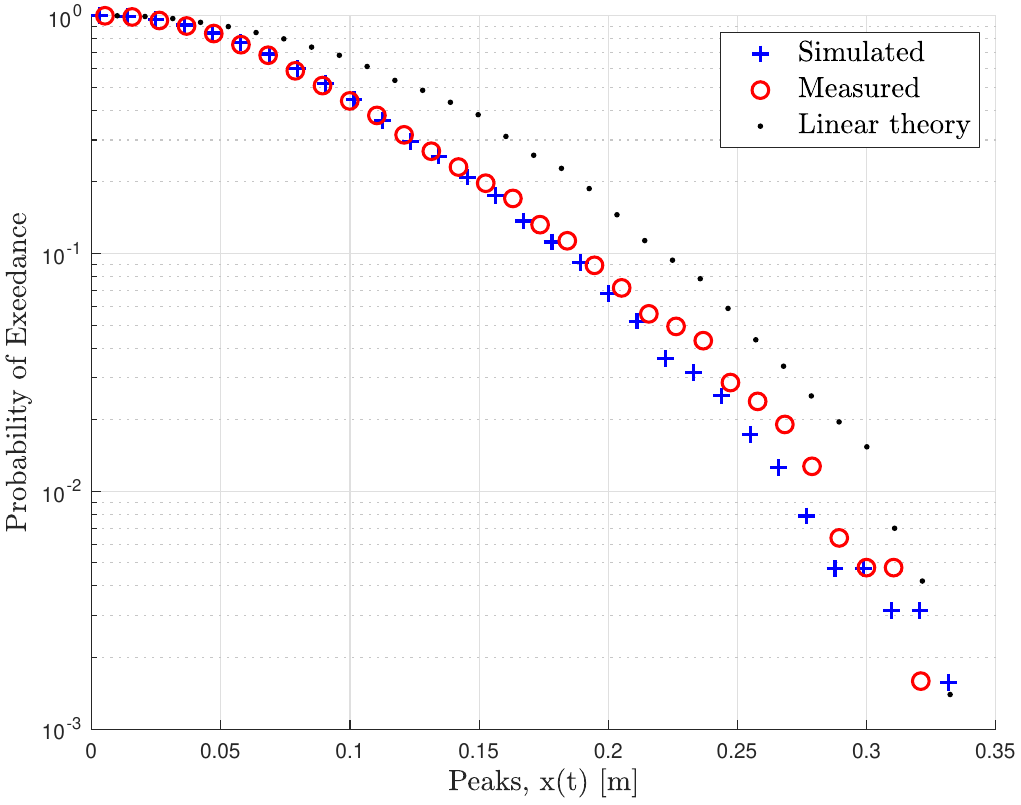}\label{fig:stat:Tp15}}%
	\subfloat[80103; $T\_p=2.0$\,s, $H\_s=0.20$\,m (see \autoref{fig:spec:Tp20}).]{\includegraphics[width=.5\columnwidth]{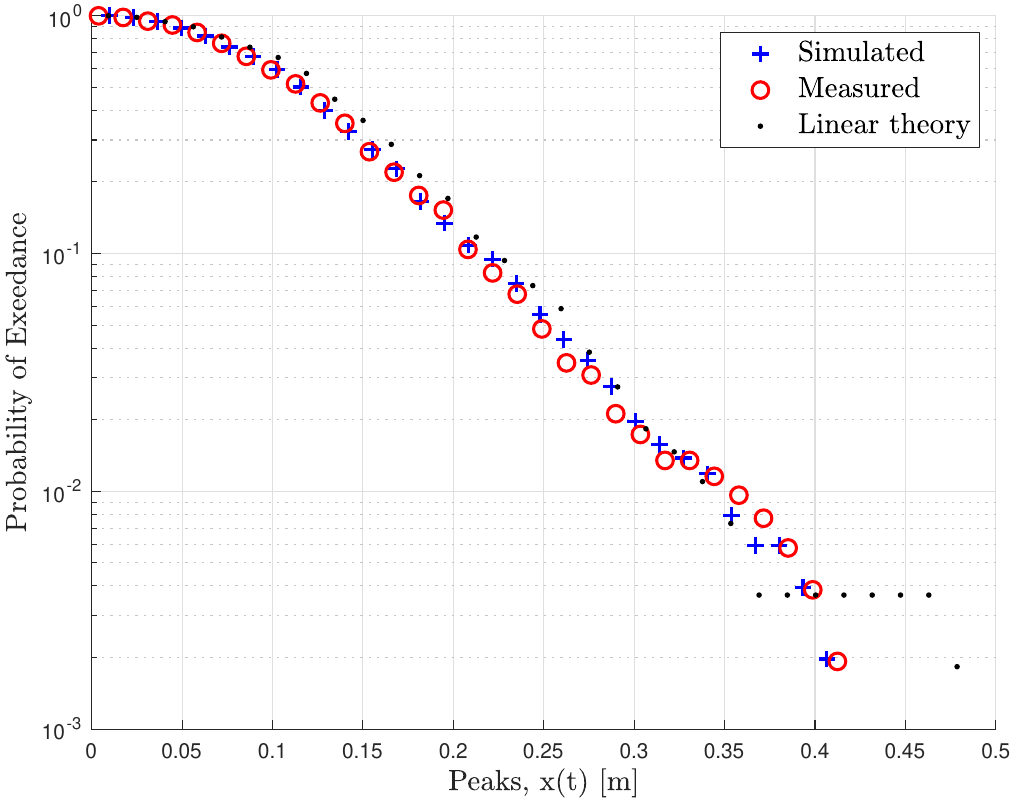}\label{fig:stat:Tp20}}\\
	\subfloat[81004; $T\_p=3.0$\,s, $H\_s=0.10$\,m (see \autoref{fig:spec:Tp30}).]{\includegraphics[width=.5\columnwidth]{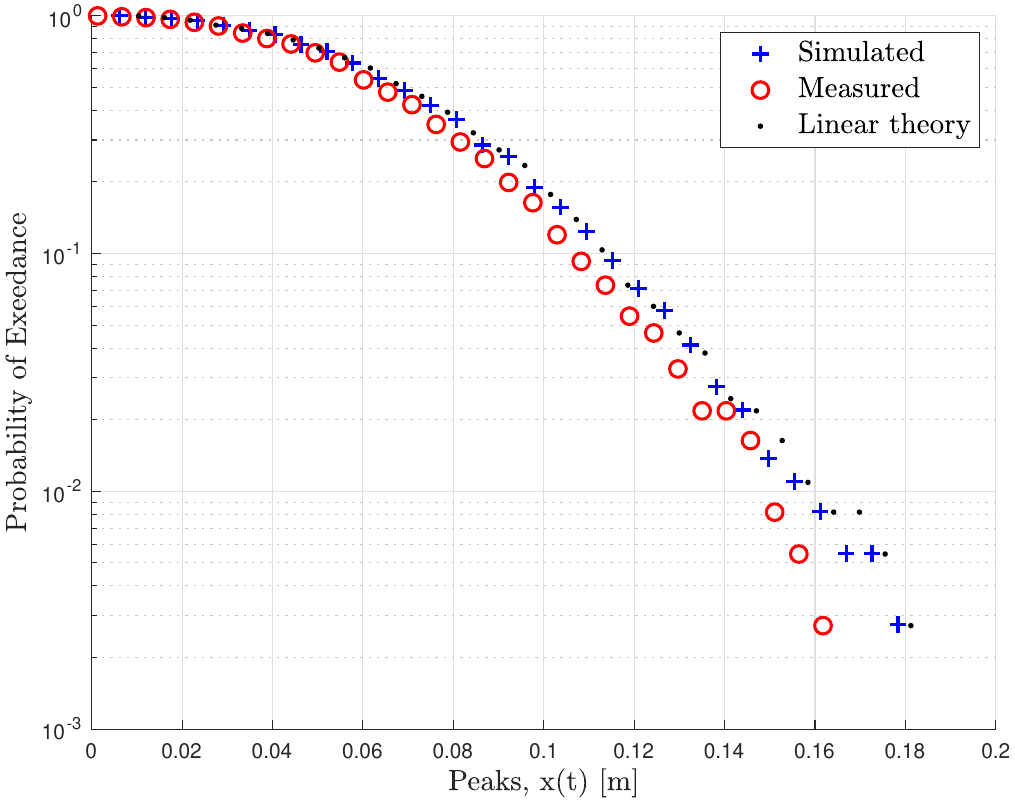}\label{fig:stat:Tp30}}%
	\subfloat[81084; $T\_p=2.0$\,s, $H\_s=0.30$\,m, stabilising damping (see \autoref{fig:specSteep}).]{\includegraphics[width=.5\columnwidth]{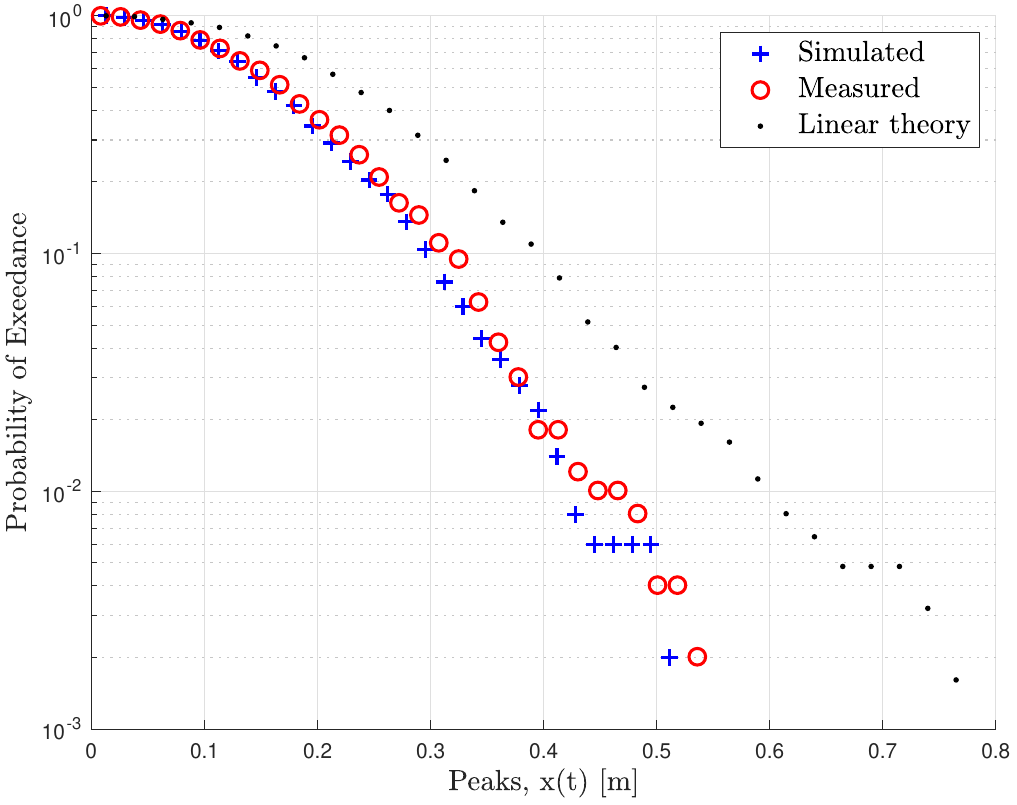}\label{fig:stat:Tp20Steep}}%
	\caption{Wave height distribution of four calibration tests as measured with the centre wave gauge at $x\approx90$\,m.}
	\label{fig:stat:heightDist}
\end{figure}

In this section, we consider the model's ability to recreate wave height statistics.
A full statistical comparison should encompass a large number of test realisations in order to establish a reliable distribution of extreme waves.
Lacking this,  we here only consider single realisations, focusing on comparative trends. 
\autoref{fig:stat:heightDist} presents the wave height distributions of four calibration tests, comparing experimental results with both numerical simulations and linear theory predictions from the same wavemaker signals. 
The latter will converge to a Rayleigh distribution upon realisation averaging.
Although we are used to seeing nonlinear crest distributions exceeding the Rayleigh distribution, we here observe the opposite trend because the wave spectra evolve before reaching the measuring location (see \autoref{fig:spec}). 
Distributions exhibit good agreement approximately up to the 95th percentile, beyond which the event sample size is insufficient.

The quality of prediction, compared to linear theory, is evident for the steeper wave cases in \autoref{fig:stat:Tp15} and \ref{fig:stat:Tp20Steep}, the later simulated with crest-stabilising modal damping.
For less steep wave conditions, the distributions more closely resemble the Rayleigh distribution, and in the mild case shown in \autoref{fig:stat:Tp30}, the energy discrepancy seen in \autoref{fig:spec:Tp30} becomes visible in the corresponding wave height statistics. 
The analysis algorithm allocates distribution bins based on the maximum observed wave height, which is seen to differs in each ensemble. 
Bins without observations appear as isolated points along horizontal lines.

\section{Closing remarks}
\label{sec:closingRemarks}
The idea of the conformally mapped model is to achieve superior efficiency by only evaluating surface dynamics in an explicit manner. 
In terms of performance, simulation time is strongly dependent on resolution, period, time stepping conditions, and modal damping, making precise statements on efficiency difficult. 
Broadly speaking, beyond real-time computation is observed in simulations presented here, except for the shortest periods considered. 
Computations are performed on a high-end laptop from the time of writing.

Simulations presented in \autoref{sec:validation} and \ref{sec:exp:recular} can be said to run in the order of a minute, while the calibration tests of \autoref{sec:exp:spec} required five to twenty minutes using 50 points per wavelength.
Simulation parameters have not been optimised for efficiency, and there is considerable potential for improving performance compared to the author's code, which is implemented in the MATLAB language. 
Foremost, with flap wavemakers, a major part of the simulation time  is devoted to interpolation on the maps in \eqref{eq:zmapTh} and \eqref{eq:signalDerivatives}. 
Specialised interpolation routines or polynomial map approximations are likely to increase efficiency. 
The modal stabilisation, which affects computation time by influencing the dynamic time stepping, can also be optimised for efficiency.
 \\

Wave calibration tests discussed in \autoref{sec:exp:spec} were emulative in nature, comparing wave spectra  generated by identical wavemaker motions.
This enabled a direct comparison of spectral content, independent of the calibration algorithm itself.  
In addition to this, a few numerically calibrated waves---computed entirely a priori---have  been tested in a separate campaign. 
When tested experimentally, these wavemaker signals met acceptability criteria without requiring further calibration.  
Although  omitted here for reasons of brevity and because the findings remain preliminary, these results further support the feasibility of calibration based solely on simulation.  
Even if simulation-based calibration should prove unreliable in some cases, it is likely to significantly reduce the number of iterations needed during physical calibration, thereby conserving valuable laboratory time.
\\

Although briefly considering wave height distributions, extreme event statistics have not been addressed here. 
The reason is the large number of test realisations needed to establish reliable tail-end statistics. 
Because it is generally infeasible to generate enough experimental data to encompass return periods of up to $10\,000$ years, estimates based on simulation alone are common. 
Today, several models of varying degrees of accuracy are often combined for such estimates, the work of \citet{gramstad2023_eventMatchingHOS} being a good example. 
However, as the formation of freak waves appears to be a highly nonlinear process, one should be cautious relying solely on approximate modelling---the conformally mapped model provides full nonlinearity but lacks three-dimensional effects. 
\citet{chalikov2009_freakWaves} as already studied freak waves using the present conformally mapped model, and identified that these occur are rapid and unpredictable events.
Although such studies are less relevant in the context of numerical wave tanks, physical model tests are commonly targeted towards extreme responses, making statistical similarity a valuable property for wave tanks.
\\

The ability to precisely predict phase-resolved experimental measurements at a distance is an ambitious objective. 
Clearly, if exact phase-resolved predictions were attainable, then both spectral and statistical properties would be inherently captured. 
However, even though direct signal comparisons are common in numerical model studies such as this, one may question the extent to which such comparisons are truly meaningful. 
At increasing distances, even slight modulations of the wave envelope can significantly alter the crests observed. 
These modulations may stem from a variety of sources, including diffractions from features in the basin and wavemaker, beach reflections, wave breaking dynamics, and the emergence of three-dimensional modulational instabilities. 
In fact, the latter were shown by \citet{mclean1981_3DBenjaminFair} to be an inherent characteristic of \emph{all} wave fields. 
Perhaps it is more appropriate to interpret direct comparisons as a form of process monitoring rather than as a strict benchmark. 
Ultimately, deciding whether it is the signal from experiment or the signal from simulation which is `more correct' becomes a matter of perspective and objective.

\section{Summary}
\label{sec:summary}

A wavemaker mapping scheme has been presented which enables the conformally mapped model framework in \citet{AHA2025_CMM}
to function as a numerical wave tank. 
The resulting formulation is shown to satisfy the kinematic boundary conditions along both the wavemaker and the free surface 
while confining numerical evaluations to the surface alone. 
Wavemaker characteristics are accurately captured, including the generation of spurious waves arising from the mismatch between the imposed paddle motion and the velocity field of progressive waves. 
The model was also shown to capture the appropriate return flow.

The model has been evaluated against experimental data for both regular and irregular waves. 
Phase-resolved signals show good agreement, even at considerable distances from the wavemaker. 
However, slight shifts in phase and carrier envelope are observed. 
These seem nearly unavoidable given the complexity of a three-dimensional physical wave flume. 
Additional discrepancies occur near crests that have undergone light breaking, although these appear localized to the affected crests and may be better captured through the inclusion of dedicated wave breaking modelling.

The numerical wave tank accurately predicts the wave spectral energy at the measurement location, as well as the spectrum's evolution along the flume. 
This serves as a testament to the fully nonlinear nature of the model and highlights a unique opportunity for numerical wave calibration in support of experimental campaigns. 
Statistical trends in wave height also appear to be well represented, although the present study has not considered large ensembles capturing the statistics of extreme event.

\section*{Acknowledgements}
This work has been internally financed by SINTEF Ocean.
The author is grateful to colleague S{\'e}bastien Lafl{\`e}che for sharing experimental data. 
Input from Kontorbamse was also appreciated. 
The language model of the AI tool ChatGPT has been used to improve the phrasing of this document.

	\bibliographystyle{abbrvnat} 
	\bibliography{sintef_bib}

\begin{thebibliography}{27}
\providecommand{\natexlab}[1]{#1}
\providecommand{\url}[1]{\texttt{#1}}
\expandafter\ifx\csname urlstyle\endcsname\relax
  \providecommand{\doi}[1]{doi: #1}\else
  \providecommand{\doi}{doi: \begingroup \urlstyle{rm}\Url}\fi

\bibitem[Akselsen(2025{\natexlab{a}})]{AHA2025_CMM}
A.~H. Akselsen.
\newblock A precise conformally mapped method for water waves in complex
  transient environments.
\newblock \emph{Journal of Computational Physics}, page 113848,
  2025{\natexlab{a}}.

\bibitem[Akselsen(2025{\natexlab{b}})]{AHA2025_multihinge}
A.~H. Akselsen.
\newblock Second-order theory for multi-hinged directional wavemakers,
  2025{\natexlab{b}}.
\newblock URL \url{https://doi.org/10.48550/arXiv.2502.09586}.
\newblock Under review in Coast.\ Eng.

\bibitem[Bonnefoy et~al.(2006)Bonnefoy, Le~Touz{\'e}, and
  Ferrant]{bonnefoy2006A_BM}
F.~Bonnefoy, D.~Le~Touz{\'e}, and P.~Ferrant.
\newblock A fully-spectral {3D} time-domain model for second-order simulation
  of wavetank experiments. {P}art {A}: {F}ormulation, implementation and
  numerical properties.
\newblock \emph{Applied Ocean Research}, 28\penalty0 (1):\penalty0 33--43,
  2006.

\bibitem[Bonnefoy et~al.(2010)Bonnefoy, Ducrozet, Le~Touz{\'e}, and
  Ferrant]{bonnefoy2010}
F.~Bonnefoy, G.~Ducrozet, D.~Le~Touz{\'e}, and P.~Ferrant.
\newblock Time domain simulation of nonlinear water waves using spectral
  methods.
\newblock In \emph{Advances in numerical simulation of nonlinear water waves},
  pages 129--164. World Scientific, 2010.

\bibitem[Chalikov(2009)]{chalikov2009_freakWaves}
D.~Chalikov.
\newblock Freak waves: Their occurrence and probability.
\newblock \emph{Physics of Fluids}, 21\penalty0 (7):\penalty0 076602, 07 2009.
\newblock ISSN 1070-6631.
\newblock \doi{10.1063/1.3175713}.
\newblock URL \url{https://doi.org/10.1063/1.3175713}.

\bibitem[Chalikov(2020)]{chalikov2020}
D.~Chalikov.
\newblock Numerical modeling of sea waves.
\newblock \emph{Izvestiya, Atmospheric and Oceanic Physics}, 56:\penalty0
  312--323, 2020.

\bibitem[Chalikov and Sheinin(1996)]{chalikov1996}
D.~Chalikov and D.~Sheinin.
\newblock \emph{Numerical modeling of surface waves based on principal
  equations of potential wave dynamics}.
\newblock US Department of Commerce, National Oceanic and Atmospheric
  Administration National Weather Service, 1996.

\bibitem[Chalikov and Sheinin(2005)]{chalikov2005modeling}
D.~Chalikov and D.~Sheinin.
\newblock Modeling extreme waves based on equations of potential flow with a
  free surface.
\newblock \emph{Journal of Computational Physics}, 210\penalty0 (1):\penalty0
  247--273, 2005.

\bibitem[Chalikov(2016)]{chalikov2016Book}
D.~V. Chalikov.
\newblock \emph{Numerical modeling of sea waves}.
\newblock Springer, 2016.

\bibitem[Clamond and Dutykh(2018)]{clamond2018accurate}
D.~Clamond and D.~Dutykh.
\newblock Accurate fast computation of steady two-dimensional surface gravity
  waves in arbitrary depth.
\newblock \emph{Journal of Fluid Mechanics}, 844:\penalty0 491--518, 2018.

\bibitem[Ducrozet et~al.(2020)Ducrozet, Bonnefoy, Mori, Fink, and
  Chabchoub]{ducrozet2020_timeReversalExperimentalRouge}
G.~Ducrozet, F.~Bonnefoy, N.~Mori, M.~Fink, and A.~Chabchoub.
\newblock Experimental reconstruction of extreme sea waves by time reversal
  principle.
\newblock \emph{Journal of Fluid Mechanics}, 884:\penalty0 A20, 2020.
\newblock \doi{10.1017/jfm.2019.939}.

\bibitem[Dyachenko et~al.(1996)Dyachenko, Kuznetsov, Spector, and
  Zakharov]{dyachenkoZakharov1996_confMap}
A.~I. Dyachenko, E.~A. Kuznetsov, M.~Spector, and V.~E. Zakharov.
\newblock Analytical description of the free surface dynamics of an ideal fluid
  (canonical formalism and conformal mapping).
\newblock \emph{Physics Letters A}, 221\penalty0 (1-2):\penalty0 73--79, 1996.

\bibitem[Dyachenko(2019)]{Dyachenko_2019_confMap}
S.~A. Dyachenko.
\newblock On the dynamics of a free surface of an ideal fluid in a bounded
  domain in the presence of surface tension.
\newblock \emph{Journal of Fluid Mechanics}, 860:\penalty0 408--418, 2019.
\newblock \doi{10.1017/jfm.2018.885}.

\bibitem[Gramstad et~al.(2023)Gramstad, Johannessen, and
  Lian]{gramstad2023_eventMatchingHOS}
O.~Gramstad, T.~B. Johannessen, and G.~Lian.
\newblock Long-term analysis of wave-induced loads using high order spectral
  method and direct sampling of extreme wave events.
\newblock \emph{Marine Structures}, 91:\penalty0 103473, 2023.

\bibitem[Houtani et~al.(2018)Houtani, Waseda, Fujimoto, Kiyomatsu, and
  Tanizawa]{houtani2018_experimentFromHOS}
H.~Houtani, T.~Waseda, W.~Fujimoto, K.~Kiyomatsu, and K.~Tanizawa.
\newblock Generation of a spatially periodic directional wave field in a
  rectangular wave basin based on higher-order spectral simulation.
\newblock \emph{Ocean Engineering}, 169:\penalty0 428--441, 2018.

\bibitem[Lamb(1932)]{lamb1932hydrodynamics}
H.~Lamb.
\newblock Hydrodynamics, 1932.

\bibitem[Longuet-Higgins and Phillips(1962)]{longuet1962_phaseVelocityShift}
M.~S. Longuet-Higgins and O.~M. Phillips.
\newblock Phase velocity effects in tertiary wave interactions.
\newblock \emph{Journal of Fluid Mechanics}, 12\penalty0 (3):\penalty0
  333--336, 1962.

\bibitem[Mansard and Funke(1980)]{mansard1980}
E.~P. Mansard and E.~Funke.
\newblock The measurement of incident and reflected spectra using a least
  squares method.
\newblock In \emph{Coastal Engineering 1980}, pages 154--172. 1980.

\bibitem[McLean et~al.(1981)McLean, Ma, Martin, Saffman, and
  Yuen]{mclean1981_3DBenjaminFair}
J.~McLean, Y.~Ma, D.~Martin, P.~Saffman, and H.~Yuen.
\newblock Three-dimensional instability of finite-amplitude water waves.
\newblock \emph{Physical Review Letters}, 46\penalty0 (13):\penalty0 817, 1981.

\bibitem[Milne-Thomson(1962)]{milneThomson1962theoreticalHydrodynamics}
L.~Milne-Thomson.
\newblock Theoretical hydrodynamics, 1962.

\bibitem[Ruban(2004)]{ruban2004}
V.~Ruban.
\newblock Water waves over a strongly undulating bottom.
\newblock \emph{Physical Review E—Statistical, Nonlinear, and Soft Matter
  Physics}, 70\penalty0 (6):\penalty0 066302, 2004.

\bibitem[Ruban(2005)]{ruban2005}
V.~P. Ruban.
\newblock Water waves over a time-dependent bottom: Exact description for 2d
  potential flows.
\newblock \emph{Physics Letters A}, 340\penalty0 (1-4):\penalty0 194--200,
  2005.

\bibitem[Sch{\"a}ffer(1996)]{schaffer_1996}
H.~A. Sch{\"a}ffer.
\newblock Second-order wavemaker theory for irregular waves.
\newblock \emph{Ocean Engineering}, 23\penalty0 (1):\penalty0 47--88, 1996.
\newblock \doi{10.1016/0029-8018(95)00013-B}.

\bibitem[Viotti et~al.(2014)Viotti, Dutykh, and Dias]{viotti2014conformal}
C.~Viotti, D.~Dutykh, and F.~Dias.
\newblock The conformal-mapping method for surface gravity waves in the
  presence of variable bathymetry and mean current.
\newblock \emph{Procedia IUTAM}, 11:\penalty0 110--118, 2014.

\bibitem[West et~al.(1987)West, Brueckner, Janda, Milder, and
  Milton]{west1987_originalHOS}
B.~J. West, K.~A. Brueckner, R.~S. Janda, D.~M. Milder, and R.~L. Milton.
\newblock A new numerical method for surface hydrodynamics.
\newblock \emph{Journal of Geophysical Research: Oceans}, 92\penalty0
  (C11):\penalty0 11803--11824, 1987.

\bibitem[Zakharov et~al.(2002)Zakharov, Dyachenko, and
  Vasilyev]{zakharov2002_conformalHOS}
V.~E. Zakharov, A.~I. Dyachenko, and O.~A. Vasilyev.
\newblock New method for numerical simulation of a nonstationary potential flow
  of incompressible fluid with a free surface.
\newblock \emph{European Journal of Mechanics-B/Fluids}, 21\penalty0
  (3):\penalty0 283--291, 2002.

\bibitem[Zakharov et~al.(2006)Zakharov, Dyachenko, and
  Prokofiev]{zakharov2006_conformalHOS}
V.~E. Zakharov, A.~I. Dyachenko, and A.~O. Prokofiev.
\newblock Freak waves as nonlinear stage of stokes wave modulation instability.
\newblock \emph{European Journal of Mechanics-B/Fluids}, 25\penalty0
  (5):\penalty0 677--692, 2006.

\end{thebibliography}

\appendix

\section{Flap wavemaker using  Schwarz--Christoffel transformation}
\label{sec:map:SC}

An alternative to the projection kernel approach for the wavemaker mapping described in \autoref{sec:map:flapProjKernel} is to make use of the Schwarz--Christoffel transformation theorem. 
In \citet{AHA2025_CMM}, a modification of this transformation formed the basis for a family of complex, sharply angled bathymetries.
To represent the corner angles of a flap-type wavemaker as illustrated in \autoref{fig:paddleBasic}, we introduce a bending of the $\z$-plane ordinate by an angle $\theta$ at $\zz = \pm \ii\DD$, and $-2\theta$ at $\zz = 0$. 
Then, to achieve a flat bottom boundary along $\z = -\ii H$, we repeat these angular modifications periodically at $\zz = 2\ii j \HH$, $j \in \mathbb{N}$.
According to the Schwarz--Christoffel theorem, the derivative of the resulting conformal map is then given by
\begin{equation}
	\zmap_{0,\zz}(\zz,t) 
	=\alpha \prod_{j=-\infty}^{\infty}\rbr{1+\frac{\DD^2}{(\zz+2\ii j \HH)^2}}^{-\theta/\pi}
	= \alpha \rbr{1+\frac{\sin^2\frac{\pi \DD}{2\HH}}{\sinh^2\frac{\pi\zz}{2\HH}}}^{-\theta/\pi}.
\label{eq:dfSC}%
\end{equation}
This can now be  numerically integrated:
\begin{equation*}
	\zmap(\zz+\Delta \zz)\simeq\zmap(\zz) +  \zmap_\zz(\zz+\Delta \zz/2)\,\Delta\zz,
\end{equation*}
with particular care taken to preserve numerical accuracy near  hinge-point singularities by appropriate choice of integration path.

As with the previous mapping approach, fixing the domain length $L=\LL$ necessitates an expansion factor $\alpha(t)$. The $\zz$-plane depth $\DD$, which maps to the desired hinge depth $D$, is also unknown.
Thus, an iterative procedure is again necessary to determine $\alpha$ and $\DD$ in order to get the correct domain length and hinge depth.
During each iteration step, \eqref{eq:dfSC} is integrated from $\zz=-\ii\HH$ to $-\ii\DD$ to evaluate $D$, and from $\zz=-\ii\HH$ to $\LL-\ii\HH$ to evaluate $L$.
Finally, the resulting mapping is shifted as described in \eqref{eq:ffPaddleShift}.
\\

The advantage of the Schwarz--Christoffel-type mapping over \eqref{eq:ffPadleNull} lie in its ability to avoid the noise near the bending points that arises in the latter due to  Gibbs' phenomenon. 
Additionally, it is not subject to the convergence limitations of \eqref{eq:ffPadleNull}, encountered when determining the mapping modes of $\XX$---roughly speaking, the iteration procedure in \autoref{list:iterations} is expected to converge as long as $|\theta|\lesssim 35\degree$.
The drawback of the Schwarz--Christoffel mapping is the care that must be taken during integration of \eqref{eq:dfSC} to maintain precision near hinge singularities. 
 The iteration scheme also becomes slightly more complex compared to the previous method, and the projection kernel is still needed for mapping the background  potential $\WW$ through \eqref{eq:WWFlap}.
$\WW$ will however become less noisy since noise is avoided in the projected function $\muw(\yy)$. 
That being said, noise can also be mitigated in the projection kernel approach by rounding the hinge corners of the flap profile $\X(\y)$.

\section{Code examples}	
\label{sec:listings}
\begin{lstlisting}[basicstyle=\ttfamily\small,label=list:iterations,caption={Implementation example (MATLAB syntax) of iteration loop that determines $\XX(\yy)$ for use in \eqref{eq:ffPadleNull} according to a prescribed lateral geometry $\X(\y)$. 
		The function \texttt{convKernel} is found in \autoref{list:CS} where the $y$-coordinate takes the palce of $x$. Function \texttt{interp1} performs a linear interpolation.
	}]
	function XX = findFlapCoordinates(yy,theta,DD,LL,n)
	% Assume unifor column vector yy from -HH to 0. 
	% Returns vector XX of function XX(yy).
	XX = 0*yy; HH = -yy(1);
	for i = 1:n
			alpha = 1-mean(XX([1:end,end-1:-1:2]))/LL; % eq.27, including mirror plane.
			D     = DD+(alpha-1)*HH;                   % eq.28
			X     = interp1([-100*HH,-D,0],[0,0,D*tan(theta)],yy); % shifting abscissa
			yMap  = alpha*yy+imag(convKernel('S',yy,XX,L,0)); % eq.26
			XX    = interp1(yy,X,yMap); % shift coordinates vertically
	end
\end{lstlisting}

\renewcommand{\^}{\slashHat}
\begin{lstlisting}[basicstyle=\ttfamily\small,label=list:CS,caption={Example function in MATLAB syntax returning projection mapping \eqref{eq:mapSC} (convolution operators) with option for derivation or integration. Mirroring assumed and highest frequency mode ignored (see \citet{AHA2025_CMM} otherwise).}]
	function convMu = convKernel(type,h,mu,z,n)
	% In:  mu: [nx,1]; z: [nx,ny]; h,n: [1,1]; type: char.  
	% Out: convMu: [nx,ny].
	% Grid assumed uniform in x and n is the derivative order.
	nx = size(nu,1); x  = real(z); y = imag(z); dx = x(2)-x(1);
	nx2 = 2*(nx-1); % Size of mirrored doamin
	y = y([1:nx,nx-1:-1:2],:);   % eq.16
	mu = mu([1:nx,nx-1:-1:2],:); % eq.16
	dk = 2*pi/(nx2*dx);
	k  = [0:nx2/2-1,-nx2/2:-1]'*dk;
	if type == 'C'
			sign = +1;
	elseif type == 'S'
			sign = -1;
	end
	kernel = 2./(exp(k.*y)+sign*exp(k.*(2*h+y))); kernel(1,:) = (n>=0); % eq.8
	kernel(nx,:) = 0; % ignoring highest frequency
	convMu = ifft(fft(mu).*(1i.*k).^n.*kernel);% evaluation including derivative
	if n < 0
		convMu = convMu+mean(convMu,1).*z.^(-n)/factorial(-n);% integrating zero-mode
	end
	convMu = convMu(1:nx,:); % remove mirror plane 
\end{lstlisting}
\renewcommand{\^}[1]{^\mr{#1}}

\end{document}